\documentclass[aps,prc,superscriptaddress,reprint,nofootinbib,preprintnumbers]{revtex4-1}
\pdfoutput=1

\usepackage{epsfig}
\usepackage{dcolumn}
\usepackage{bm}
\usepackage{amssymb}
\usepackage{amsmath}
\usepackage{hyperref}
\usepackage[normalem]{ulem}

\usepackage{graphicx}
\usepackage[dvipsnames]{xcolor}
\usepackage{color}
\usepackage{float}

\usepackage{rotating}
\usepackage{bigints}
\usepackage{lineno}
\usepackage[normalem]{ulem}  
\usepackage{ulem}

\renewcommand\sout{\bgroup\color{blue} \ULdepth=-.5ex \ULset}

\begin{document}
\title{Description of the first order phase transition region\\ of an equation of state for QCD with a critical point}

\author{J. M. Karthein}
\email[Corresponding author: ]{jmkar@mit.edu}
\affiliation{Center for Theoretical Physics, Massachusetts Institute of Technology, Cambridge, MA 02139, USA}

\author{V. Koch}
\affiliation{Nuclear Science Division, Lawrence Berkeley National Laboratory, 1 Cyclotron Road, Berkeley, CA 94720, U.S.A.}

\author{C. Ratti}
\affiliation{Department of Physics, University of Houston, Houston, TX 77204, USA}

\preprint{MIT-CTP/5770}

\date{\today}

\begin{abstract}
We map the mean-field Ising model equation of state onto the QCD phase diagram, and reconstruct the full coexistence region in the case of a first order phase transition. Beyond the coexistence line, we maintain access to the spinodal region in the phase diagram, thus providing a description of metastable and unstable phases of matter as well.
In this way, our approach includes the super-heated hadronic phase and the super-cooled quark-gluon plasma, which are useful for hydrodynamic simulations of the fireball created in a heavy-ion collision at low collision energy, where a first order phase transition is expected. 
We discuss the features of the pressure and other thermodynamic observables as functions of temperature and baryonic chemical potential, in particular their behavior in the coexistence region. Finally, we compare our equation of state to other 3D-Ising model ones available in the literature.
\end{abstract}

\maketitle

\section{Introduction}
\label{sec:intro}

One of the open questions in the field of strongly interacting matter under extreme conditions is whether there is a critical point on the phase diagram of Quantum Chromodynamics (QCD), separating a crossover at low density \cite{Aoki:2006we} from a postulated first order phase transition at high baryon chemical potential $\mu_B$ (for recent reviews, see \cite{Bzdak:2019pkr,Du:2024wjm}). The search for the QCD critical point has been one of the main goals of the Second Beam Energy Scan (BES-II) at the Relativistic Heavy-Ion Collider (RHIC), which concluded its runs in 2021 and recently presented results on the net-proton cumulants at low collision energies. While these results do not clearly rule out or confirm the existence of the critical point, they hint at an interesting behavior for energies $\sqrt{s_{NN}}\leq$ 20 GeV, which calls for theoretical interpretation and understanding.

An essential component in the search for the QCD critical point is an equation of state featuring critical behavior, with flexible parameters (such as the location and strength of the critical point) which can be varied over broad ranges, eventually leading to Bayesian analyses with posterior distributions constraining these parameters. Guided by this principle, the BEST collaboration proposed a family of equations of state with a critical point in the 3D Ising model universality class \cite{Parotto:2018pwx,Karthein:2021nxe,An:2021wof}, which was recently extended to higher chemical potential in Ref. \cite{Kahangirwe:2024cny}. Similar approaches have been developed e.g. in Refs. \cite{Nonaka:2004pg,Plumberg:2018fxo,Kapusta:2022pny}. All these equations of state are based on the universal scaling of thermodynamic observables in the vicinity of the critical point, which in the case of QCD follows the 3D Ising model class \cite{Pisarski:1983ms,Rajagopal:1992qz}.
These thermodynamic properties are the first ingredient in a comprehensive simulation of the evolution of the fireball created in heavy-ion collisions at low collision energy. They are used as input in hydrodynamic simulations, which need to be modified for the propagation of critical fluctuations as well \cite{Stephanov:2017ghc,Herold:2022laa,Pradeep:2022mkf,Pihan:2022xcl}. For this purpose, it is crucial to provide a way to describe the thermodynamic observables on the coexistence line in the $(T,\mu_B)$ plane, which opens up into a metastable region in the $(T,n_B)$ plane, where $n_B$ is the net-baryonic density, the typical variable used in hydrodynamic simulations.

The coexistence line corresponds to the first order phase transition line, occurring at temperatures below the critical point one: $T<T_c$.
In addition to signals for the critical point, evidence for a first order phase transition would indicate a nontrivial structure within the QCD phase diagram. 
Moreover, while not a direct measurement of the critical point, the presence of a first order transition would imply its existence.
This motivates the need to understand signatures of this first order transition in addition to directly studying the effect of a critical point. 
A first order phase transition gives rise to a mechanically unstable, so-called spinodal region, where the system rapidly breaks into the two phases via spinodal decomposition \cite{Binder1984}, leading to clumping of matter into droplets of characteristic size \cite{Chomaz:2003dz}. 
Given that the system created in a heavy ion collision expands rapidly, this rapid breakup is a welcome feature. 
Indeed, previous studies have shown that fluctuations due to spinodal decomposition could provide signatures of a first order transition \cite{Sasaki:2007db,Skokov:2009yu,Randrup:2009gp,Steinheimer:2012gc,Steinheimer:2019iso}. 
In addition, fast expansion of the system also helps in reaching the spinodal region, as has been demonstrated in \cite{Randrup:2009gp,Steinheimer:2012gc}.
Needless to say that,
in order to study these fluctuations, we need an equation of state that is able to access and characterize the features of the first order phase transition.

For this reason, in this paper we go beyond the coexistence line itself and study the coexistence region of the phase diagram, with its metastable and spinodal branches.
In doing so, we characterize the first order phase transition region with a thermodynamic approach.
We show that there is a unique path the system takes through the spinodals, alternatively to the Maxwell construction.
In fact, the finite-lifetime heavy-ion-collision system can be described by the progression through the coexistence and subsequent spinodal points, as opposed to a system in complete thermodynamic equilibrium which proceeds through the coexistence region by slowly varying the baryon density along the first order line.
The mean field Ising equation of state is utilized in our approach to study these features of the phase coexistence region, since the spinodals lie on the real axis in this case, while in the 3D Ising model they obtain complex values \cite{An:2017brc}.
This spinodal region has also been of interest in the nuclear equation of state at low temperature and finite density \cite{Wellenhofer:2015qba}.
Furthermore, spinodal breakup has been utilized to explore the nuclear liquid gas transition in experiment  (see \cite{Chomaz:2003dz} for a review).
For studies on the spinodal decomposition in a rapidly expanding quark-gluon plasma, see e.g. Refs. \cite{Randrup:2009gp,Steinheimer:2012gc,Kapusta:2024nii}.

The present paper is organized as follows: we first present the mean field Ising model in Sec. \ref{sec:Ising_EoS}, with specific emphasis on characterizing isothermal trajectories and also including a study of the Ginzburg criterion. 
We study how the features of the first order phase transition from the Ising model map onto the QCD phase diagram in Sec. \ref{sec:mapping} and study, in particular, how the spinodals depend on the parameters of the mapping. 
In Sec. \ref{sec:fullthermo} we reconstruct the full pressure in the QCD phase diagram and calculate further thermodynamics from the pressure including entropy and baryon density.
Finally, we conclude our work in Sec. \ref{sec:concl}.

\section{The Ising Equation of State}\label{sec:Ising_EoS}

\subsection{Mean Field Ising Model}
We map the Ising model phase diagram (in terms of reduced temperature $r=(T-T_c)/T_c$ and magnetic field $h$) onto the QCD one (in terms of temperature $T$ and baryonic chemical potential $\mu_B$), with the goal of introducing a critical point and first-order phase transition on top of the equation of state from lattice QCD, similarly to what was done in Refs. \cite{Parotto:2018pwx,Karthein:2021nxe,Kapusta:2022pny,Kahangirwe:2024cny}. 
Our main goal is 
to identify the spinodal lines in the QCD phase diagram as mapped from the Ising model, and provide a full description of the equation of state around the first-order phase transition.

When the transition is first order, the spinodal region is a distinctive feature of the QCD phase diagram, where quantities like entropy density, baryon density or energy density are multi-valued functions of the temperature at fixed chemical potential. Typically, only one of these branches is stable, while the others are metastable or unstable.
When considering isothermal trajectories in the phase diagram, the spinodal points, or spinodal singularities, demarcate the limits of metastability \cite{Fisher:1967,Binder1987}.
Furthermore, the spinodals are related to the Lee-Yang edge singularities in the Ising model, as studied in Ref. \cite{An:2017brc}. 
In the mean field Ising model, the spinodals are found at real values of the magnetic field $h$, i.e. they lie on the real $h-$axis.
On the other hand, beyond mean field they are displaced from the real $h-$axis by an amount \cite{An:2017brc}:
\begin{equation} \label{eq:Delphi}
    \Delta \phi = \pi (\beta \delta - \frac{3}{2}),
\end{equation}
where $\beta$ and $\delta$ are the critical exponents in the Ising model.
When the critical exponents are different from their mean field values of $\beta=1/2,~ \delta=3$, this phase is non-zero.
As is the case in the 3D Ising model, the spinodal points are pushed into the complex plane, owing to the importance of fluctuations which are neglected in the mean field approach.
The angle by which they are shifted into the complex plane is given by the difference between $(\beta \delta)_{3\mathrm{D}} = 1.5648$ and the mean field value of 3/2, yielding $0.0648\pi$.
Given the smallness in the difference of these critical exponents, we take this as a motivation to proceed with the mean field approximation. 
Later in this section, we shall introduce the Ginzburg criterion and discuss the range of applicability of mean field.

The mean field equation of state is given by the relationship between the free energy (Helmholtz $F$ or Gibbs $G$), magnetization ($M$), and temperature ($r=(T-T_c)/T_c$).
The external magnetic field ($h$), used as a control parameter, is given in terms of $M$ and $r$.
In addition, the pressure ($P$) is equivalent to the Gibbs free energy up to a minus sign.
The Helmholtz free energy, magnetic field, and pressure are defined as:

\begin{align} 
    \centering
    F(M,r) &= h_0 M_0 \Big{(} \frac{1}{2}a r M^2 + \frac{1}{4}b M^4 \Big{)} \label{eq:MF_EoS_F}  \\
    h(M,r) &= \Bigg{(}\frac{dF}{dM}\Bigg{)}_r = h_0 (a r M + b M^3) \label{eq:MF_EoS_h} \\
    P(M,r) &= - G(M,r) = M h - F(M,r)   \label{eq:MF_EoS_P} \\ &= h_0 M_0 \Big{(}\frac{1}{2} a r M^2 + \frac{3}{4} b M^4 \Big{)} \nonumber,
\end{align}
where $a=3,~ b=1$.
This is consistent with the linear parametric model of the scaling EoS in terms of the parametric variables $(R,\theta)$ \cite{Bzdak:2019pkr,Schofield:1969zza,Guida:1996ep}. 
The normalization constants for the magnetization and the magnetic field are determined by imposing $M(r = - 1, h=0^+) = 1$ and $M(r=0, h) \propto \mathrm{sgn}(h) |h|^{1/\delta}$  and found to be $h_0= \frac{1}{3\sqrt{3}},~ M_0= \frac{1}{\sqrt{3}}$ \cite{Bzdak:2019pkr,LandauLifshitz,Nonaka:2004pg}.
We note here that the choice of mean field Ising model allows us to forgo the formulation of the equation of state in terms of parametric variables in order to solve the system analytically with rational exponents.
Furthermore, this gives us direct access to the Ising variables $(r,h)$, such that we have only a single mapping procedure to perform to QCD variables $(T,\mu_B)$, unlike with the 3D Ising model scaling equation of state which relies on the $(R,\theta)$ parametrization \cite{Parotto:2018pwx,Karthein:2021nxe,Kahangirwe:2024cny}.

\begin{figure*}[t]
    \centering 
    \begin{tabular}{c c} \includegraphics[width=0.5\textwidth]{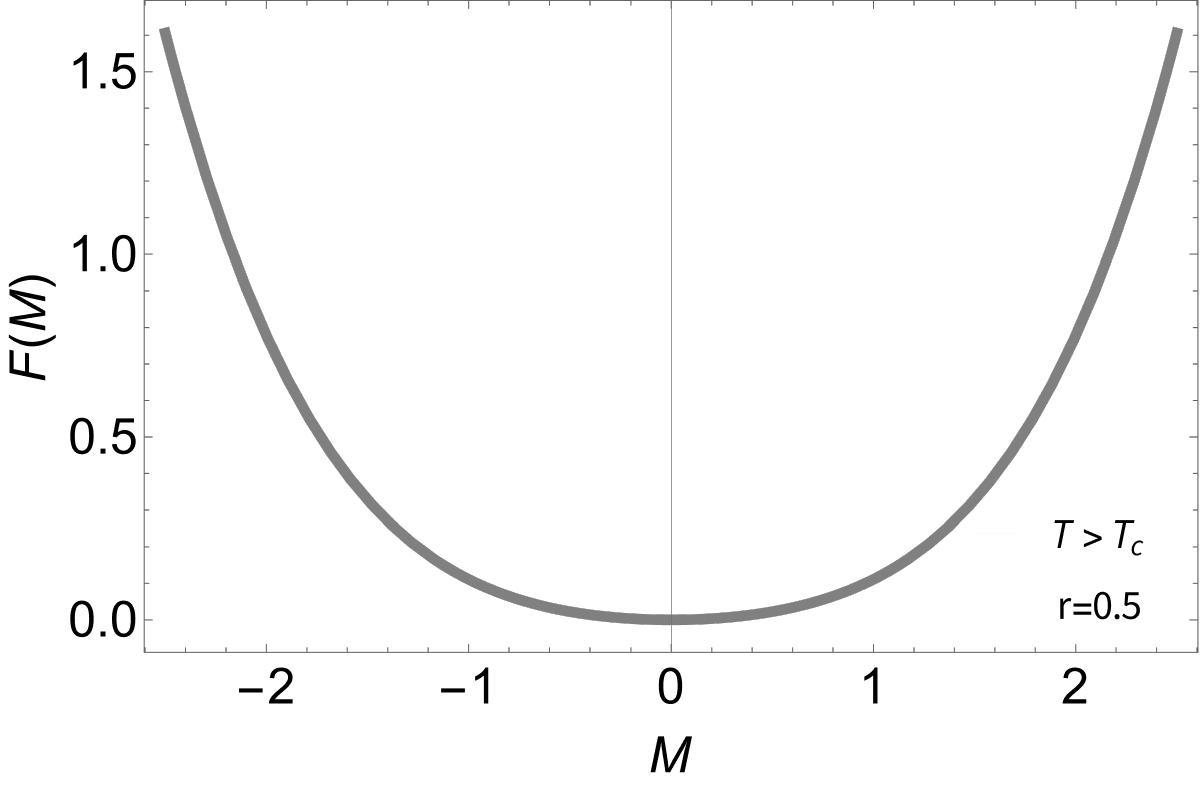}  &\includegraphics[width=0.5\textwidth]{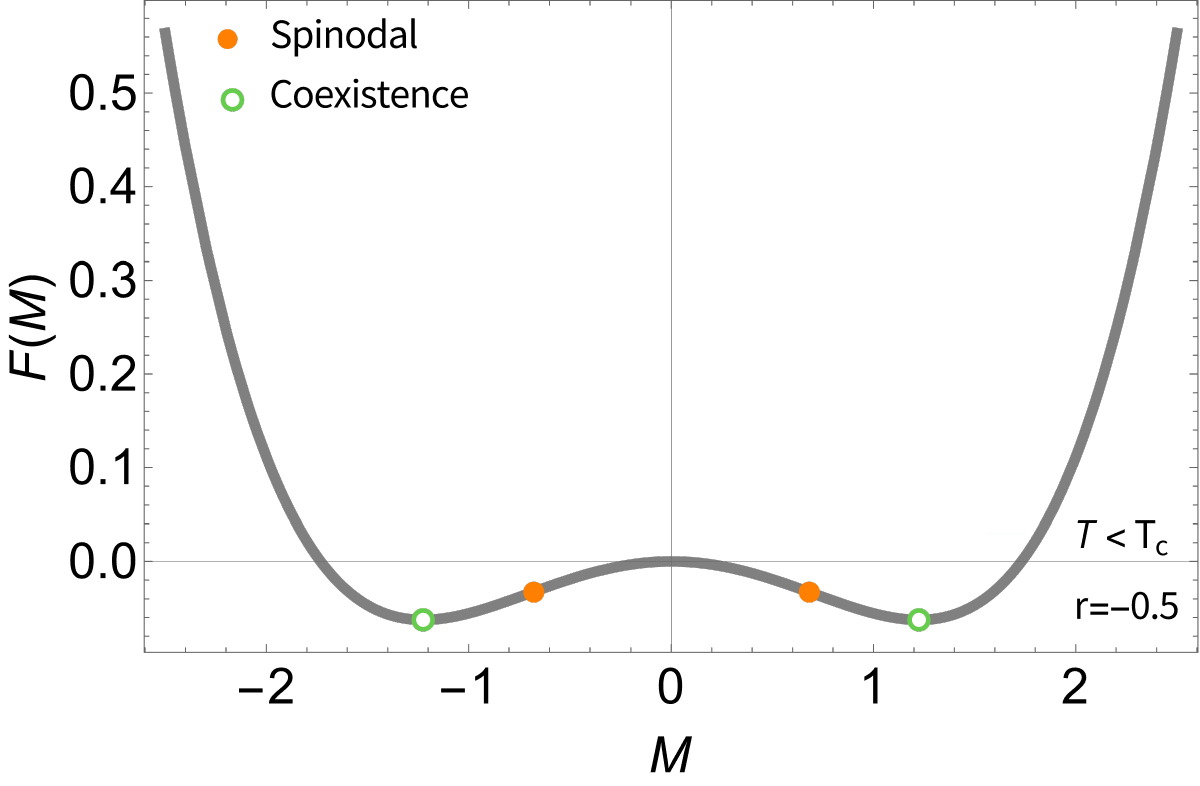} 
    \end{tabular}
    \caption{The free energy of the mean field Ising model from Eq. (\ref{eq:MF_EoS_F}) for isothermal trajectories $r=\pm 0.5$ above and below the critical point, shown on the left and right, respectively. Below the critical point on the first order phase transition side, the coexistence points and spinodals are highlighted in open green and filled orange circles, respectively.}
    \label{fig:IsingFreeEner}
\end{figure*}

Figure \ref{fig:IsingFreeEner} shows the Ising free energy for a given temperature $r$ as a function of the fixed magnetization from Eq. \eqref{eq:MF_EoS_F}.
The left-hand plot shows the free energy for temperatures above the critical point one, i.e. $r>0$.
This corresponds to the crossover transition in the Ising model.
In this case, we have a simple parabolic shape as both quadratic and quartic terms have positive coefficients. 
As the critical point is approached, the curvature at the minimum is reduced until it vanishes and the curve becomes completely flat.
From this zero curvature feature at the critical point, the coexistence region emerges below the critical temperature, $T< T_c$ or $r<0$.
The right-hand plot in Fig. \ref{fig:IsingFreeEner} shows these features for $r<0$, corresponding to the first order phase transition.
Here, the double-minimum character is apparent, due to the negative sign of the quadratic term in Eq. \eqref{eq:MF_EoS_F} when $r<0$.
Furthermore, we can characterize the typical features of an isothermal curve passing through a first order phase transition: the coexistence and spinodal points.
The coexistence points are the minima, $(\partial F/ \partial M)_r =0$, separated by a free energy barrier, while the spinodals are the inflection points $(\partial^2 F/ \partial M^2)_r =0$.
Moreover, since $h=(\partial F/ \partial M)_r$, it is clear that the coexistence points correspond to $h=0$.
Thus, the spinodals are defined as $(\partial h/ \partial M)_r=0$.

The two minima of equal magnitude of the free energy on the first order phase transition side imply that there are two phases which are equally probable and thus coexist. The value of magnetization $M$ and temperature $T$ of these minima define the coexistence line in the $(M,~T)$ plane.
Phase coexistence arises at the point in the system evolution where one phase begins to change into the other.
Because these phases occur together, this is known as the coexistence region.
In a typical approach, one might employ a Maxwell construction between the coexistence points in order to work with a system in complete thermodynamic equilibrium.
This would correspond to a straight line connecting the two coexistence points, along which the order parameter flips sign.
However, we have further information in this mean field Ising approach: the spinodal points.
As shown here, the spinodals are the inflection points, with $\partial^2 F/\partial^2 M = 0$.
By further studying the relationship between the magnetization and magnetic field, we shall discover the properties leading to the spinodal points.

\begin{figure*}[t]
    \centering 
    \begin{tabular}{c c} \includegraphics[width=0.5\textwidth]{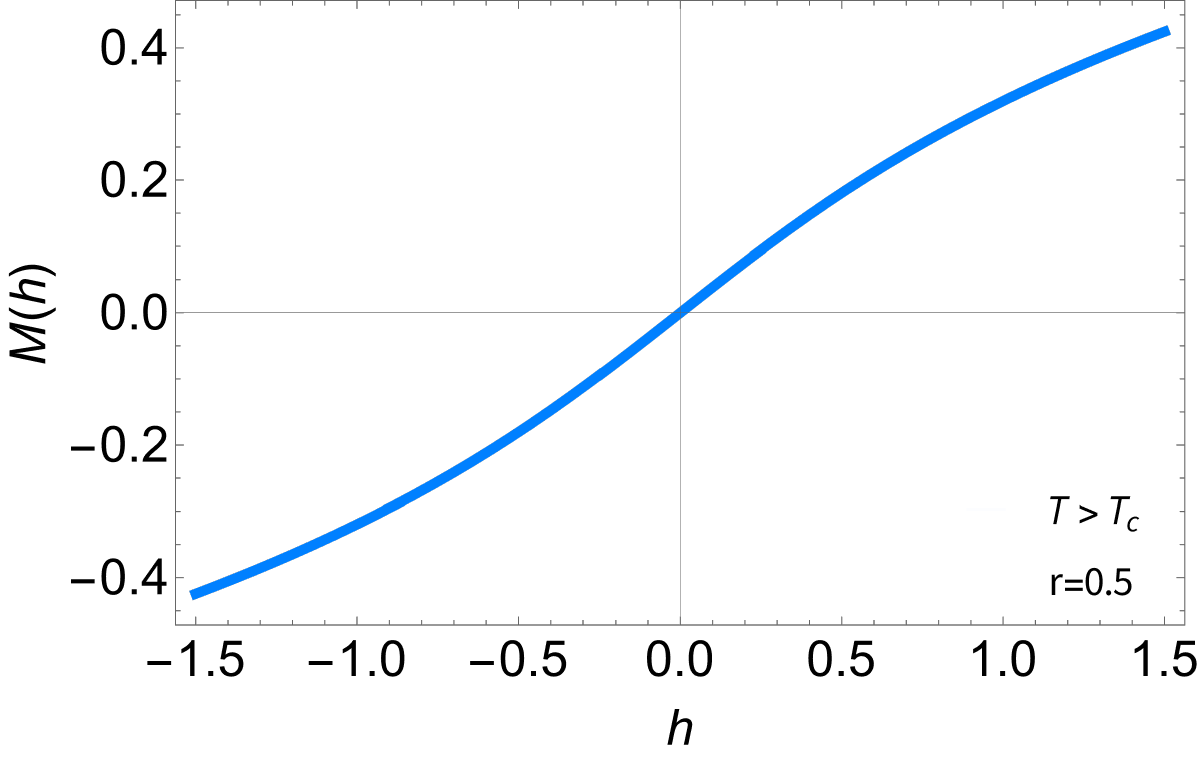}  &\includegraphics[width=0.5\textwidth]{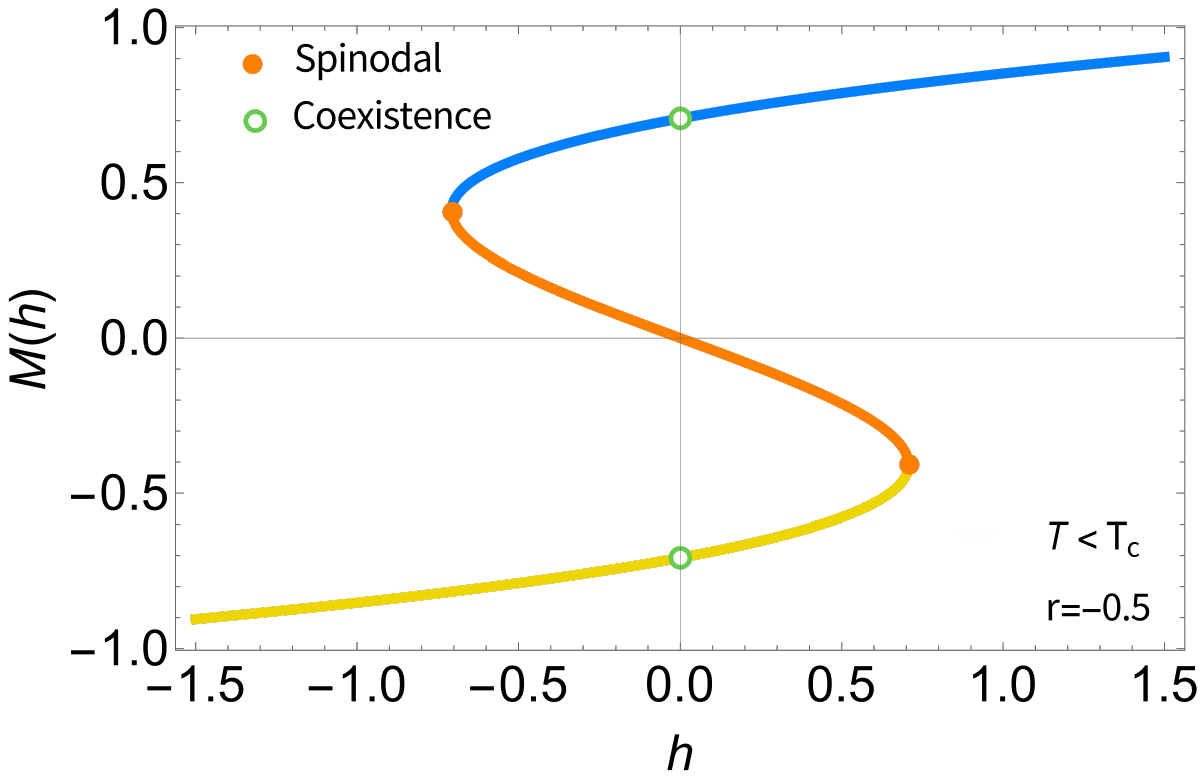} 
    \end{tabular}
    \caption{The Ising magnetization as a function of the external magnetic field as shown in Eq. \eqref{eq:Mofh} above and below the critical point for reduced temperature $r=\pm 0.5$, shown on the left and right, respectively. On the crossover side, only one real solution exists, while along the first order phase transition three real solutions exist. The coexistence and spinodal points are shown on the first order side in open green and filled orange circles, respectively.}
    \label{fig:IsingMvh}
\end{figure*}

\begin{figure*}
    \centering 
    \begin{tabular}{c c} \includegraphics[width=0.5\textwidth]{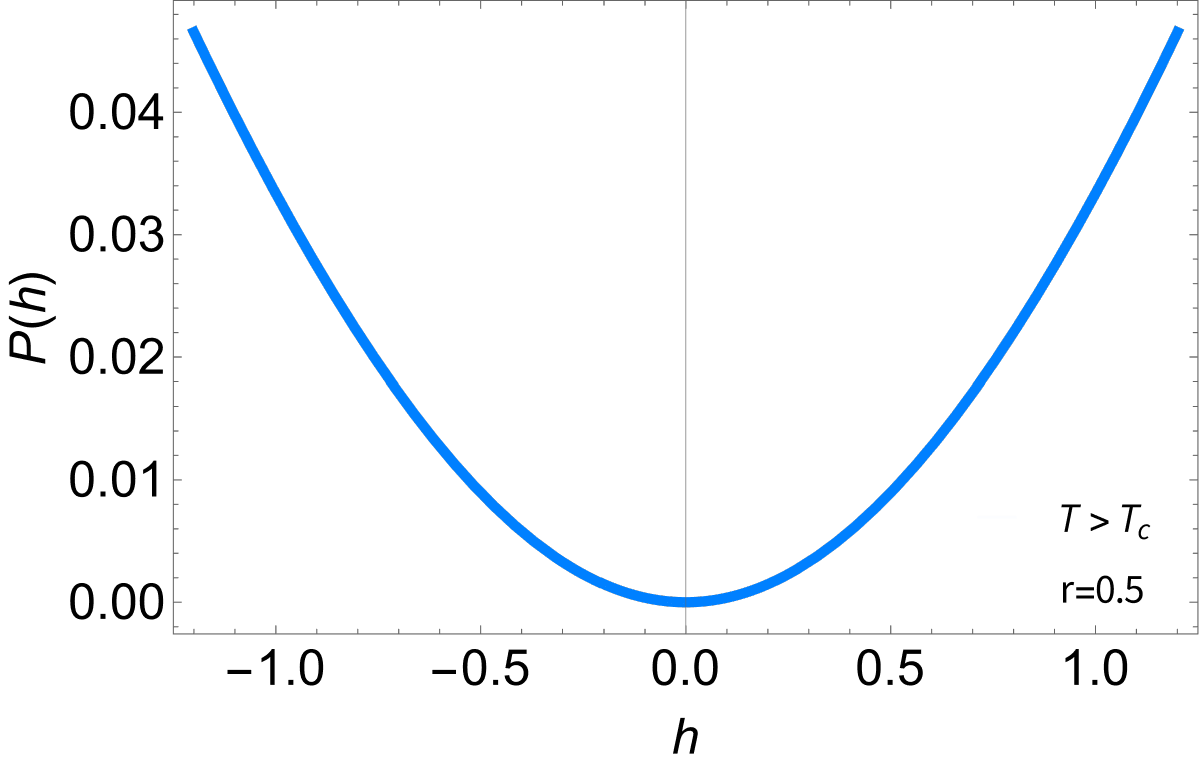}  &\includegraphics[width=0.5\textwidth]{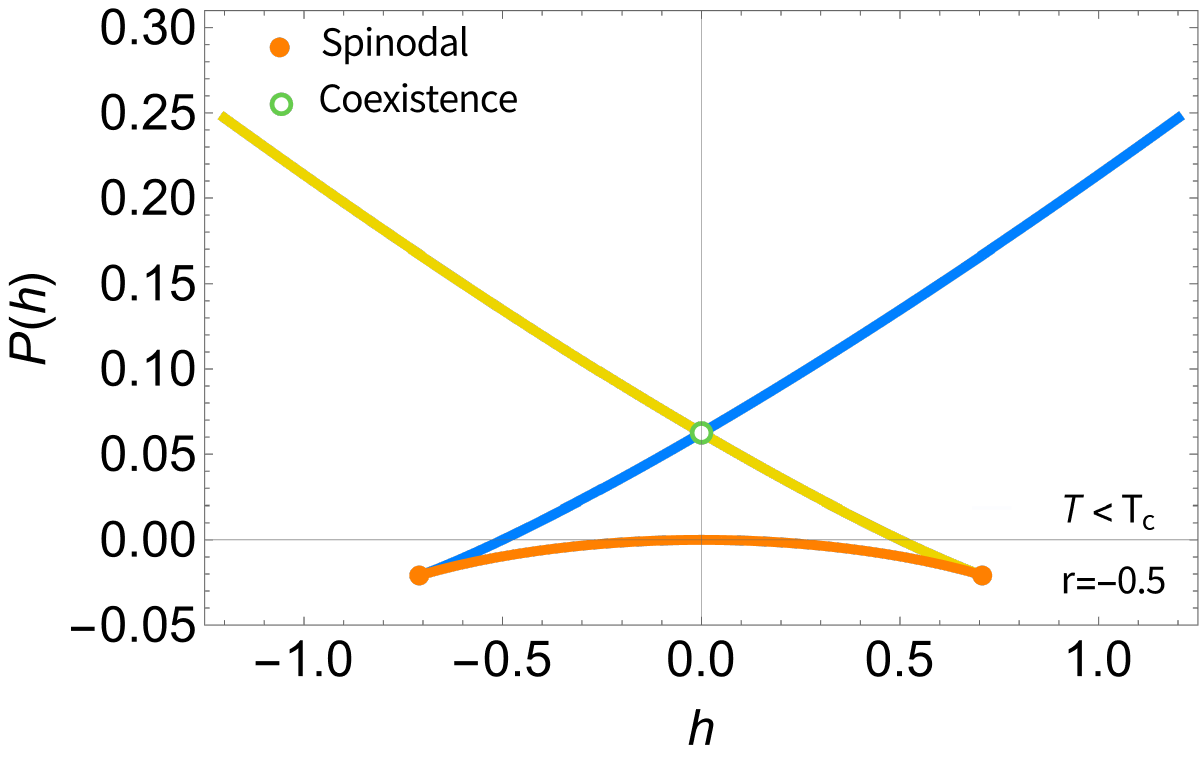} 
    \end{tabular}
    \caption{The pressure in the mean field Ising model for isothermal trajectories above and below the critical point for reduced temperature $r=\pm 0.5$, shown on the left and right, respectively. On the first order phase transition side, three real solutions are found, corresponding to the various colors of the curves. The coexistence and spinodal points are shown on the first order side in open green and filled orange circles, respectively.}
    \label{fig:IsingPvh}
\end{figure*}

Indeed, we shall further characterize the mean field equation of state by the behavior of the magnetization $M(h)$.
In the case of the mean field Ising equation of state, as shown in Eq. \eqref{eq:MF_EoS_h}, $h$ is a function that is defined in terms of $M$ and $r$.
Thus, we must invert the equation $h(M,r)$, leading to three solutions for this cubic function.
To do so, we write $h(M,r) = M (M + i \sqrt{3r})(M - i \sqrt{3r})$ yielding one real solution and an additional  pair of complex solutions for $M(h,r)$:
\begin{equation}    \label{eq:Mofh}
    \centering
    M(h,r) = 
    \begin{cases}
    \dfrac{2^{1/3}}{\eta^{1/3}} r -\dfrac{\eta^{1/3}}{2^{1/3}}  \\
            - \dfrac{(1 \pm i \sqrt{3})r}{2^{2/3} \eta^{1/3}} + \dfrac{(1 \mp i \sqrt{3}) \eta^{1/3}}{2^{4/3}},
    \end{cases}
\end{equation}
where we define $\eta=\sqrt{h^2 + 4r^3} - h$.
In order to access these features in the QCD phase diagram, we require a map between $(r,h) \rightarrow (T,\mu_B)$.
Therefore, it is necessary for us to invert $h(M,r)$ in favor of $M(h,r)$.
Thus, we ensure that our method allows us to map $h$ as a variable onto the QCD phase diagram.
In our approach, we choose to work within the $T-\mu_B$ phase diagram in order to utilize the first-principles input to the equation of state for strongly-interacting matter from lattice QCD \cite{Bellwied:2015rza}.
The alternative choice would be the $T-n_B$ phase diagram \cite{Randrup:2010ax,Vovchenko:2015vxa}.
The analogous variables in the Ising system would, thus, be reduced temperature $r$ and magnetization $M$. 
Given this choice, $M$ is retained as a control variable allowing one to use the equation of state defined in Eq. \eqref{eq:MF_EoS_h}, which provides a well-defined, single-valued function for $h(M,r)$.
The behavior of this quantity can be inferred from Fig. \ref{fig:IsingMvh}.
By rotating the figure, or equivalently switching the axes, one can see $h(M)_r$ is a single-valued function. 
To reiterate, we remain with the choice of $h$ as it is analogous to $\mu_B$ and allows us to utilize the constraints on the equation of state at vanishing $\mu_B$ from lattice QCD.

In Fig. \ref{fig:IsingMvh}, the behavior of $M(h)$ at fixed $r$ above and below the critical point is shown on the left- and right-hand plots, respectively.
As can clearly be seen in the left-hand plot, only one solution is real on the crossover side where $r>0$, corresponding to the solution which is always real in Eq. \eqref{eq:Mofh}.
On the other hand, the interesting features of the first order phase transition are apparent for $r<0$, where the complex pair of solutions become real for certain values of $h$ at fixed $r$.
Here we see the multi-valued behavior for this quantity that we expect from these three real solutions.
Each of the three solutions for the magnetization from Eq. \eqref{eq:Mofh} is shown in a different color.
The blue corresponds to the solution that is real across all temperatures, while the yellow and orange are the complex conjugate pair of solutions which become real when $r<0$.
Both the coexistence points and spinodal points defined above as $h=0$ when $(\partial h/ \partial M)_r>0$ and $(\partial h/ \partial M)_r=0$, respectively, appear naturally here.
By looking at this quantity in particular, we see that away from the coexistence points, i.e. away from $h=0$, the system is either in one phase or the other, given by the sign of the magnetization.
On the other hand, as the system approaches $h=0$, moving through the coexistence points and then proceeding further to the spinodal points, matter becomes either super-heated or super-cooled.
Between the coexistence and spinodal points we have this extreme environment for the phases where the system is in a metastable state.
Then, beyond the spinodal points the system becomes unstable due to the negative slope, i.e. $(\partial M/ \partial h)<0$.
This is analogous to the thermodynamic instability $(\partial n_B/ \partial \mu_B)<0$ as encountered, for example, in the well-known liquid-gas phase transition.

The third solution exists only in this unstable region occurring between the two spinodal points along an isotherm.
Therefore, we see here that the yellow and blue solutions describe the system up to metastability, and beyond the limit of metastability the final solution shown in orange is needed, corresponding uniquely to the instability $(\partial M/ \partial h)<0$.
Upon further consideration of this isotherm, we note that the Maxwell construction corresponds to proceeding from one phase to the other via the coexistence points at $h=0$.
For example, in the case of negative magnetization, the system follows the bottom curve shown here in yellow until it reaches the coexistence point given by the open green circle, after which it proceeds through the phase transition along the $h=0$ line until the next coexistence point where it ends up on the solution shown here in blue.
The phase has changed due to the change in sign of the magnetization, but this approach has completely disregarded the multi-valued nature of this quantity.
We, thus, make use of the full information in the coexistence region by describing the metastable and unstable regions.
For the rapidly expanding strongly-interacting system in heavy-ion collisions, it is possible to reach these meta- or un-stable regions.

We then turn to the behavior of the pressure in the Ising model.
The quantity of interest for our purpose of studying these features in the QCD phase diagram is, in fact, the pressure.
We, thus, utilize the expressions for the magnetization from Eq. \eqref{eq:Mofh} and obtain the pressure, Eq. \eqref{eq:MF_EoS_P},  as a function of the external variables $r$ and $h$.
Moreover, this allows us to track the spinodals in a straightforward way and study how they map onto the QCD phase diagram in temperature and baryon chemical potential.
In this way, by utilizing the mean field approach, we preserve the information on multi-valued observables.

The pressure in the crossover ($r>0$) and first order ($r<0$) regions is shown in Fig. \ref{fig:IsingPvh}.
Because the pressure relies on the form of $M(h)$, the presence of a single or three real solutions above or below the critical point is manifest in the pressure itself.
In the case of the first order phase transition (Fig. \ref{fig:IsingPvh}, right), the coexistence point, where the pressure of the two phases is the same, corresponds to this location where the two solutions meet. 
The spinodal points, on the other hand, are located where the super-heated/super-cooled phases terminate, with the subsequent onset of the unstable region.
The third solution describes this region between the spinodal points.
Generally, the phase transition proceeds by either nucleation or spinodal decomposition, when the system is in the metastable or unstable state, respectively \cite{Binder1987,Cahn1959,Abyzov2007}.
Thus,
between the spinodals the system likely undergoes spinodal decomposition \cite{Chomaz:2003dz,Randrup:2009gp,Steinheimer:2012gc,Kapusta:2024nii}, or possibly nucleation \cite{Cahn1959,Mishustin:1998eq,Csernai:1992as} but only when the nucleation rate is large enough compared to the expansion rate of the medium.

\begin{figure}[h!]
    \centering
    \includegraphics[width=0.49\textwidth]{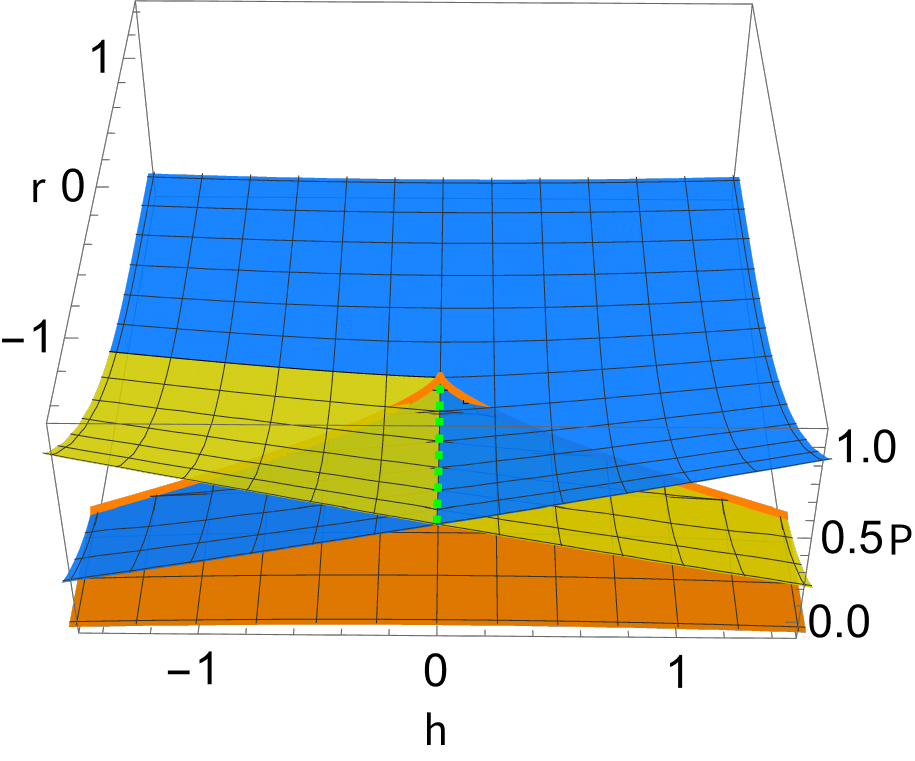}
    \includegraphics[width=0.49\textwidth]{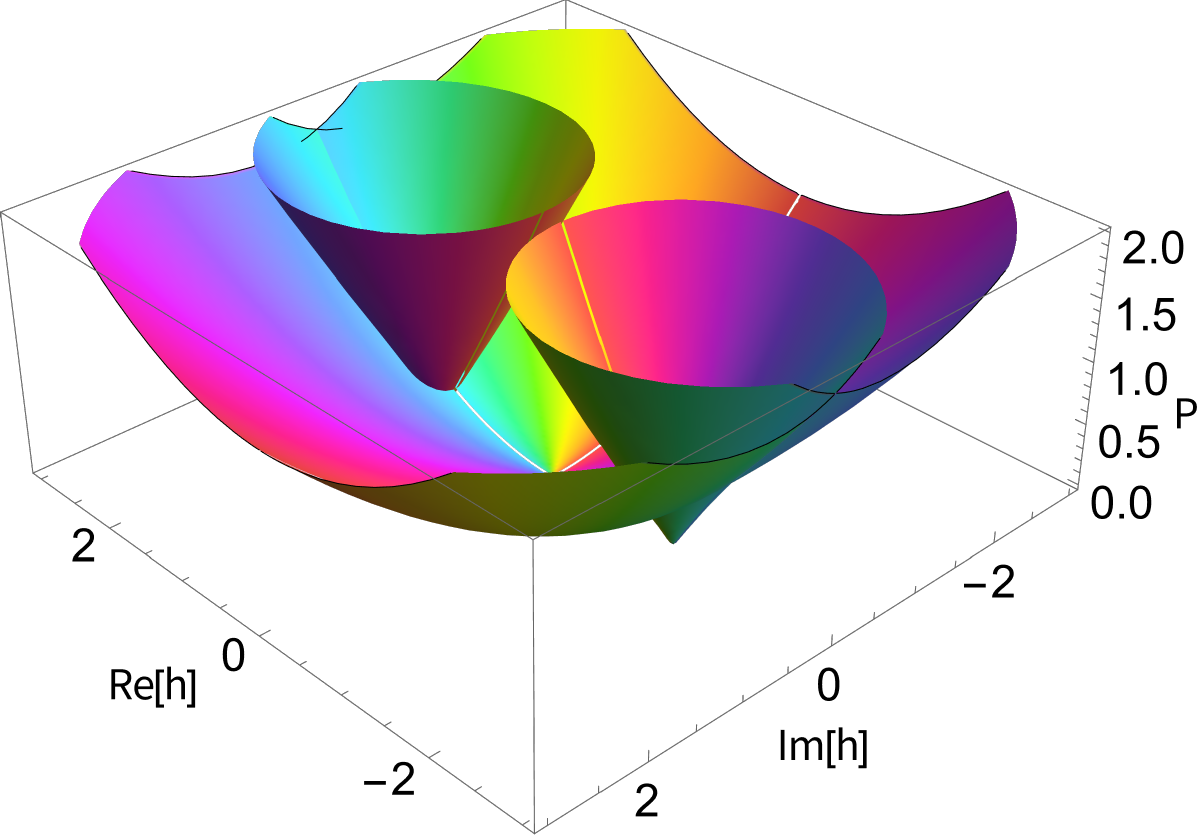}
    \includegraphics[width=0.49\textwidth]{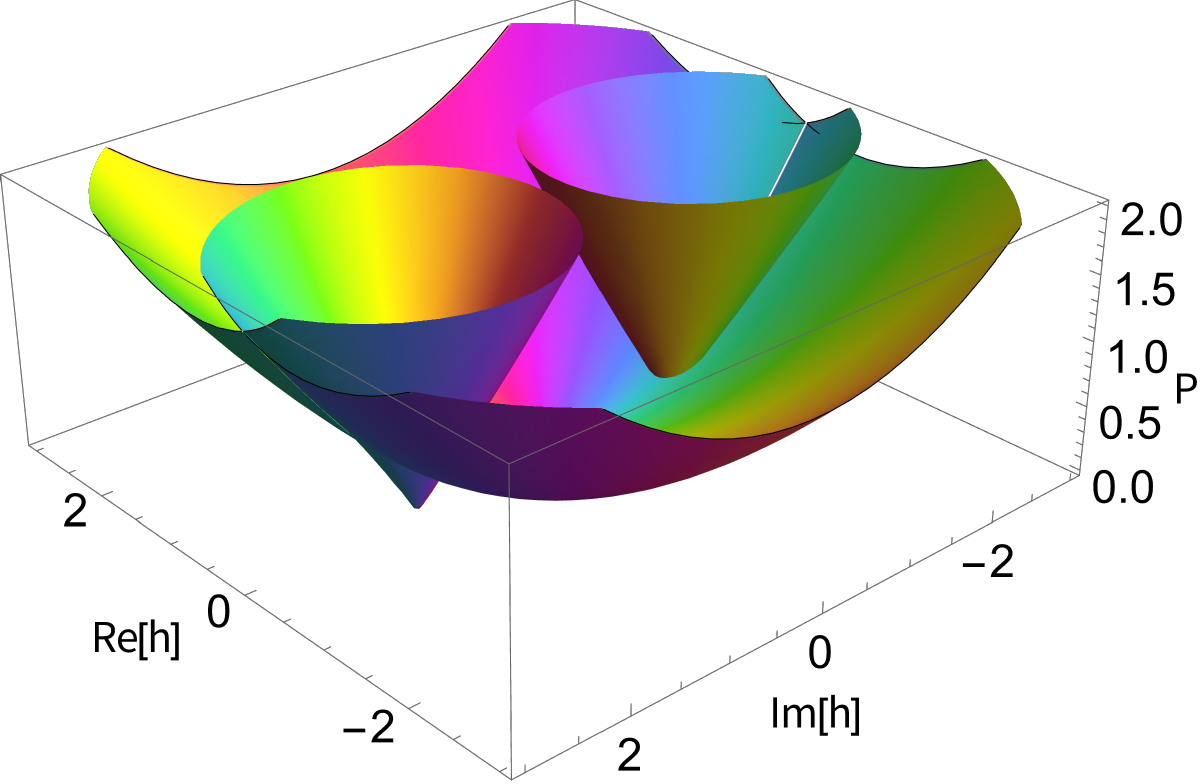}
    \caption{Behavior of the pressure across the low and high temperature sheet (top panel), as well as its behavior in the complex $h$ plane for low (middle panel) and high (bottom panel) temperature.}
    \label{fig:Ising_complex_sheets}
\end{figure}

We study the pressure in the full Ising phase diagram in $h$ and $r$ in order to see its structure more clearly.
The three solutions for the pressure are shown in terms of the Ising variables $(r,h)$ in the top panel of Fig. \ref{fig:Ising_complex_sheets}.
This allows us to visualize the behavior of the pressure on the low ($r<0$) and high ($r>0$) temperature sheets.
On the high temperature sheet, where the transition is a smooth crossover, only one solution exists which corresponds to the real root of $h(M,r)$, while on the low temperature sheet, where the transition becomes first order, the two complex roots also become real giving rise to three distinct planes for the pressure.
In this figure, the three solutions for the pressure following from $P(h,r)$ that one obtains applying Eqs. \eqref{eq:Mofh} to \eqref{eq:MF_EoS_P} are shown in different colors with the same scheme applied as in the isotherms.
The coexistence curve corresponds to the line along which the first two solutions, shown in yellow and blue, meet for $h=0,~ r<0$, shown here as the dashed green line. 
On the other hand, the spinodals are located where the first two solutions meet the third one, shown here as the solid orange lines.
Thus, as in the isotherm shown in Fig. \ref{fig:IsingPvh}, the spinodals are the lower edges of the triangle outlining the coexistence region.
When crossing the $r=0$ line, i.e. the critical isotherm, the pressure collapses to one solution above $T_c$, i.e. on the high temperature sheet. 
Therefore, when moving along a trajectory in the $h-r$ plane for which $r$ is not fixed (which we will see later corresponds to a QCD isotherm), we will end up crossing the critical isotherm and going onto the low temperature sheet. 
It is important to note that there is no phase transition associated with crossing the critical isotherm.

If we consider the behavior in the complex $h$-plane  on each of the low and high temperature sheets separately, we can understand the presence of the spinodal lines in the phase diagram.
In the middle panel of Fig. \ref{fig:Ising_complex_sheets}, we show the pressure in the complex $h$ plane for $r<0$.
We see that the Lee-Yang edge singularities are located along the real $h$-axis in this case.
Conversely, for the high temperature $r>0$ sheet, these manifestations corresponding to the spinodals on the other sheet lie completely along the imaginary axis (bottom panel of Fig. \ref{fig:Ising_complex_sheets}). 
Beyond mean field, these spinodal singularities no longer lie along the real or imaginary axes, but rather move into the complex plane.

In the present study, we are concerned with characterizing the thermodynamical aspects of the first order phase transition via the equation of state.
Moreover, because the equation of state serves as crucial input to the hydrodynamic simulations of heavy-ion collisions, we require full information on the baryon density and energy density as functions of the temperature and chemical potential.
In the first order region, the baryon and energy densities become multi-valued solutions of $T$ and $\mu_B$, where the phases coexist.
Therefore, the simulations need the ability to distinguish the different branches of the baryon and energy density in to order establish where precisely the system lies in terms of $T$ and $\mu_B$ within the coexistence region.
This is exactly the behavior that we capture by utilizing the mean field Ising equation of state, as shown by the magnetization and pressure in the Ising model in Figs. \ref{fig:IsingMvh} and \ref{fig:IsingPvh}.
By having explicit control over the three real solutions appearing for $r<0$, we are able to characterize the nature of the phase coexistence.
Thus, by translating these features into QCD variables, this equation of state framework allows the hydrodynamical simulations to track the phases in the first order region.
This is particularly necessary when modeling low energy collisions where a first order transition may be present.
However, it is important to note that the spinodals are valid in the regime of metastability which has a finite lifetime, and as such, they do not belong to a system in thermal equilibrium \cite{Langer2000,Binder1987}.
They are, nonetheless, manifest from the mathematics as properties of the equation of state, belonging to the broader class of Lee-Yang edge singularities \cite{Lee1952}.
Furthermore, spinodal decomposition (together with nucleation) helps to describe the dynamics of real physical systems near a first order phase transition.
From an applied perspective, spinodal decomposition is ubiquitous in industrial applications \cite{Jones2002}.
The importance of understanding the spinodals is, therefore, clear, and we proceed to produce and study an equation of state for QCD that contains these features of a first order phase transition.
The scope of this work is to study how these features map onto the QCD phase diagram, and as such we do not consider dynamical aspects owing to the finite lifetime of metastability.

\subsection{Ginzburg Criterion}
Before mapping the features of the Ising model onto QCD, we first discuss the range of validity of the mean field approximation which we are utilizing in this study. 
By applying a mean field approach, one inherently assumes that the average behavior is sufficient to describe the system and that fluctuations can be neglected.
The well-known Ginzburg criterion establishes the regime of validity for mean field where the fluctuations must be smaller than the magnitude of the order parameter itself \cite{Ginzburg1960}.
This idea was taken a step further by Binder \cite{Binder1987} who showed that, in the case of the spinodal singularity, the interaction range, $R$, must be sufficiently large for systems with dimensionality below the upper critical dimension, to apply mean field treatment: 
\begin{equation}
    1 \ll R^d (h_c - h)^{(6-d)/4}.
\end{equation}
Indeed, the real space Ginzburg criterion as shown by Als-Nielsen and Birgeneau \cite{Als-Nielsen1977} can be written as:
\begin{equation}
    \chi_M \ll V M^2,
\end{equation}
where $\chi_M = \partial h/ \partial M$ is the magnetic susceptibility and $V$ is the volume, maximally determined by the size of the correlation length $\xi^d$.
By utilizing the well-known critical scaling \cite{ZinnJustin} for the susceptibility $\chi_M \sim r^{-\gamma}$, order parameter, $M \sim r^{\beta} $, and correlation length, $\xi \sim \xi_0 r^{-\nu}$, the Ginzburg criterion becomes:
\begin{equation}
    r^{-\gamma + d \nu - 2 \beta} \ll \xi_0^d.
\end{equation}
Inside this region, the mean field approximation begins to breakdown as fluctuations become important.
This is consistent with what is found in Ref. \cite{An:2017brc} in terms of the behavior of the scaling-invariant variable $w=h t^{-\beta \delta}$.
In the case of the 3D Ising model, when considering the current level of precision of the determination of the critical exponents,
the left-hand side becomes $r^0$.

This leads to the conclusion that the size of the system determined by the size of the correlation length must be much greater than 1.
This large (diverging) correlation length indicates the presence of the critical point singularity. 
If the correlation length is taken to be in units of fm, a correlation length of several fm has been shown to be large enough to see the critical effects in a system with finite size and lifetime \cite{Berdnikov:1999ph}. 
Thus, in the regime of large correlation lengths, corrections to mean field may become relevant.
On the other hand, this formulation of the Ginzburg criterion shows that mean field is an appropriate description when the correlation length is small.
Additionally, we will compare the overall equation of state between the 3D and mean field Ising models in Sec. \ref{sec:fullthermo}.

\section{Ising-QCD Mapping} \label{sec:mapping}

\begin{figure}
    \centering
    \includegraphics[width=\linewidth]{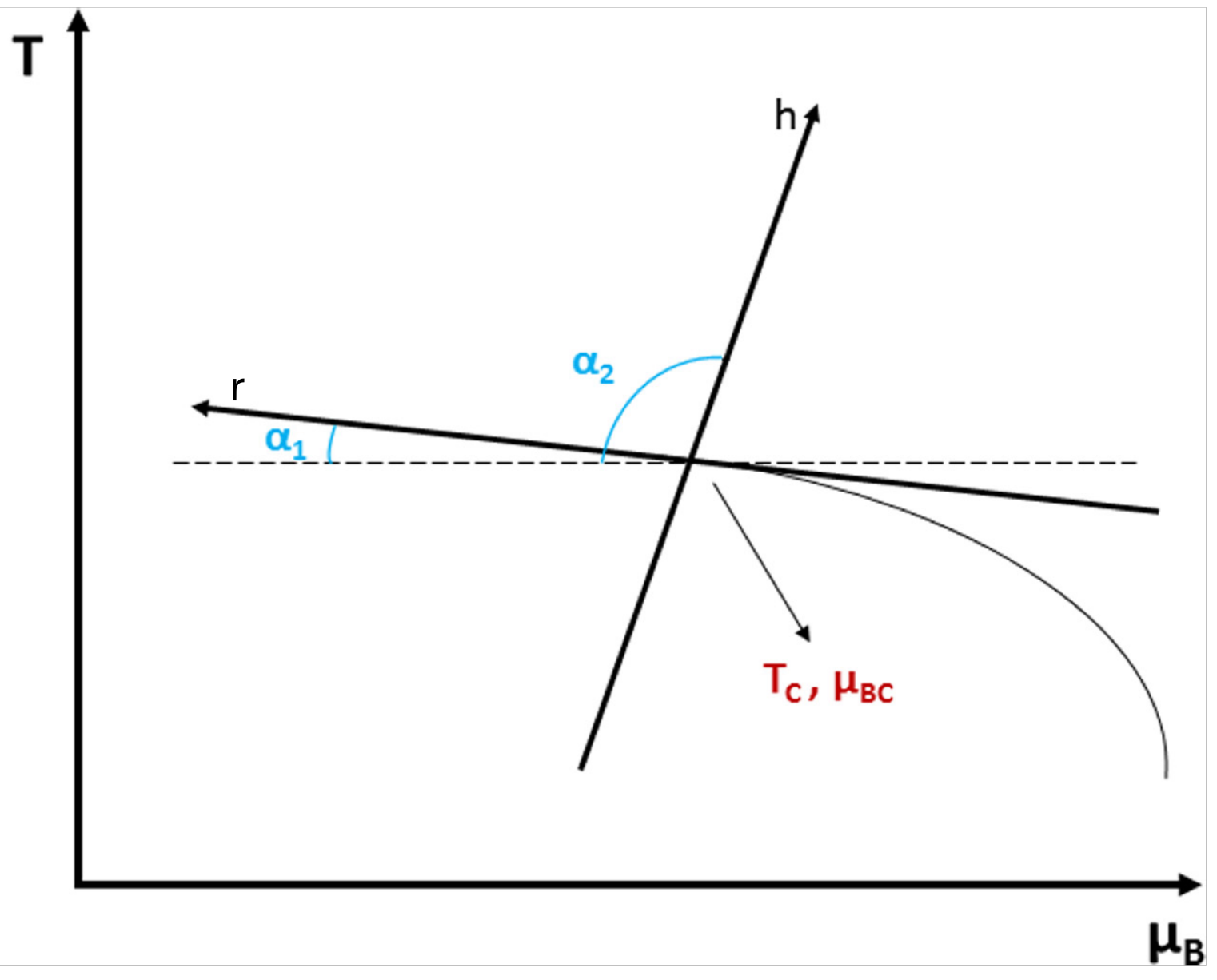}
    \caption{A sketch of the mapping between Ising and QCD variables. The Ising axes $h,r$ are mapped via the angles $\alpha_2$ and $\alpha_1$, respectively, onto the QCD phase diagram with the critical point located at $(T_c,\mu_{B,c})$.}
    \label{fig:mapTmuB}
\end{figure}

While universality determines the behavior of the system in the critical region, so far there is no universal method of mapping between the Ising model and QCD coordinate systems.
Therefore, in order to study the effect of a critical point and corresponding first order phase transition line in the QCD phase diagram, we must rely on a non-universal mapping between Ising and QCD coordinates: $(r,h) \rightarrow (T,\mu_B)$. Furthermore, we will utilize the first-principles constraints on the equation of state at vanishing $\mu_B$ from lattice QCD.
We choose a linear mapping between these variables, which has been studied in detail in Refs. \cite{Parotto:2018pwx,Pradeep:2019ccv,Karthein:2021nxe, Mroczek:2022oga,Pradeep:2024cca}:
    \begin{equation}
    \begin{split} \label{eq:mapTmuB}
        \frac{T-T_c}{T_c} &= \omega(\rho r \sin{\alpha_1} + h \sin{\alpha_2}) \\
        \frac{\mu_B-\mu_{B,c}}{T_c} &= \omega(-\rho r \cos{\alpha_1} - h \cos{\alpha_2})
    \end{split}
    \end{equation} 
where $(T_c , \mu_{B,c})$ are the coordinates of the critical point, and ($\alpha_1$, $\alpha_2$) are the angles between the axes of the QCD phase diagram and the Ising model ones as illustrated in Fig. \ref{fig:mapTmuB}. 
Finally, $\omega$ and $\rho$ are the scaling parameters between  Ising and QCD coordinates: $\omega$ determines the overall scale of both $r$ and $h$, while $\rho$ determines the relative scale between them.
Thus, in total there are 6 free parameters in this non-universal map.
A reduction in the freedom of the mapping can be obtained by choosing the location of the critical point to be on the chiral crossover transition line as determined from lattice QCD simulations, $T^{\rm{latt}}_c(\mu_B)$ \cite{ Bonati:2015bha,Bellwied:2015rza,Bonati:2018nut,Bazavov:2018mes,Borsanyi:2020fev}.
In this case, the curvature of the transition line constrains the angle $\alpha_1$ at which to place the transition line from the Ising model given by the $r$-axis.
In addition, naturally, when choosing a value of $\mu_{B,c}$ the transition line $T^{\rm{latt}}_c(\mu_B)$ will determine the critical temperature $T_c$.
Thus, the critical point sits on the chiral phase transition line, and the $r$-axis of the Ising model is tangent to the transition line of QCD at the critical point.    

\subsection{Effect of Non-universal Mapping Parameters on Spinodal Location }\label{sec:spinodals_location}

In order to determine the location of the spinodal points within the QCD phase diagram after the mapping procedure, we derive the dependence of the spinodal location on the mapping parameters.
Firstly, we begin with the spinodal definition in the mean field Ising model, consistent with what has been shown in Ref. \cite{An:2017brc}
    \begin{equation} \label{eq:IsingSpinodals}
        h_{\mathrm{sp}} = \pm\frac{2 r^{3/2}}{3 \sqrt{3}},
    \end{equation}
which is obtained by solving the equation $(\partial h/ \partial M)_r=0$.
Here, we note that we only consider the spinodals in their region of validity, i.e. only for $r<0$, which leads to the absorption of the ``$i$" appearing in the more general formulation of Lee-Yang edge singularities in Ref. \cite{An:2017brc}.
Now, plugging in the mapping equations $h(T,\mu_B), ~r(T,\mu_B)$ with orthogonal angles $\alpha_1, ~\alpha_2$ leads to the following expression:
    \begin{align} \label{eq:MappedSpinodals}
        &\frac{(\mu_B-\mu_{B,c}) \sin \left(\alpha _1\right) + (T-T_c) \cos \left(\alpha _1\right)}{ T_c   w} = \\ &\pm \frac{2}{3 \sqrt{3}} \Bigg(\frac{\left(-(\mu_B-\mu_{B,c}) \cos (\alpha _1) + (T-T_c) \sin \alpha _1 \right)}{\rho 
   T_c w }\Bigg)^{3/2}. \nonumber
    \end{align}
We choose orthogonal angles to eliminate the additional dependence on the angle $\alpha_2$ and study the effect of the angle $\alpha_1$. 
We leave the study of relaxing the orthogonality condition and exploring the effect of $\alpha_2$ to future work. 
In order to study the spinodal location in $\mu_B$ in a more closed-form expression, we can consider the two limiting cases of the angle $\alpha_1=0^\mathrm{o},~90^\mathrm{o}$, which correspond to directly mapping the Ising phase transition line $h=0$, i.e. the $r$-axis, along lines parallel to the QCD $\mu_B$-axis or $T$-axis, respectively. 
The former case is more appropriate for QCD at small values of the chemical potential since we know that the chiral phase transition line is very flat in this region, giving rise to a small curvature or angle \cite{Bellwied:2015rza,Borsanyi:2020fev}. 
On the other hand, at large values of the chemical potential, the transition line should approach the $\mu_B$-axis orthogonally \cite{Halasz:1998qr}, corresponding to the latter case.
It is also important to note that the case of $\alpha_1 = 90^{\mathrm{o}}$ corresponds to mapping analogous quantities onto one another, i.e. $h \rightarrow \mu_B, ~ r \rightarrow T$.
Conversely, for vanishing $\alpha_1$ it is the opposite, $h \rightarrow T, ~ r \rightarrow \mu_B$.
Thus, when $\alpha_1=0$, fixing the QCD temperature $T$ corresponds to constant magnetic field lines rather than constant temperature in Ising variables.
In other words, an isotherm in the Ising model now corresponds to a trajectory at fixed baryon chemical potential $\mu_B$.
This also applies to the case of small $\alpha_1$ and will be important for understanding the behavior of the pressure mapped to $T$ and $\mu_B$.

We begin with $\alpha_1=0^\mathrm{o}$ and the positive solution for $h_\mathrm{sp}$ from Eq. \eqref{eq:IsingSpinodals}. 
In this case, the location of the spinodal in $\mu_B$ corresponds to:
    \begin{equation} \label{eq:muBSpinodalAlpha0}
        \mu_{B,\mathrm{sp}}^{\alpha_1=0}(T) = \mu_{B,c} + \frac{3}{2^{2/3}} \rho  w^{1/3} T_c^{1/3} (T - T_c)^{2/3}.
    \end{equation}
In fact, this expression is the same regardless of which $h_\mathrm{sp}$ solution we begin with. Thus, this already shows that we will only find a single spinodal point in $\mu_B$ for this choice of angle. In order to further illuminate this situation for the choice of small angle, it is important to note that the critical isotherm ($r=0$) rather than the coexistence line ($h=0$) is being crossed along the QCD isothermal trajectories. In other words, if we considered constant $\mu_B$ trajectories we would find the two solutions we expect corresponding to each of the $h_\mathrm{sp}$ solutions. 
    
On the other hand, if we consider the case with $\alpha_1=90^\mathrm{o}$, we find the spinodal locations to be:
    \begin{equation} \label{eq:muBSpinodalAlpha90}
        \mu_{B,\mathrm{sp}}^{\alpha_1=\pi/2}(T) = \mu_{B,c} \pm \frac{2 \sqrt{3}}{9} \rho^{-3/2}  w^{-1/2} T_c^{-1/2} (T_c - T)^{3/2}.
    \end{equation}
Here, the sign corresponds to the sign of the $h_\mathrm{sp}$ solution. Thus, we are able to determine the location of both spinodals for a given QCD isothermal trajectory in the case of $\alpha_1=90^\mathrm{o}$.

These limiting cases help to understand what we can expect to see for the spinodals in terms of $T$ and $\mu_B$.
We see that because the spinodals in the Ising model are defined by $(\partial h/ \partial M)_r=0$, at fixed Ising temperature $r$, the way $r$ maps to $T$ is very important for isothermal trajectories.
By mapping the Ising coexistence line tangential to the chiral crossover line, there will generally be a mixing between the Ising variables as mapped onto $T$ and $\mu_B$, since the angle $\alpha_1\ne 90^\mathrm{o}$.
Thus, isotherms from the Ising phase diagram become straight lines in the QCD phase diagram characterized by some value of $\mu_B/T$.

\subsection{Spinodal Behavior Along Isothermal Trajectories}
\label{sec:spinodalisotherms}
Next, we study isothermal curves for the Ising pressure as mapped onto the QCD phase diagram.
As can be seen from the derivation of the spinodal location in $\mu_B$ when the angle $\alpha_1$ is small, a QCD isotherm will only capture one of the spinodals while the other lies in a region of the phase diagram not reached by that isotherm.
The behavior of the Ising pressure in its own phase diagram, as shown in Fig. \ref{fig:Ising_complex_sheets}, becomes important for understanding this effect.
As can be seen in this 3D representation, for any Ising isotherm $r<0$, both edges of the spinodal lines will be reached by such an isothermal curve.
On the other hand, once the Ising variables are traded for the QCD ones, an isotherm for $T<T_c$ may not necessarily capture both sides of the triangular structure for all mapping parameters.
This is especially true for small angles, where the mapping is such that $h \rightarrow T$ and $r \rightarrow \mu_B$, as discussed previously.
Furthermore, as the mapping in Eq. (\ref{eq:mapTmuB}) shows, $h$ and $r$ can mix in the expressions for $T$ and $\mu_B$.
This means that these triangular structures in the pressure can become distorted along the isotherms.
In order to elucidate this isothermal behavior in QCD variables, Fig. \ref{fig:SpinodalPhDiag} shows the spinodal and coexistence curves as mapped onto the QCD phase diagram.
From these figures, it is clear that for smaller angles, or equivalently smaller values of $\mu_{B,c}$, a line of constant $T$ cannot capture both spinodal lines.
This is due to the fact that the spinodal line above the phase transition line, which represents super-heated hadronic matter, opens up into the phase diagram towards higher temperatures.  
These results will be discussed in detail in Sec. \ref{sec:spinodals_phase_diag}.

\begin{figure}[t!]
    \centering
    \includegraphics[width=0.49\textwidth]{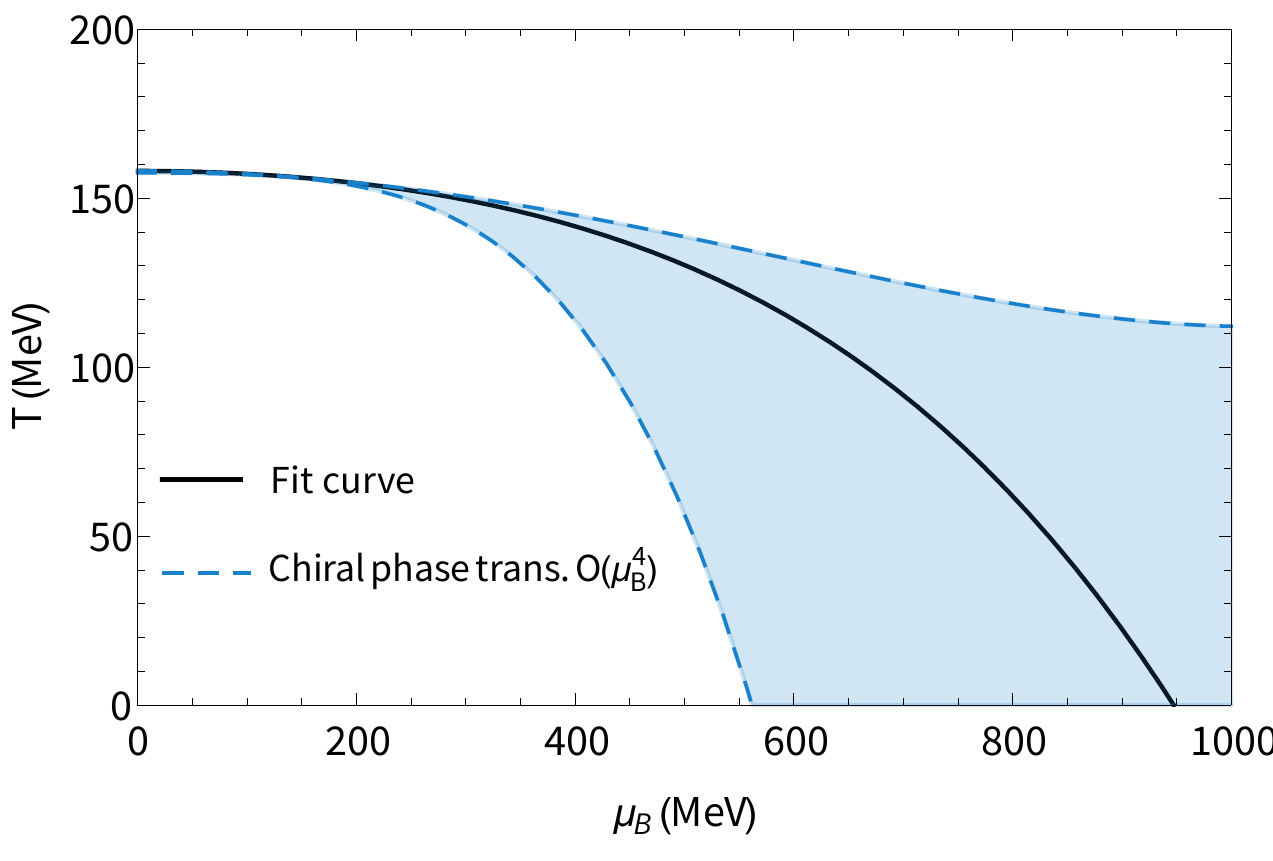}
    \caption{Results for the chiral phase transition line to $\mathcal{O}(\mu_B^4)$ from Ref. \cite{Borsanyi:2020fev}, given by the blue band, along with the specific curve utilized in this analysis shown in black. This curve was determined by fitting the lattice data from Ref. \cite{Borsanyi:2020fev} that obey the implicit equation shown in Eq. \eqref{eq:chiral_phase_trans_OmuB4}.}
    \label{fig:ChPhaseTransChoice}
\end{figure}

We attempted to obtain access to the spinodal region for the same choice of mapping parameters from Ref. \cite{Parotto:2018pwx}, which determine the location of the critical point ($\mu_{B,c}$ from which we obtain $T_c$ and $\alpha_1$) as well as the size and shape of the critical region ($w,~\rho,~\alpha_2$).
The location in this example choice is $\mu_{B,c}=350$ MeV with scaling parameters $w=1,~ \rho=2$ and orthogonal Ising axes such that $\alpha_2=\alpha_1 + 90^\mathrm{o}$.
In Ref. \cite{Parotto:2018pwx}, the chiral phase transition line was utilized up to $\mathcal{O}(\mu_B^2)$, which results in a very small mapping angle of $\alpha_1=3.8^\mathrm{o}$.
In this work, we utilize updated results from the Wuppertal-Budapest collaboration on the chiral phase transition line up to $\mathcal{O}(\mu_B^4)$ from lattice QCD \cite{Borsanyi:2020fev}:
\begin{equation}    \label{eq:chiral_phase_trans_OmuB4}
    \frac{T_c(\mu_B/T_c)}{T_c(\mu_B=0)} = 1 - \kappa_2\Big{(}\frac{\mu_B}{T_c(\mu_B)}\Big{)}^2- \kappa_4\Big{(}\frac{\mu_B}{T_c(\mu_B)}\Big{)}^4.
\end{equation}
In Fig. 6 we show the results for this curve with all errors included as a band in blue.
As the chemical potential increases, the lattice results become less constraining.
As such, we choose a parametrization of the chiral phase transition line within the range of values estimated by the lattice calculations, shown as the black, solid line in Fig. \ref{fig:ChPhaseTransChoice}.
This choice is motivated by obtaining a curve that bends steeply enough to yield larger angles $\alpha_1$ as $\mu_B$ increases.
Any such parametrization would be acceptable, as long as the curve terminates at baryon chemical potentials larger than that of the proton mass where we know confined hadronic matter exists.
Additionally, the transition line should bend down to touch the $\mu_B$-axis and become orthogonal at $T=0$  due to the third law of thermodynamics \cite{Halasz:1998qr}.
As in Eq. \eqref{eq:chiral_phase_trans_OmuB4}, we use a fourth order polynomial in order to parametrize our fit curve.
As such, we do not expect to capture the singularity at $T=0$, where $\partial T/\partial \mu_B$ becomes infinite.

\begin{figure} 
    \centering
    \includegraphics[width=0.48\textwidth]{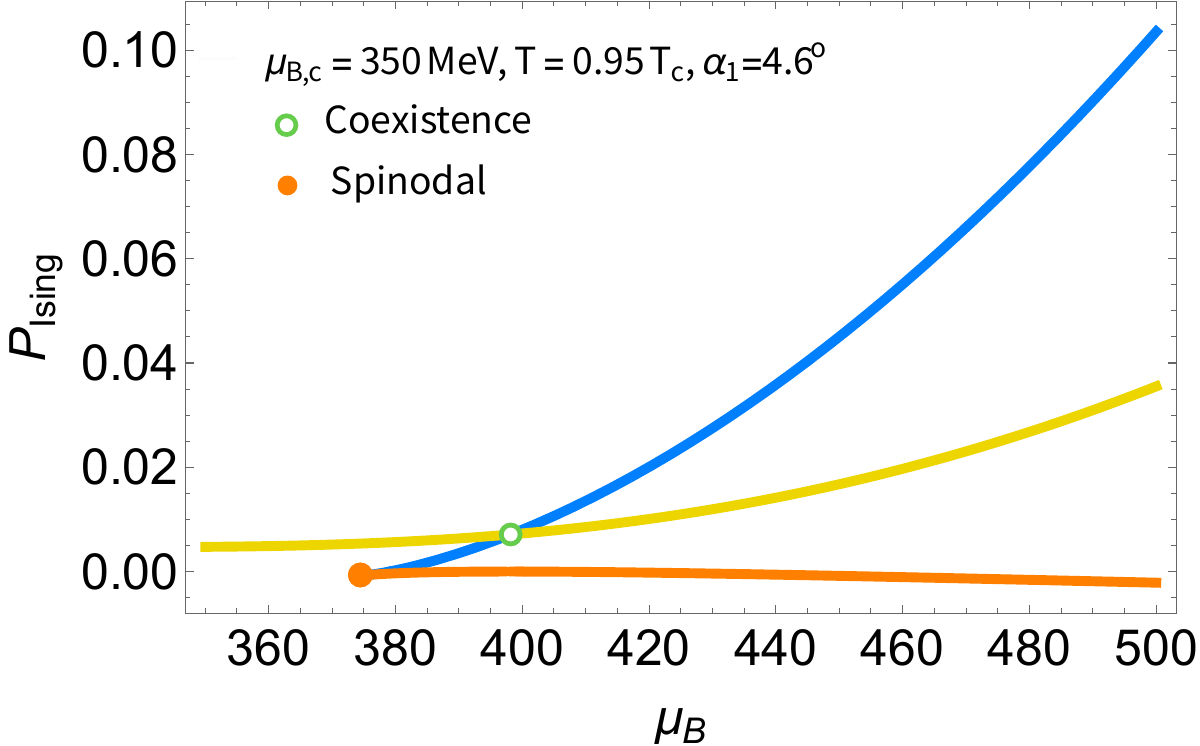}
    \caption{Ising pressure isotherm as mapped onto the QCD phase diagram for a critical point chosen to be the same as that of Ref. \cite{Parotto:2018pwx,Karthein:2021nxe} where $\mu_{B,c}=350$ MeV.
    }
    \label{fig:SpinodalIsotherms_OrigBEST}
\end{figure}

Given our parametrized fit curve, we find a different angle $\alpha_1$ for $\mu_{B,c}=350$ MeV than Ref. \cite{Parotto:2018pwx}. 
In addition, the value of $T_c$ for a given choice of the critical chemical potential, $\mu_{B,c}$, will also be adjusted.
In this case, our updated parameters which are fixed by the placement of the critical point along the chiral phase transition line are $\alpha_1=4.6^{\rm{o}}$ and $T_c = 146$ MeV for the choice  $\mu_{B,c}=350$ MeV.

Fig. \ref{fig:SpinodalIsotherms_OrigBEST} shows a QCD isotherm for the pressure in the case of such a shallow mapping of $\alpha_1=4.6^{\rm{o}}$ along the chiral phase transition line.
First, the mapping of the features of the first order phase transition from the Ising model to QCD should be understood.
In this figure, the three solutions for the pressure, coming from the three solutions for $M(h)$ as shown in Eq.~\eqref{eq:Mofh}, are shown in different colors along with the value of temperature to which this isotherm corresponds.
The coexistence region begins where the first two solutions, shown here in yellow and blue, meet at $\mu_B>\mu_{B,c}$, while the spinodal points are located where the first two solutions meet the third one shown in orange (see Fig. \ref{fig:IsingPvh}).
In this case of a small angle, we do not capture the point where the second and third solutions meet, since this is at a different temperature (see discussion in Sec. \ref{sec:spinodals_location}).
Here, we chose an isotherm that is close to the critical point, $T=0.95 T_c=140$ MeV, in order to depict the behavior of the spinodals, particularly to point out that the shape of the isotherms becomes distorted (compare to Fig. \ref{fig:IsingPvh}).
In fact, the rightmost spinodal becomes stretched out, moving through the phase diagram to another value of $T$ and $\mu_B$.

\begin{figure}
    \centering
    \includegraphics[width=0.48\textwidth]{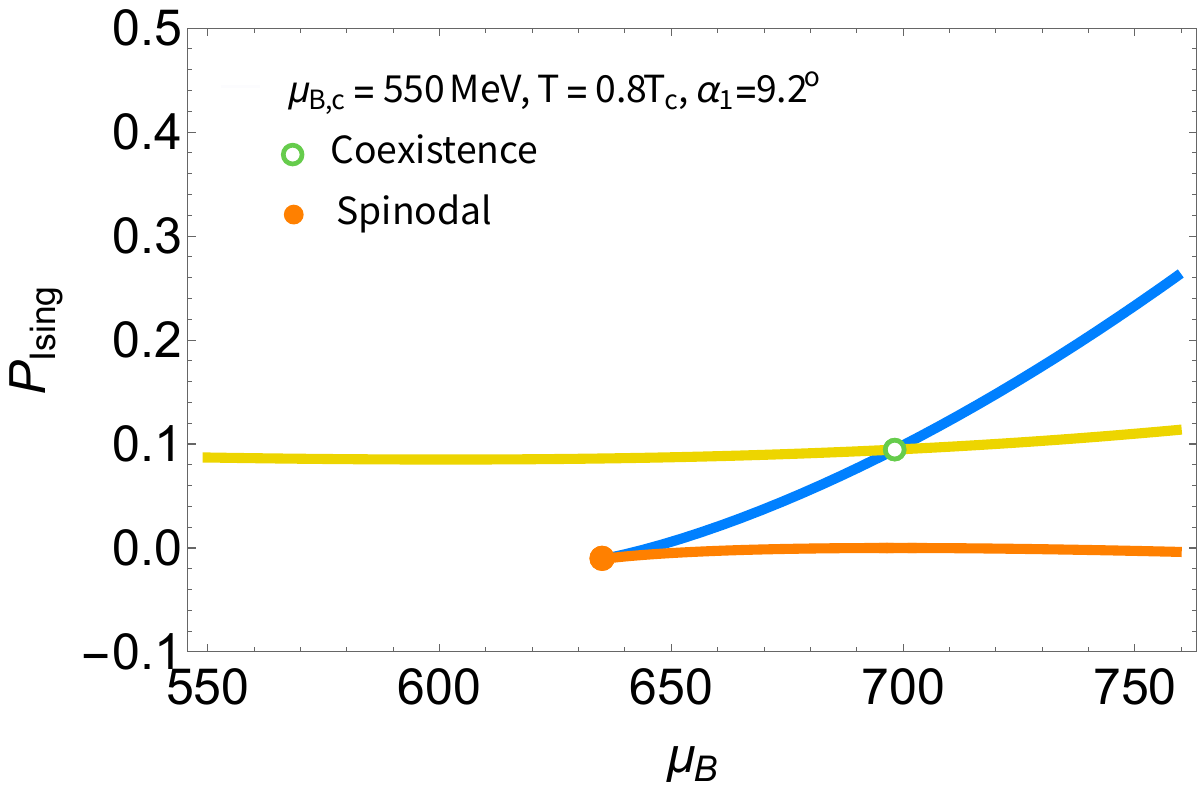}
    \includegraphics[width=0.49\textwidth]{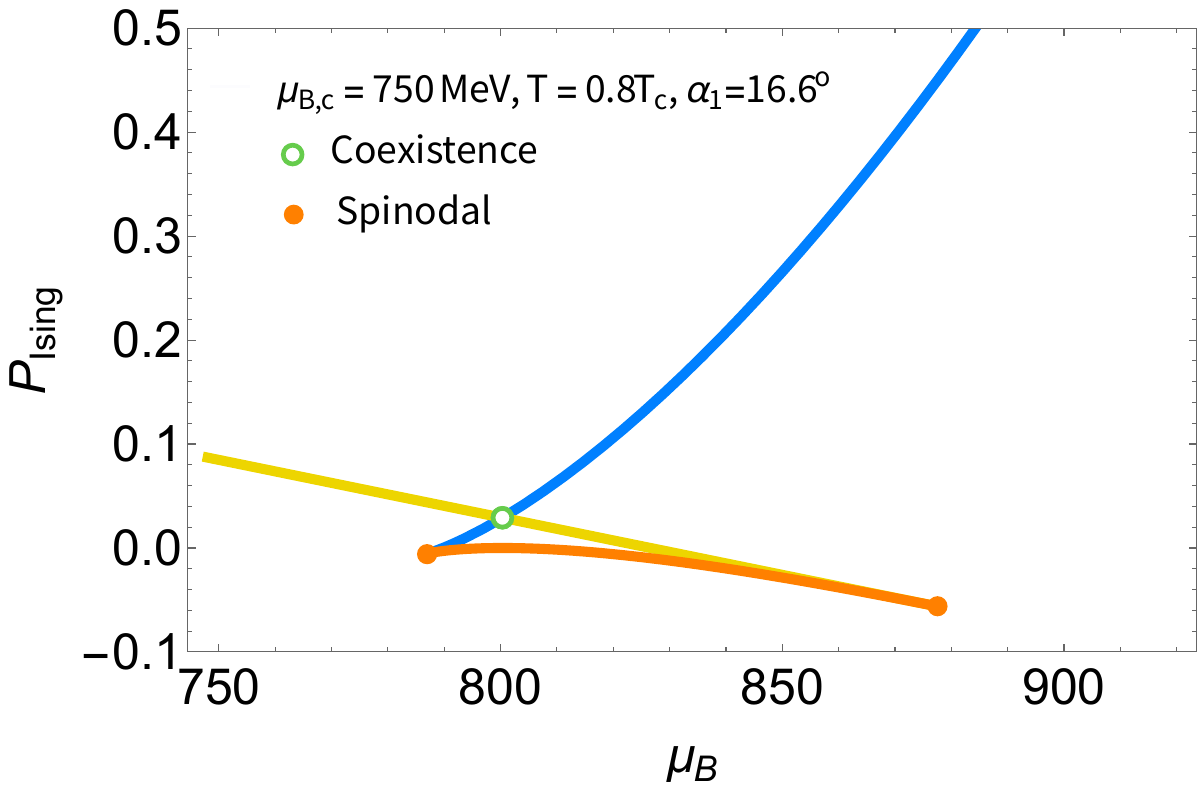}
    \includegraphics[width=0.49\textwidth]{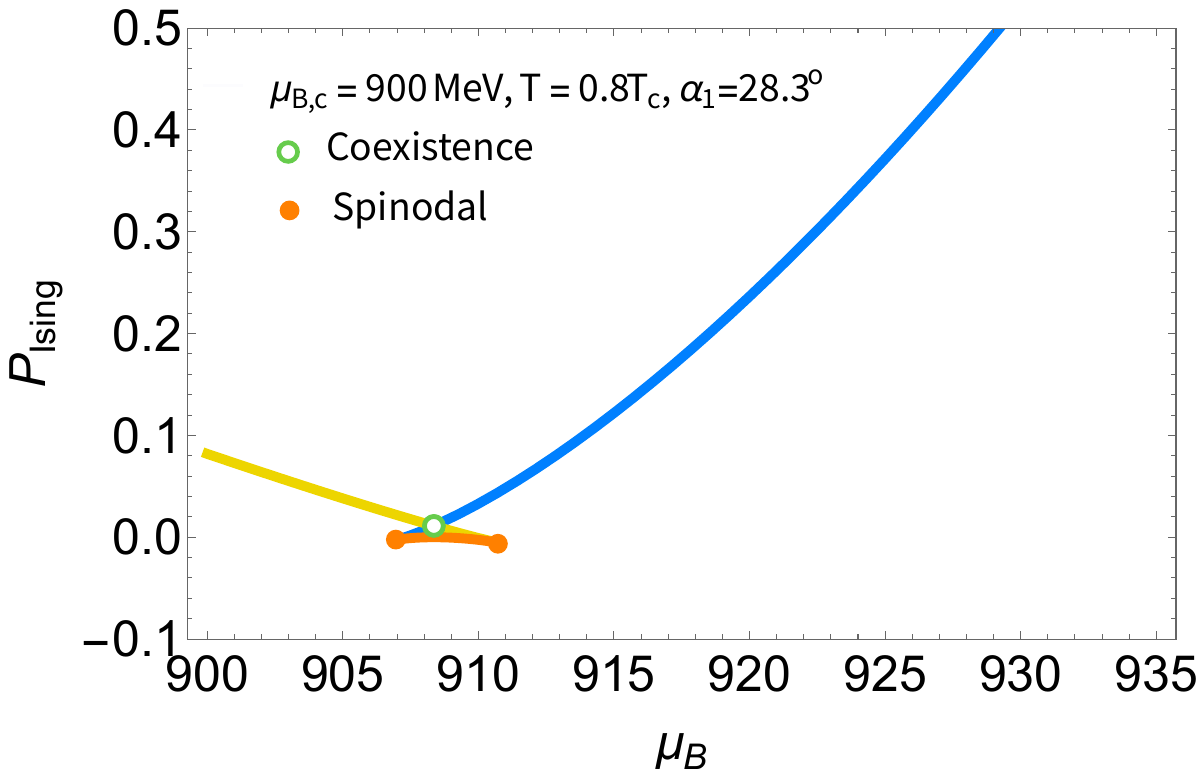}
    \caption{Pressure as a function of the chemical potential at fixed temperature $T=0.8T_c$ in the mean field Ising model as mapped onto the QCD phase diagram, for various choices of $\mu_{B,c}$.}
    \label{fig:SpinodalIsotherms}
\end{figure}

After showing the isothermal behavior for small angles $\alpha_1$, we explore other choices for the location of the critical point, which consequently change $\alpha_1$.
The angle of the mapping tangent to the chiral phase transition line increases with increasing $\mu_B$ as the parabola approaches its terminus at $T=0$.
Results for the isotherms for several different options for the location of the critical point at $\mu_{B,c}=550, \, 750, \, 900$ MeV are shown in Fig. \ref{fig:SpinodalIsotherms}.
As discussed in Sec. \ref{sec:mapping}, when choosing the critical point along the chiral phase transition line we obtain a constraint on the angle $\alpha_1$.
These options for placing the critical point lead to mapping angles of $\alpha_1= 9.2^{\rm{o}}, \, 16.6^{\rm{o}}, \, 28.3^{\rm{o}}$, respectively.
We show that the spinodal features seen in the isothermal trajectories  are still distorted for an angle of $9.2^{\rm{o}}$, at least at this temperature relatively close to the critical point.
On the other hand, for the subsequent choices corresponding to $\alpha_1 = 16.6^{\rm{o}}, 28.3^{\rm{o}}$, we capture the typical isothermal behavior.
In particular, we can now identify the metastable phases beyond the coexistence point.
We see that both these choices for the placement of the critical point give rise to super-heated and super-cooled phases, where the blue and yellow solutions track past the coexistence point and begin to describe a metastable state.
In between the spinodal points lies the instability and the unique unstable solution shown in orange.
From this, we demonstrate the strong influence that the angle $\alpha_1$ has on the mapping of the spinodal points onto the QCD phase diagram. 
This is, however, not the only parameter affecting the spinodals in the QCD phase diagram.
We show the effect of additional parameters in Appendix \ref{sec:appendix}.

\subsection{Spinodals in the QCD Phase Diagram}\label{sec:spinodals_phase_diag}

Beyond a single isothermal curve for the pressure as a function of baryon chemical potential, we determine the spinodal curves in the QCD phase diagram.
From Eq. \eqref{eq:MappedSpinodals} we obtain the $T$- and $\mu_B$-dependence of the spinodals in order to determine the extent of the spinodal region in the phase diagram for a given choice of mapping parameters.
We choose the same three locations for the critical point as those shown in Fig. \ref{fig:SpinodalIsotherms}, which are $\mu_{B,c}=550,750,900$ MeV.

\begin{figure}
    \centering
    \includegraphics[width=0.48\textwidth]{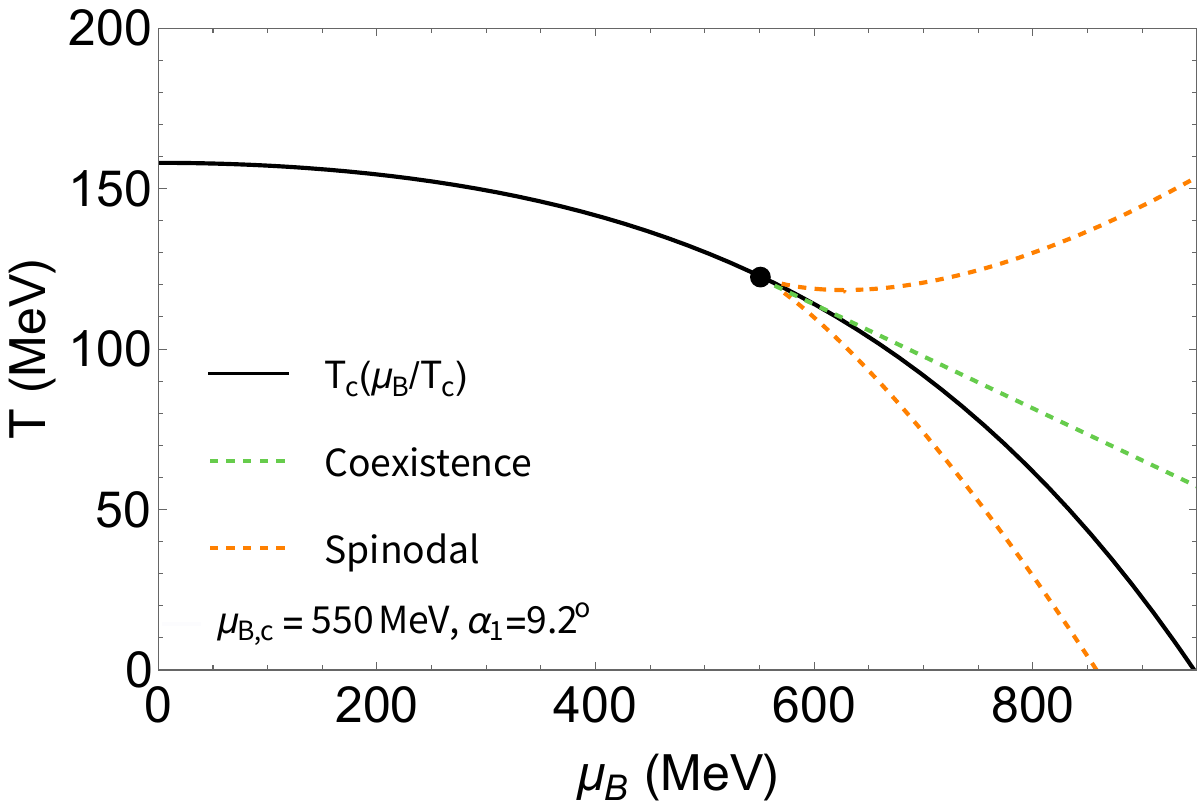}
    \includegraphics[width=0.49\textwidth]{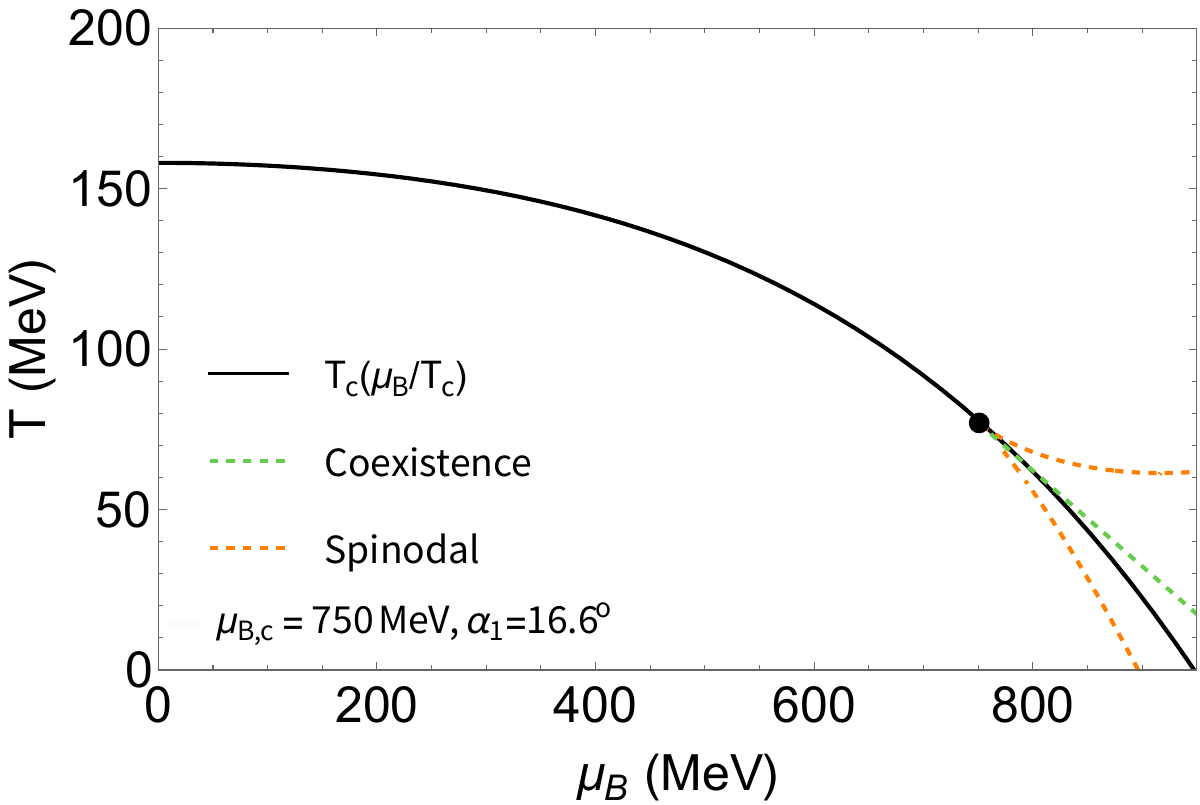}
    \includegraphics[width=0.49\textwidth]{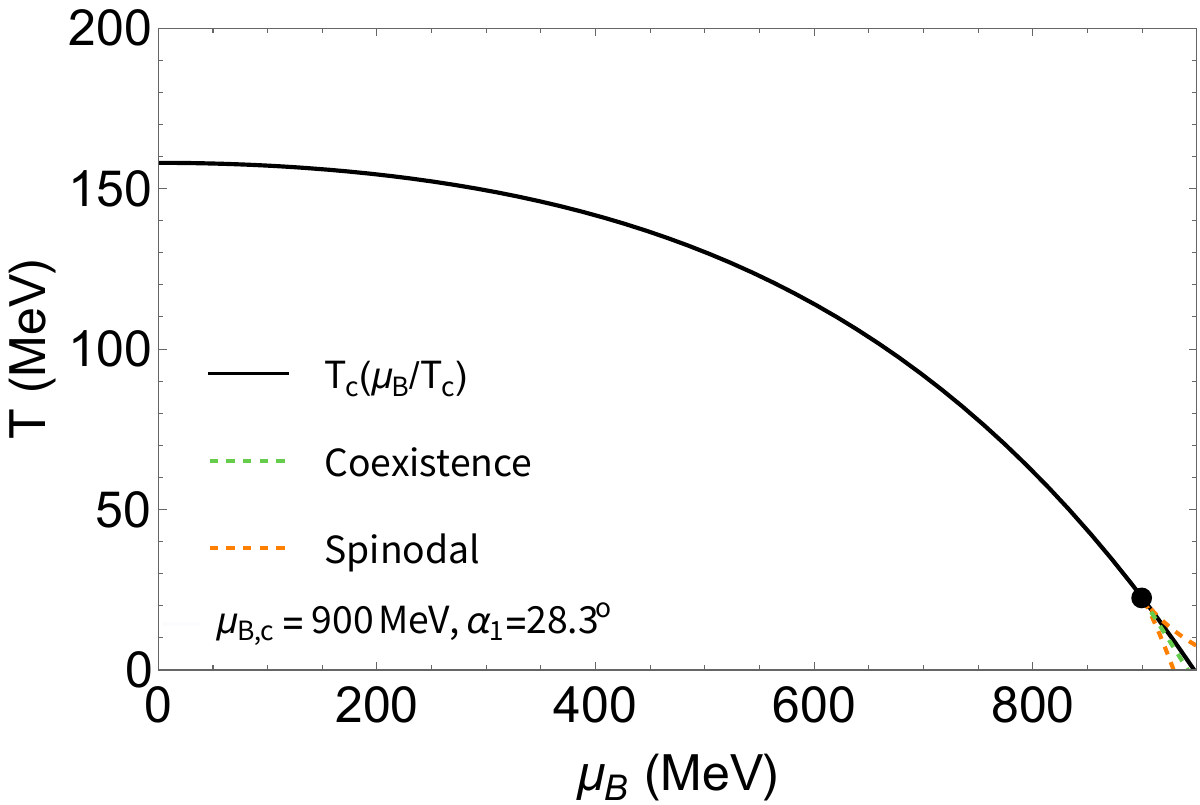}
    \caption{Behavior of the spinodal curves in the QCD phase diagram for $\mu_{B,c}=550$ MeV (upper panel), $\mu_{B,c}=750$ MeV (central panel) and $\mu_{B,c}=900$ MeV (lower panel).
    }
    \label{fig:SpinodalPhDiag}
\end{figure}

The results for the spinodal lines in the QCD phase diagram are shown in Fig. \ref{fig:SpinodalPhDiag} for $\mu_{B,c}=550$ MeV (top), $\mu_{B,c}=750$ MeV (middle), $\mu_{B,c}=900$ MeV (bottom), all with the choice of scaling parameters $w=1,~ \rho=2$.
In each of these plots, our parametrization of the chiral phase transition line is shown along which sits the critical point, given as the filled circle, for each of the three choices.
We see that, for smaller values of $\mu_{B,c}$ and consequently of the angle $\alpha_1$, the metastable region covers more of the phase diagram.
We also show the coexistence line as mapped onto the phase diagram given our choice of the mapping shown in Eq. \eqref{eq:mapTmuB}.
In this case, the transition line ($h=0$) from the Ising model is a straight line tangent to the chiral phase transition line ($T_c(\mu_B/T_c)$) due to the fact that we employ this linear map.
The coexistence line should rather curve down toward the $\mu_B$-axis, and when $T=0$ it should meet the axis at $90^{\rm{o}}$, as discussed in Sec. \ref{sec:spinodalisotherms}.
For the largest angle choice, we see that the linear approximation follows closely the chiral phase transition line, as the tangent line becomes a better approximation as the angle increases.
As we are focused mainly on the spinodal implementation in the QCD phase diagram, we leave the study of an alternative mapping procedure to future work.

For the first choice of $\mu_{B,c}$, the high-temperature spinodal never reaches the $T=0$ axis, but rather moves away from it at large chemical potential. 
At this point, we compare our result for the spinodal lines in the phase diagram to those obtained in the holographic model in Refs. \cite{Grefa:2021qvt,Hippert:2023bel}. The holographic model provides an excellent description of lattice QCD thermodynamics, it naturally incorporates the ideal fluidity of the Quark-Gluon Plasma and it predicts a critical point on the QCD phase diagram. Starting from the critical point, at chemical potentials $\mu_B>\mu_{B,c}$, the first-order phase transition (coexistence) line departs, surrounded by the two spinodals. Entropy density, energy density and net-baryon density become multi-valued solutions of the temperature at fixed chemical potential in this region, and the high-temperature spinodal line runs almost parallel to the $T=0$ axis at large $\mu_B$, similarly to what happens in the middle panel of Fig. \ref{fig:SpinodalPhDiag} in our case.
Additionally, critical exponents were extracted from this holographic approach and were found to be that of mean field \cite{DeWolfe:2010he},
thus further linking these two sets of results.
From this we also see that, if the parameters of the mapping can be constrained by experimental efforts in the search for the QCD critical point, we will be able to determine the size of the spinodal region in the QCD phase diagram.

\subsection{Pressure-versus-density isotherms}

Since QCD is symmetric for the exchange of matter with anti-matter, the charge conjugation symmetry must be obeyed in our equation of state across the $\mu_B=0$ line.
Furthermore, as can be seen from the isothermal trajectories, without symmetrization the density would be negative for the solution shown in Fig. \ref{fig:SpinodalIsotherms} in yellow which has a negative slope.
The symmetrization proceeds by utilizing the information at $-\mu_B$ such that:
\begin{equation} \label{eq:P_nB_symm}
\begin{split}
    P_{\text{symm}}(T,\mu_B) &= \frac{1}{2} (P(T,\mu_B) + P(T,-\mu_B)),  \\
    n_{B\text{,symm}}(T,\mu_B) &= \frac{1}{2} (n_B(T,\mu_B) - n_B(T,-\mu_B)).
\end{split}
\end{equation}
In order to show the effect of the symmetrization and calculate further characteristic signatures of the first order phase transition, we will proceed with one of the choices of the placement of the critical point, $\mu_{B,c}=550$ MeV.
Given the results for the spinodal curves in the phase diagram shown in Fig. \ref{fig:SpinodalPhDiag}, we choose an isotherm that is sufficiently close to the critical point such that both spinodal curves are captured by that isotherm.
With this in mind, we choose an isotherm of $T=0.98 T_c=120$ MeV. 

\begin{figure*}[t]
    \centering  
    \begin{tabular}{c c}
    \includegraphics[width=0.49\textwidth]{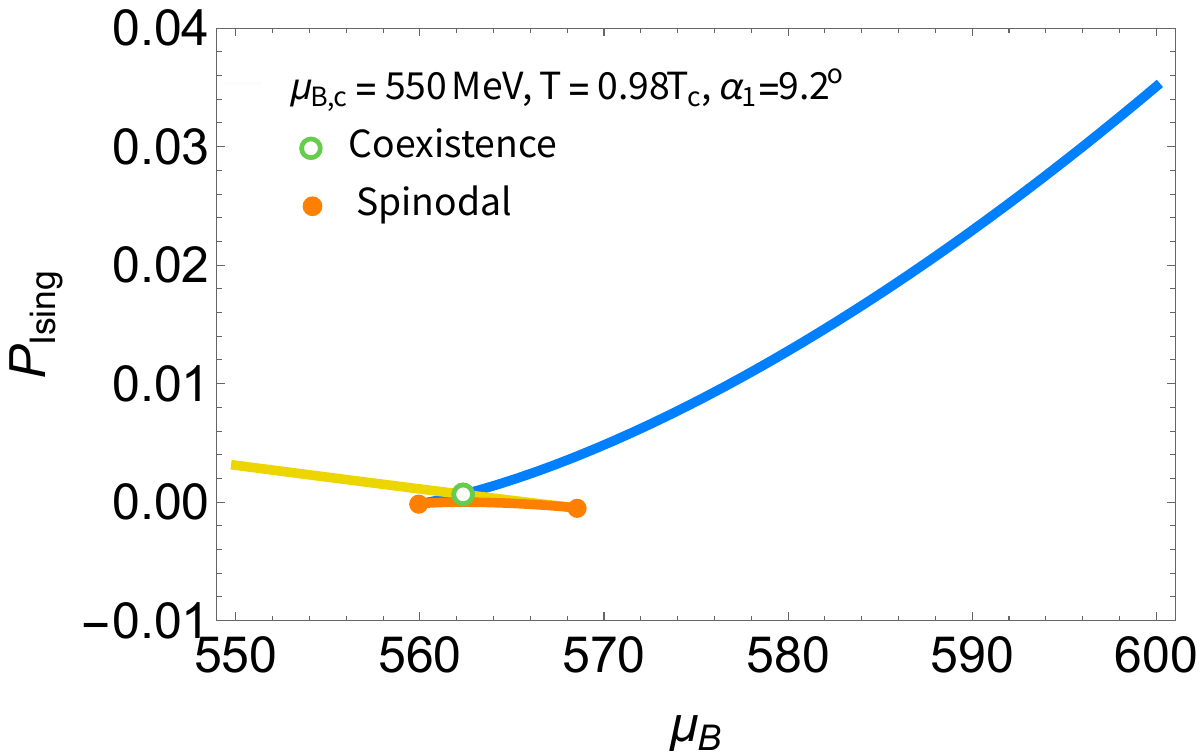} \includegraphics[width=0.49\textwidth]{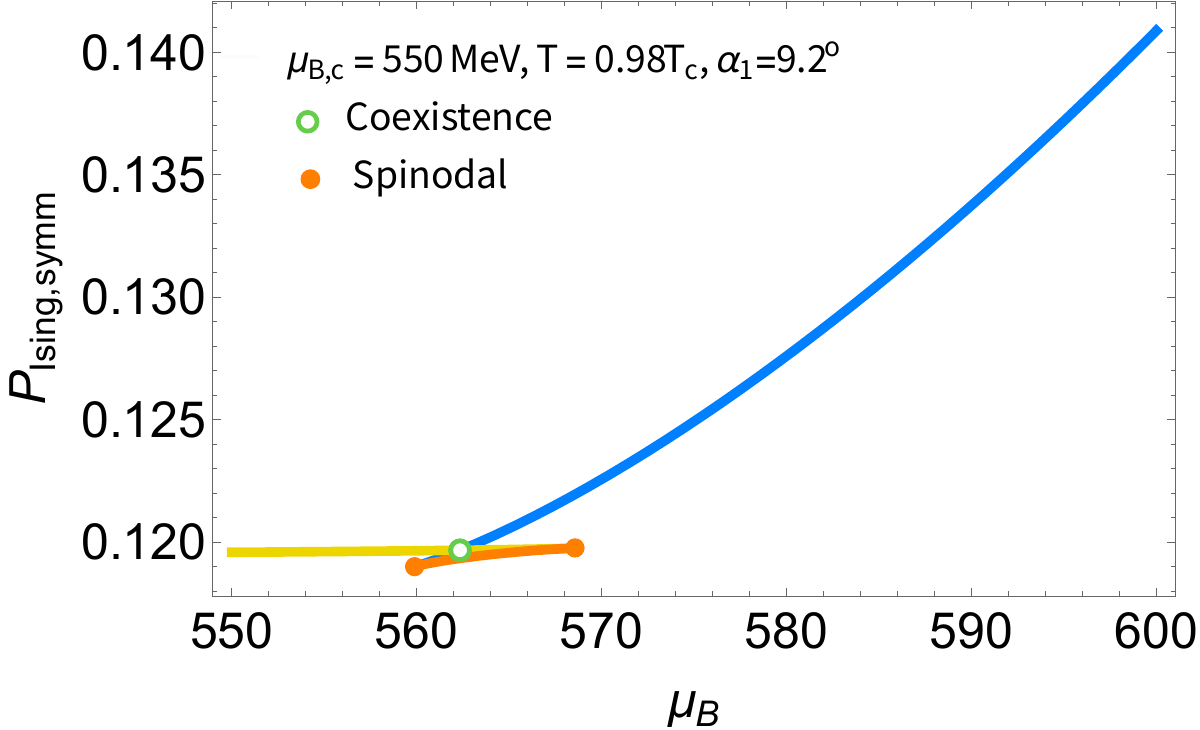}
     \end{tabular}
    \caption{Ising model pressure as a function of the chemical potential for $\mu_{B,c}=550$ MeV and $T=0.98T_c$ before symmetrization (left panel) and after symmetrization (right panel). The open green and filled orange circles indicate the location of the coexistence and spinodal points, respectively.}
    \label{fig:pIsingSymm}
\end{figure*}

In Fig. \ref{fig:pIsingSymm} we show the isothermal pressure curve for the choice motivated here of $\mu_{B,c}=550$ MeV, $T=120$ MeV, before and after the symmetrization process. 
It is evident that after symmetrization the sign of the slope changes for the left and middle solutions shown in yellow and orange. 
This is an important development as a result of the symmetrization because the slope of the curve $P(\mu_B)$ corresponds directly to the baryon density $n_B = (\partial P/\partial \mu_B)_T$.
The baryon density, thus, becomes positive, as we would expect for our system in a matter-dominant regime with positive baryon chemical potential.

\begin{figure*}
    \centering
    \begin{tabular}{c c}
    \includegraphics[width=0.49\textwidth]{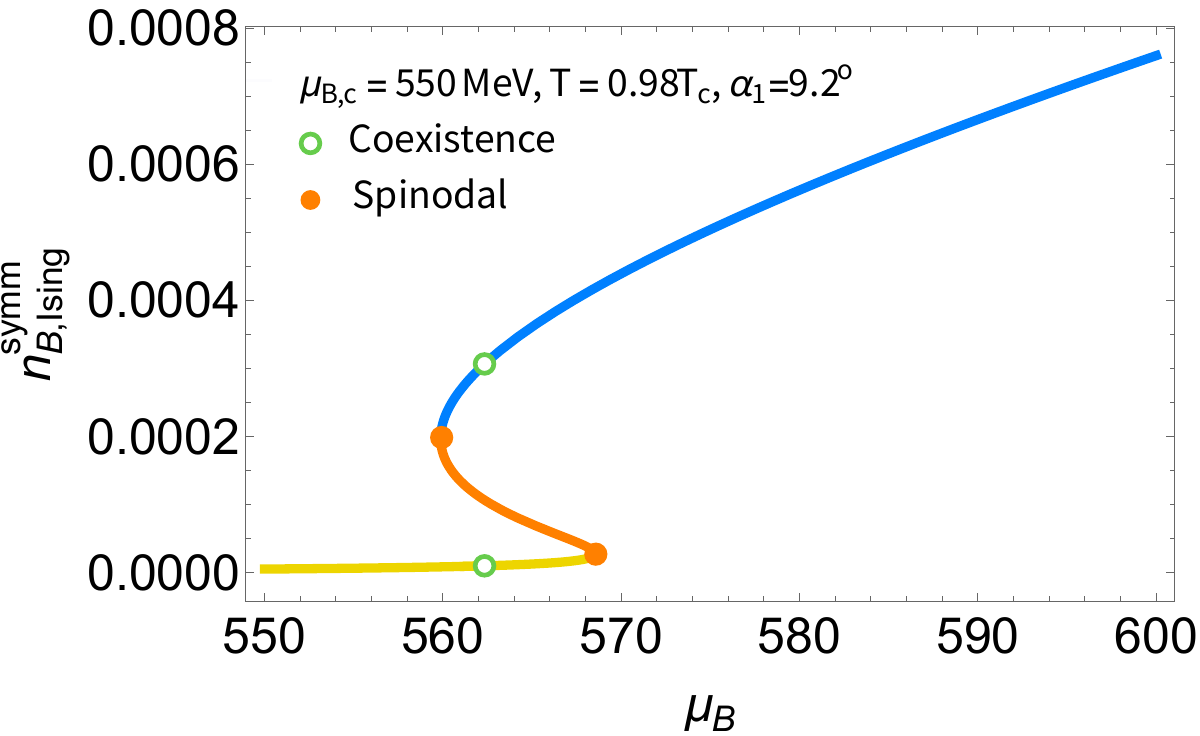} \includegraphics[width=0.47\textwidth]{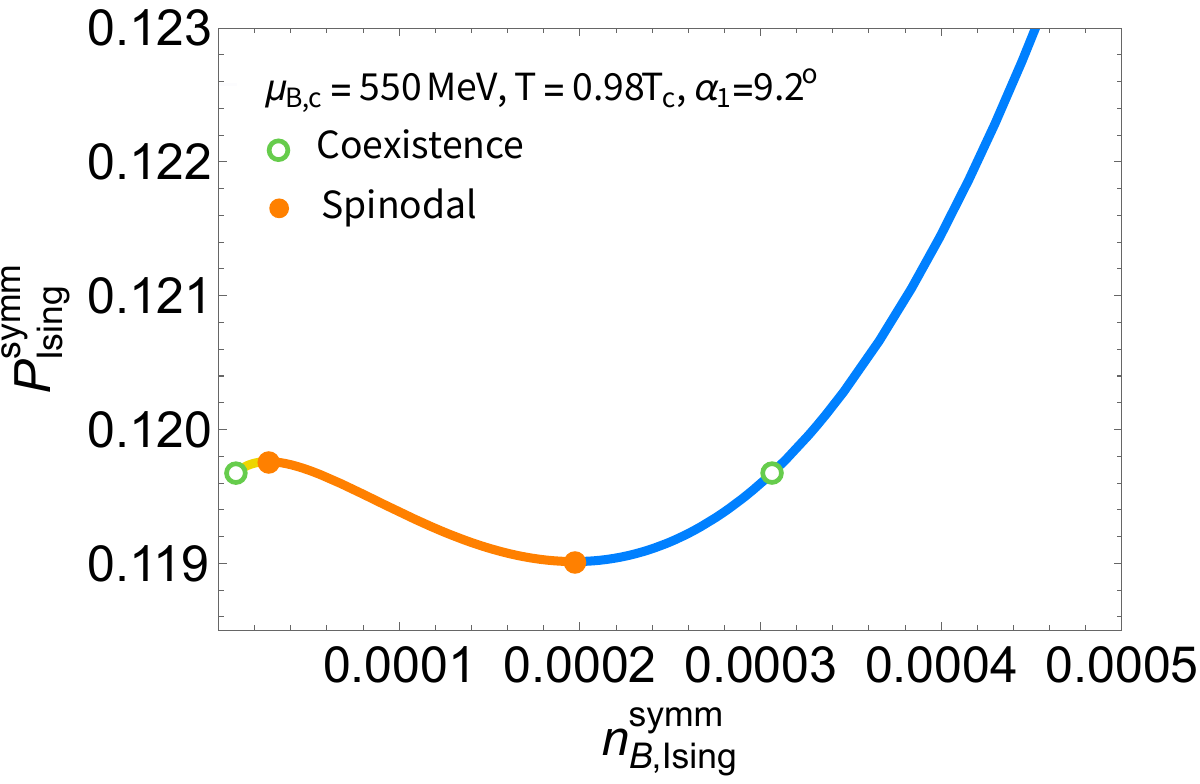}  
    \end{tabular}
    \caption{Symmetrized Ising model density as a function of the chemical potential (left) and the symmetrized pressure as a function of the symmetrized density (right). The coexistence points are indicated with the open green circles, while the spinodals are given by the filled orange circles.
    }
    \label{fig:nBIsingSymm}
\end{figure*}

After the symmetrization is complete, we can study the behavior of the baryon density itself in the first order phase transition region for $\mu_B > \mu_{B,c}$.
Figure \ref{fig:nBIsingSymm} (left panel) shows the symmetrized baryon density from the Ising model as a function of the chemical potential, as well as the behavior of the pressure as a function of the
baryon density after symmetrization (right panel).
Important features are present in the baryon density, which were not previously available in the critical equation of state mapped to QCD with the 3D Ising model \cite{Parotto:2018pwx,Karthein:2021nxe}.
Since the spinodal features lie in the complex plane for the 3D Ising model as discussed in Sec. \ref{sec:Ising_EoS}, the density would simply jump from the coexistence point given by the open green marker on the blue curve to the next coexistence point on the yellow curve during the cooling process.
On the other hand, in our mean field Ising approach, we can access the full curve by proceeding through the spinodal points as well.
In this case, both the coexistence points along the first order phase transition line and the spinodal points reaching out from them are now clearly present for each of the solutions of the critical baryon density.
Here, we show the three different solutions given by the yellow, orange, and blue curves with the coexistence and spinodal points highlighted in open and filled circles, respectively.
Similarly for the behavior of the pressure as a function of the baryon density, we achieve a result that shows the progression through the mixed phase.
The orange curve shown here depicts a mechanical instability where $\partial P/ \partial n_B < 0$, further affirming that this solution represents the unstable phase.

\section{Full Thermodynamic Results}
\label{sec:fullthermo}

We implement the critical features of the mean field Ising model as mapped onto the QCD phase diagram in such a way that the Taylor expansion coefficients of our final pressure match the ones calculated on the lattice \cite{Guenther:2017hnx} order by order:
    \begin{equation} \label{eq:Taylorcoeffmatch}
        T^4 c_n^{\rm{LAT}}(T) = T^4 c_n^{\rm{Non-Ising}}(T) + T_c^4 c_n^{\rm{Ising}}(T).
    \end{equation}
\begin{figure*}[t]
    \centering 
    \begin{tabular}{c c c} \includegraphics[width=0.325\textwidth]{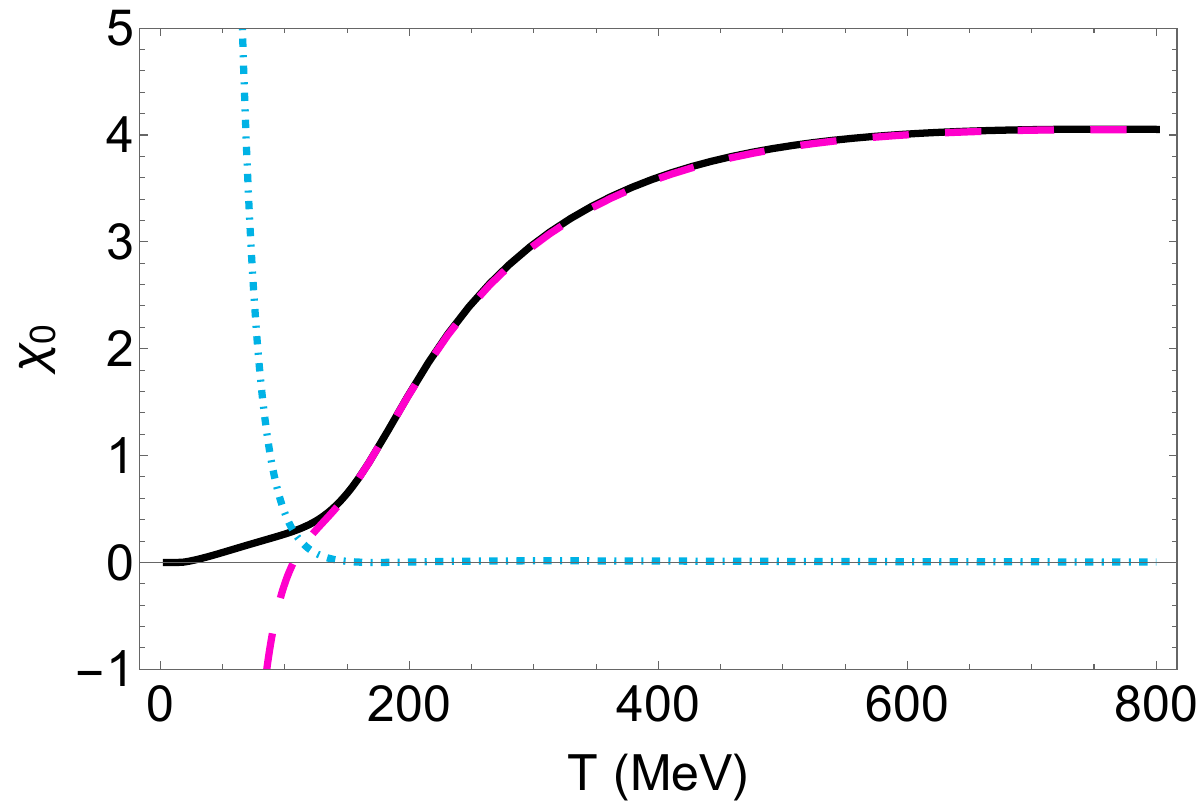}  &\includegraphics[width=0.333\textwidth]{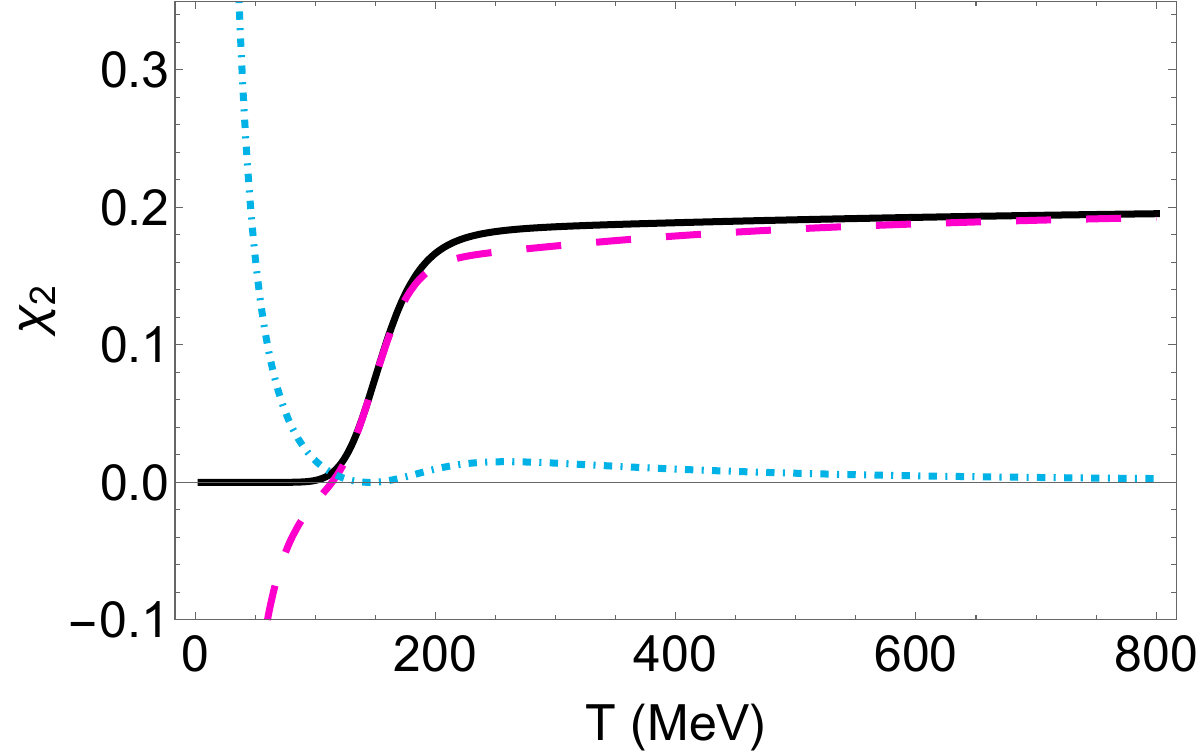} &\includegraphics[width=0.333\textwidth]{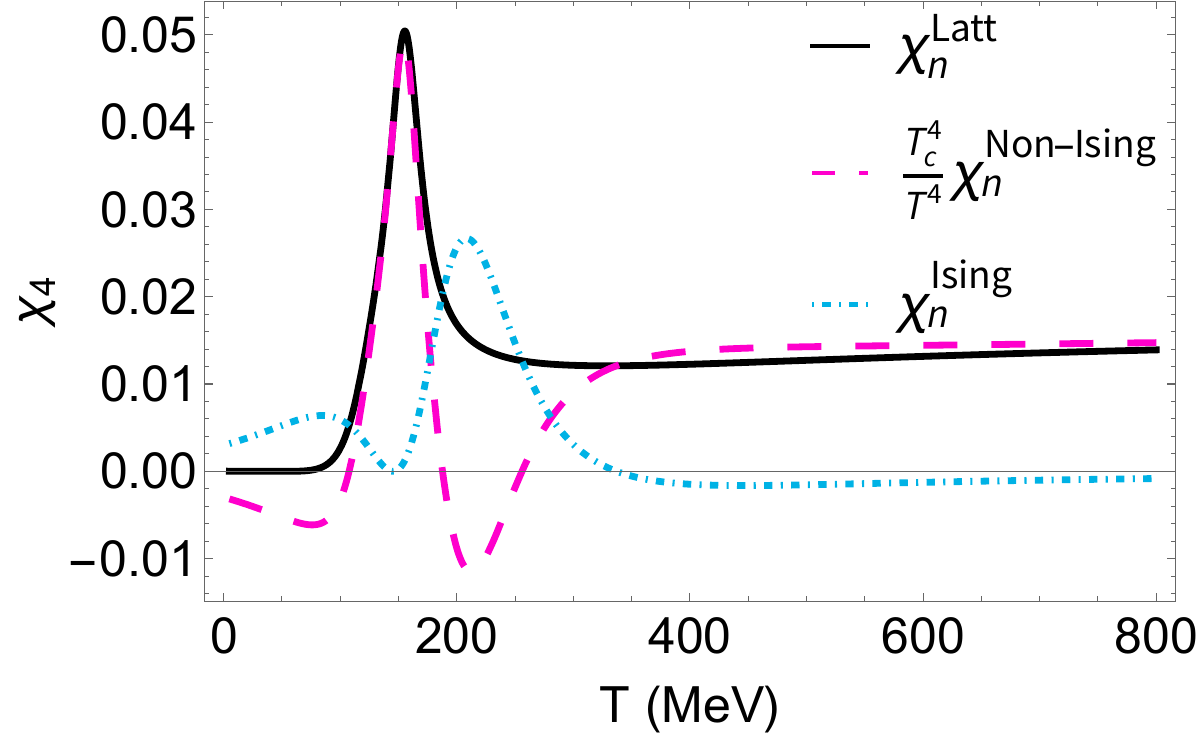}
    \end{tabular}
    \caption{Zeroth-order (left), second-order (middle) and fourth order (right) Taylor expansion coefficients. In all figures, the black full lines are parametrized lattice QCD results \cite{Guenther:2017hnx}, the blue dotted lines correspond to the Ising model result, and the dashed magenta lines are the non-Ising contribution.
    }
    \label{fig:Taylor}
\end{figure*}

\begin{figure*}[t]
    \centering 
    \begin{tabular}{c c c} \includegraphics[width=0.325\textwidth]{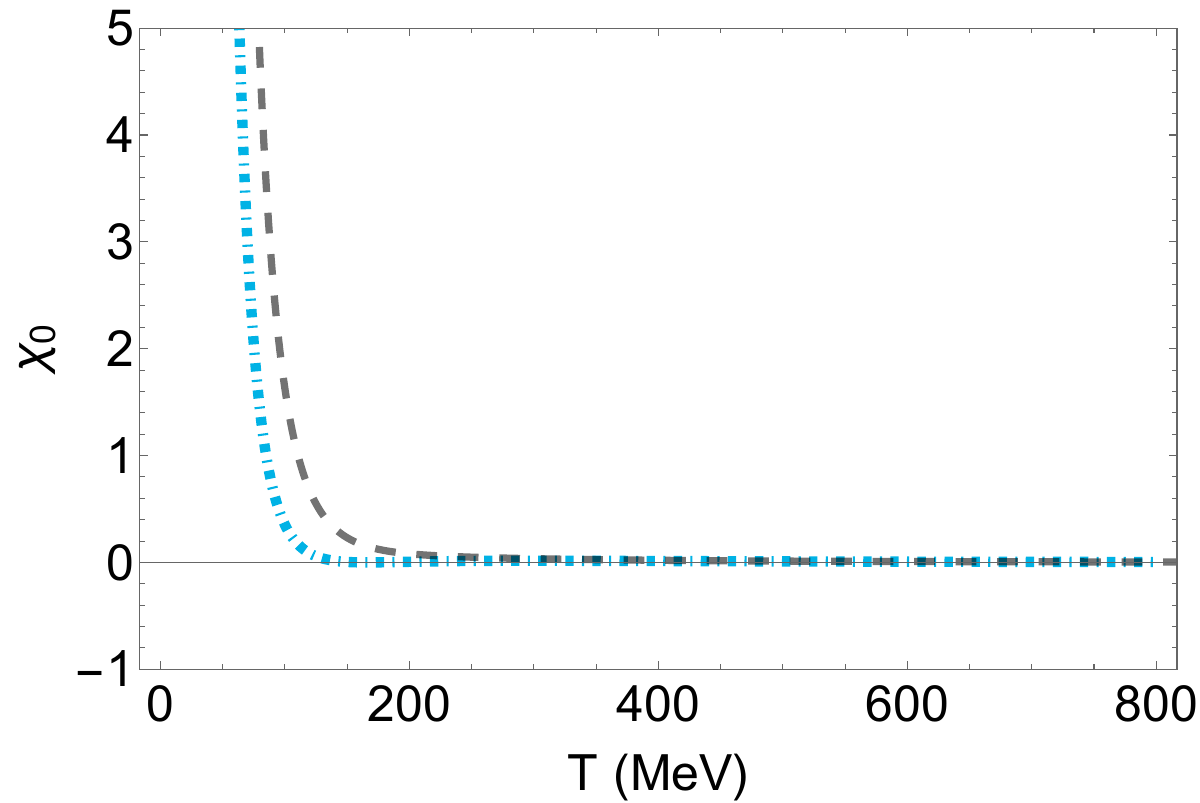}  &\includegraphics[width=0.333\textwidth]{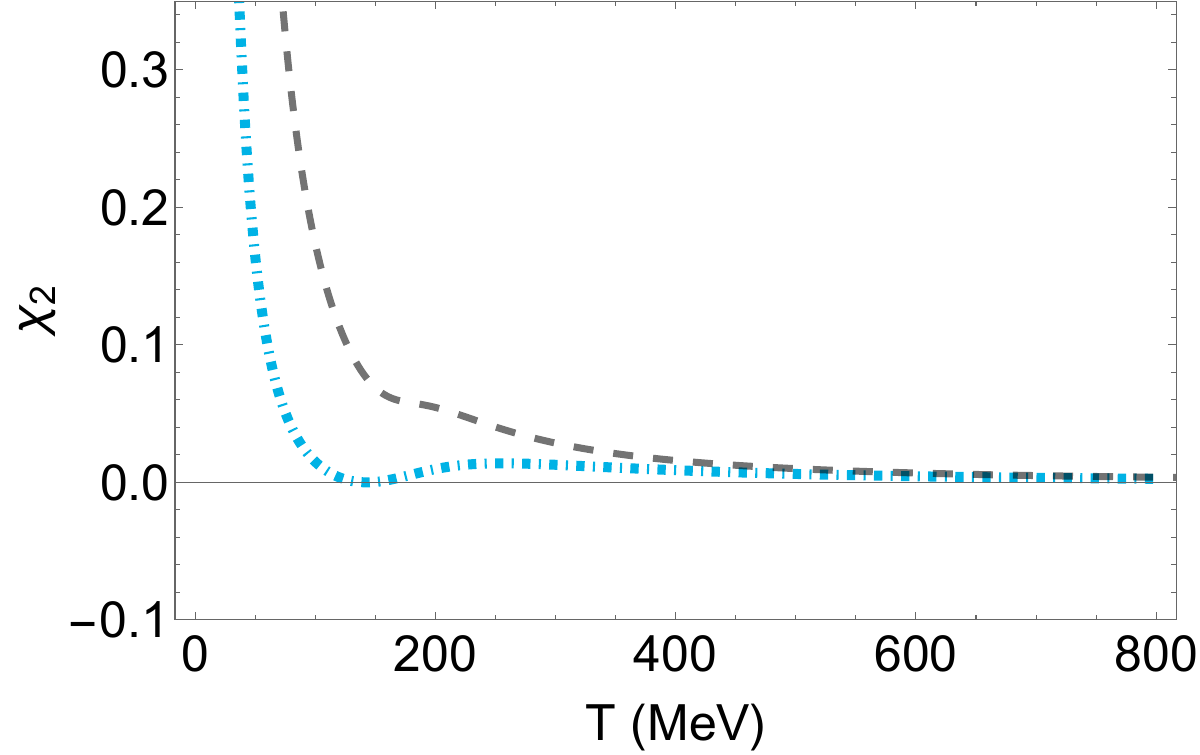} &\includegraphics[width=0.333\textwidth]{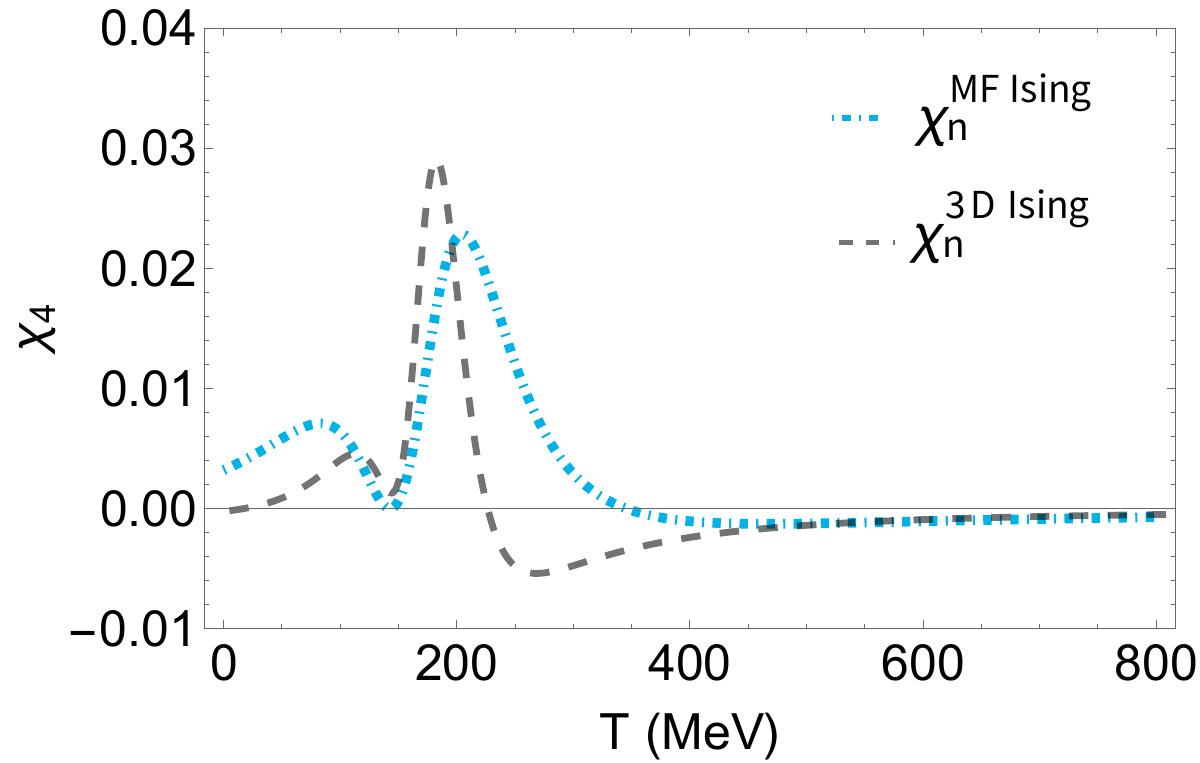}
    \end{tabular}
    \caption{Zeroth-order (left), second-order (middle) and fourth order (right) Taylor expansion coefficients. In all figures, the blue dotted lines are the same as those in Fig. \ref{fig:Taylor}, namely the mean field Ising model results. The Taylor expansion coefficients from the 3D Ising model used in Refs. \cite{Parotto:2018pwx,Karthein:2021nxe} are given by the gray dashed curves.}
    \label{fig:IsingChisMF3D}
\end{figure*}

The Taylor coefficients from lattice QCD that we choose to work with are those with phenomenologically relevant condition of strangeness neutrality.
In general, we can write the thermodynamic pressure as a sum of singular and non-singular contributions.
This procedure of matching to lattice QCD gives us our background, or Non-Ising, pressure that contributes along with the singular equation of state from the mean field Ising model.
Namely, our Non-Ising pressure is given by the difference between the lattice and Ising Taylor coefficients. 
The different contributions to the Taylor coefficients are shown in Fig. \ref{fig:Taylor}.
Note that, by following this construction, any differences at $\mu_B=0$ between the choices of mean field or 3D Ising model will be absorbed into the background Non-Ising pressure.
For completeness, however, we show the comparison between the mean field Ising Taylor coefficients and the ones from the 3D Ising model in Fig. \ref{fig:IsingChisMF3D}.
From this, we show that the general features are very similar between the choices, i.e. the number of extrema/inflection points.

In order to achieve the full pressure with both singular and non-singular contributions, we reconstruct the pressure as:
    \begin{equation} \label{eq:fullpress}
    \begin{split}
        P(T,\mu_B)=T^4 \sum_n c_n^{\rm{Non-Ising}}(T) \left(\frac{\mu_B}{T} \right)^n
        \\
        + T_c^4 P_{\rm{crit}}^{\rm{QCD}}(T,\mu_B),
    \end{split}
    \end{equation}
where $P_{\rm{crit}}^{\rm{QCD}}$ is the critical contribution to the pressure from the mean field Ising model that has been symmetrized and mapped onto the QCD phase diagram as described in Sec. \ref{sec:mapping}.

In order to obtain the appropriate low temperature behavior, we merge the reconstructed pressure with the pressure from the HRG model along a curve $T'(\mu_B)$ parallel to the transition line shown in Eq. \eqref{eq:chiral_phase_trans_OmuB4} and Fig. \ref{fig:ChPhaseTransChoice}, with an overlap region of $\Delta T=17$ MeV:
    \begin{equation} \label{eq:Pmerging}
    \begin{split}
        \frac{P_{\text{Final}}(T,\mu_B)}{T^4} = \frac{P(T,\mu_B)}{T^4} \frac{1}{2}
        \Big[1 + \tanh{\Big(\frac{T-T'(\mu_B)}{\Delta T}}\Big)\Big] \\
        + \frac{P_{\rm{HRG}}(T,\mu_B)}{T^4} \frac{1}{2}
        \Big[1 - \tanh{\Big(\frac{T-T'(\mu_B)}{\Delta T}}\Big)\Big].
    \end{split}
\end{equation}
This particular method for constructing an equation of state with a critical point was developed in Ref. \cite{Parotto:2018pwx}.
For a thorough discussion of choices made and an investigation of the parameter space, we refer the reader to that work.

\begin{figure*}
    \includegraphics[width=0.67\textwidth]{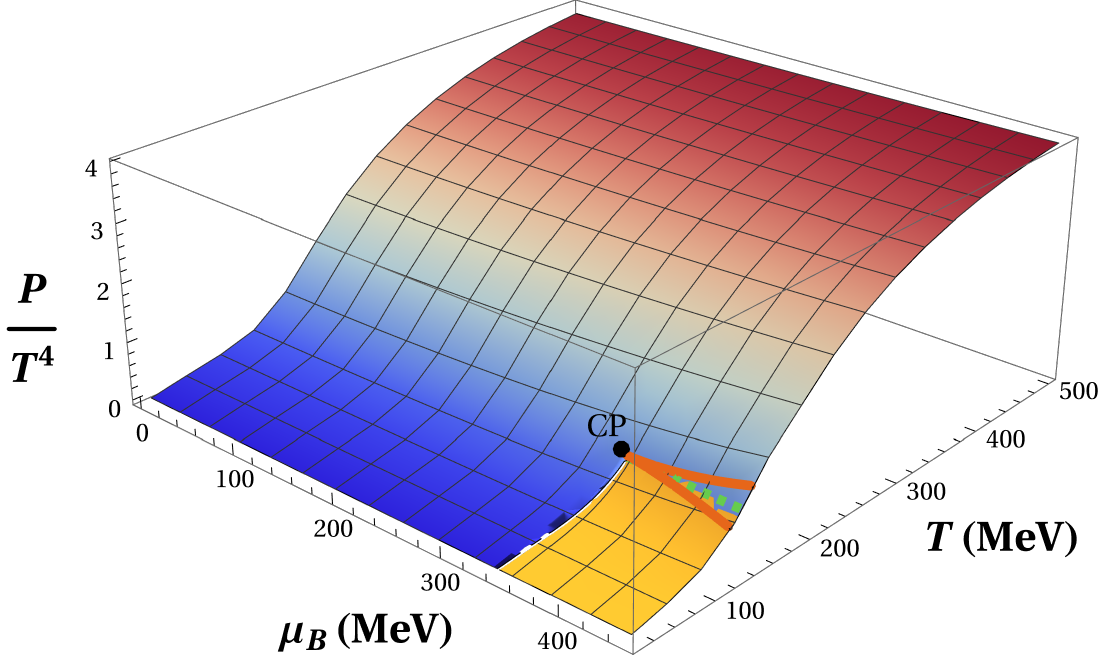} \includegraphics[width=0.52\textwidth]{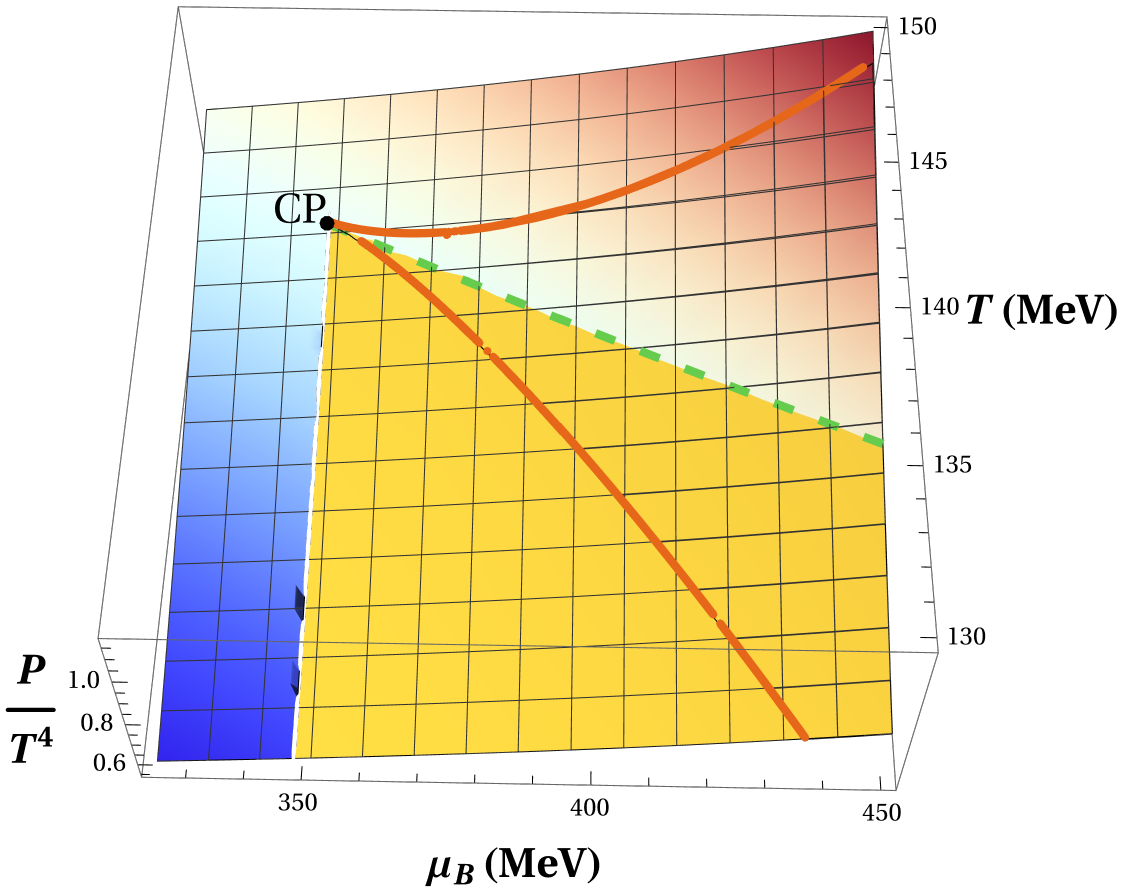} \includegraphics[width=0.46\textwidth]{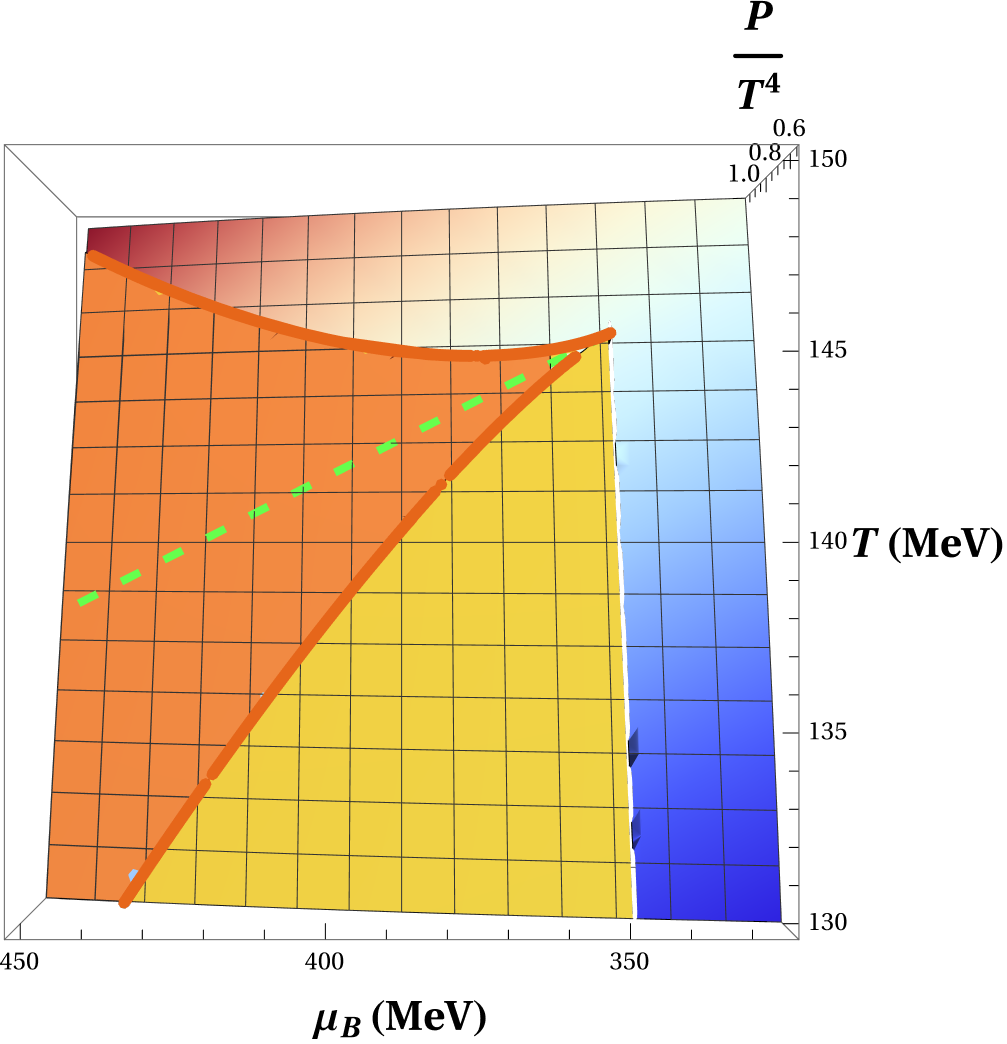}
    \caption{Full result for the pressure as a function of the temperature and baryon chemical potential matching lattice QCD at $\mu_B=0$, with a critical point placed at $\mu_{B,c}=350$ MeV and with all mapping parameters chosen similarly to Ref. \cite{Parotto:2018pwx} for comparison to that original work. The top panel shows the full coverage of the phase diagram for $(0 \leq \mu_B \leq 450)$ MeV and $(30 \leq T \leq 500)$ MeV. The two bottom panels focus in on the critical and first order regions for the same pressure as in the top panel. The mean field Ising model contributes three solutions: one which is always real (red/blue gradient) and two complex conjugate solutions that are only real for $\mu_B>\mu_{B,c}$ (yellow and orange). In the top panel, the ``always-real" solution covers most of the phase diagram, while beyond the critical point in the first order region an additional solution shown in yellow can be seen along with the spinodal (solid) and coexistence (dashed) lines. The third and final solution has a smaller pressure and as such lies beneath the other two. The bottom left panel shows a top view in order to highlight the coexistence line shown where the gradient and yellow solutions meet. Additionally, the bottom right panel shows a bottom view (a reflection of the top view panel around the vertical axis) which displays the third solution and shows where it meets the first two, corresponding to the two spinodal lines in the phase diagram.
    }
    \label{fig:pFullThermo}
\end{figure*}

After the merging with the HRG model at low temperatures, we ultimately obtain the full thermodynamic results in the phase diagram as shown in Fig. \ref{fig:pFullThermo}.
This is, thus, the full pressure that matches lattice QCD results at $\mu_B=0$ and contains the critical features from the mean field Ising model.
Here we see the multi-valued nature of the thermodynamics in the first order transition region in the full phase diagram.
The top panel shows the behavior in a broad range of $T$ and $\mu_B$, while the two lower plots are focused around the critical region and first order line, the left of which is a view from the top while the right one is the perspective from the bottom. 
Firstly, we see that in the first order phase transition region we obtain access to the additional forms of the pressure which have been complex in nature on the crossover side.
The various solutions are shaded in different colors (gradient red/blue, yellow, orange) as also presented in the isothermal trajectories in Figs. \ref{fig:SpinodalIsotherms_OrigBEST}, \ref{fig:SpinodalIsotherms}, \ref{fig:pIsingSymm} (blue, yellow, orange).
The gradient colored portion corresponds to the solution for the pressure which is always real as described in Sec. \ref{sec:Ising_EoS}.
On the other hand, the solid colored yellow section visible in all plots, as well as the orange section visible from the bottom view, correspond to the complex solutions giving rise to the additional features along the first order line.
The solution which is always real is the analog to what was previously available from such studies in the 3D Ising model \cite{Parotto:2018pwx,Karthein:2021nxe}.
In fact, the results for the pressure from this mean field Ising approach and the case of the 3D Ising model agree quite well quantitatively.
Away from the critical region the agreement is sub-percent level, as expected since this region is informed by the first-principles lattice QCD input \cite{Guenther:2017hnx}.
On the other hand, near the critical point the difference is $\lesssim 10\%$.
Thus, the resulting equation of state from this approach is very comparable to the case of the scaling equation of state with 3D critical exponents \cite{Parotto:2018pwx,Karthein:2021nxe,Kahangirwe:2024cny}. 
However, what is different between these approaches is the characterization of the first order phase transition including describing the full coexistence region with the spindodals.
Given the small difference between the two equations of state, we expect the effects on the hydrodynamic evolution away from the first order region to be almost negligible.

Finally, with the full pressure as a function of $T$ and $\mu_B$ in these 3D plots, we have access to the full coexistence line where the solution for the pressure which is always real, shown in gradient red/blue, and the complex conjugate solution in yellow meet to form the first order transition line.
In Fig. \ref{fig:pFullThermo}, the coexistence line is given by the dashed green curve, particularly highlighted in the bottom left panel. 
As in the isotherms shown in Fig. \ref{fig:SpinodalIsotherms}, the coexistence point corresponds to where the always real solution (blue) meets the first complex conjugate solution which is real on the first order side (yellow), while the last solution gives the unstable branch (orange).
The limits of metastability are shown here as well in the spinodal lines, given by the orange curves going through all three panels of Fig. \ref{fig:pFullThermo}.
The bottom right panel is a view of this 3D plot from below.
This further allows us to see the spinodal curves in the phase diagram where this unstable solution meets the other two and demarcates the onset of thermodynamic instability, as is also shown in Fig. \ref{fig:nBIsingSymm}. 
Finally, the line at $\mu_B \sim \mu_{B,c}$ where the gradient red/blue and yellow solutions meet depicts the critical isotherm, i.e. the $r=0$ line.
From the pressure in the Ising phase diagram shown in Fig. \ref{fig:Ising_complex_sheets}, we see that this is the expected behavior.
While on the low temperature sheet ($r<0$) there are three real solutions, only one real solution exists on the high temperature sheet ($r>0$).
Thus, by crossing the $r=0$ line onto the high temperature sheet, the system collapses onto the single solution that is real for all $r$, shown in blue in the Ising phase diagram (Fig. \ref{fig:Ising_complex_sheets}) and gradient red/blue in the QCD phase diagram (Fig. \ref{fig:pFullThermo}).
Furthermore, from the behavior of the Ising pressure in Fig. \ref{fig:Ising_complex_sheets}, it is clear that the pressure is continuous where the blue and yellow sheets meet along the critical isotherm.
Therefore, the critical isotherm is not associated with a phase transition, as previously discussed in Sec. \ref{sec:Ising_EoS}.
We note that the unstable sheet, corresponding to the thermodynamic instability ($\partial n_B/\partial \mu_B < 0$), shown in orange, can only be seen from below due to the fact that this solution has, relatively, the smallest value of the pressure (see e.g. Fig. \ref{fig:SpinodalIsotherms}).
Although we can display each of these solutions uniquely in this 3D plot,  by increasing the size of the critical region, it is also possible to see the multi-valued behavior explicitly by eye for a different choice of scaling parameters $w, \rho$, which we show in Appendix \ref{sec:appendix}.

\begin{figure}
    \centering    \includegraphics[width=0.49\textwidth]{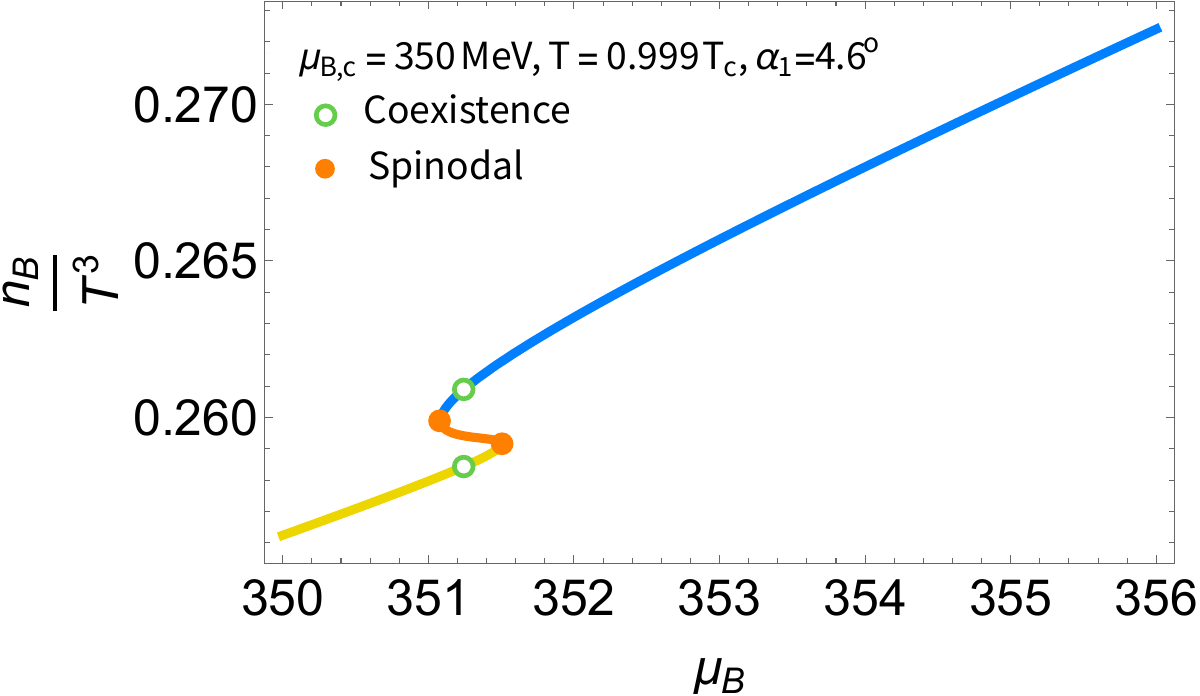} 
    \includegraphics[width=0.49\textwidth]{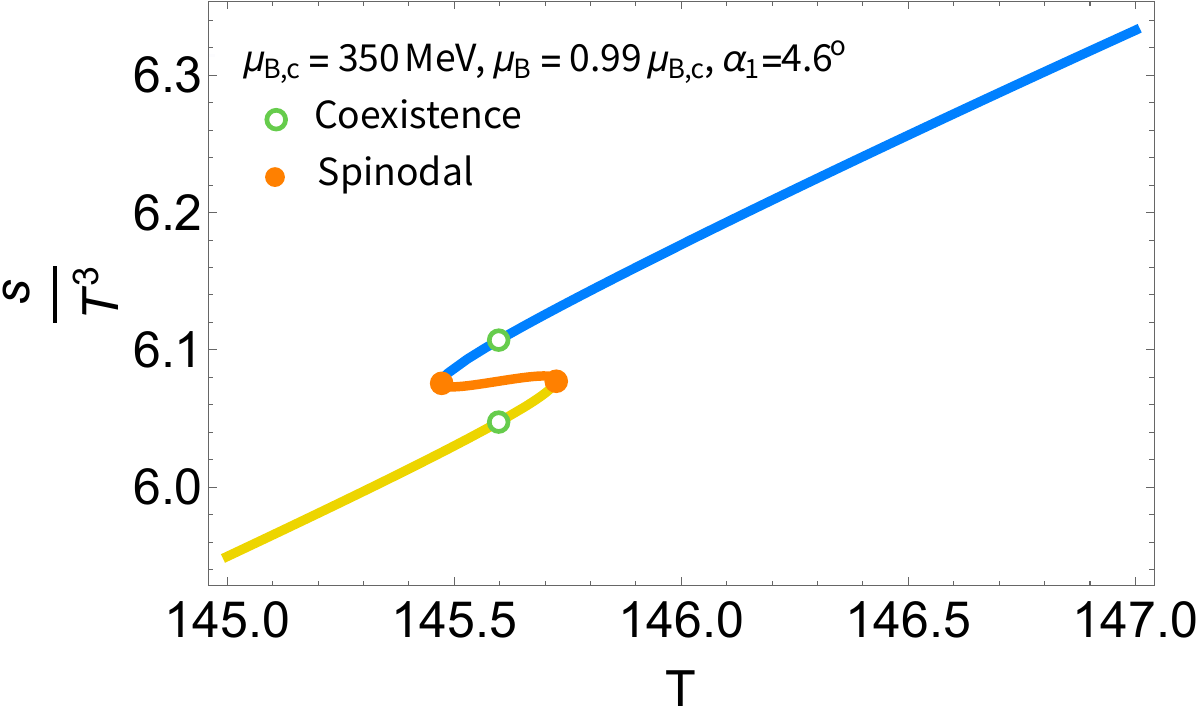}
    \includegraphics[width=0.49\textwidth]{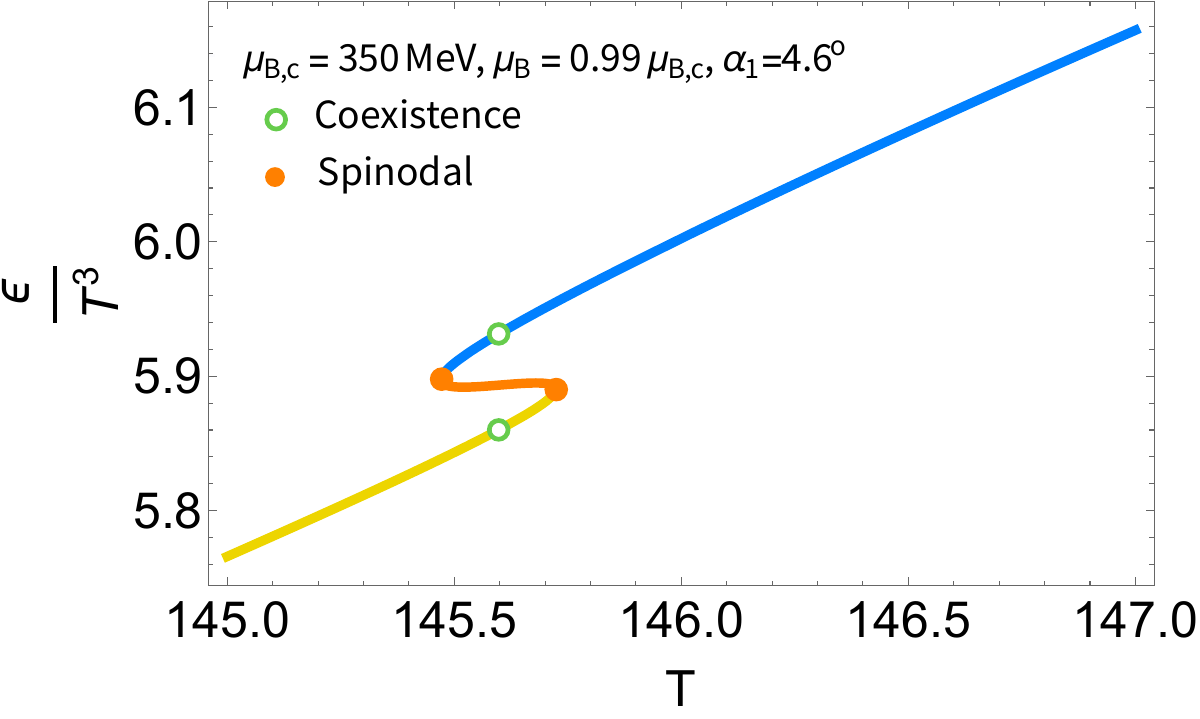}  
    \caption{Baryon density (top panel), entropy density (center panel), and energy density (lower panel) taken as derivatives of the full pressure matched to lattice QCD. An isotherm is shown for the baryon density, while constant baryon chemical potential is shown for the entropy and energy densities. The features of the first order phase transition line including the spinodal points are shown, beyond the expected jump at the coexistence points. The coexistence points are indicated with the open green circles, while the spinodals are given by the filled orange circles.}\label{fig:pDerivIsothermsFullResult}
\end{figure}
Next we consider curves for the thermodynamic derivatives of the pressure at constant temperature or chemical potential.
The additional thermodynamic quantities of interest are calculated by:
\begin{equation} \label{eq:thermoderivs}
\centering
    \begin{split}
         \,\,\,\, \frac{n_B}{T^3}&=\frac{1}{T^3} 
        \left( \frac{\partial P}{\partial \mu_B} \right)_T, \\
        \frac{s}{T^3}&=\frac{1}{T^3} 
        \left( \frac{\partial P}{\partial T} \right)_{\mu_B}, \\
        \frac{\epsilon}{T^4}&=\frac{s}{T^3} - \frac{P}{T^4} + \frac{\mu_B}{T} \frac{n_B}{T^3}. \\
    \end{split}
\end{equation}
Figure \ref{fig:pDerivIsothermsFullResult} shows the behavior of the baryon density as a function of the chemical potential, as well as the entropy and energy densities as functions of the temperature.
While isothermal trajectories are considered for the baryon density, the entropy and energy densities are studied along the same lines of constant baryon chemical potential.
We see that these thermodynamic derivatives of the pressure would exhibit discontinuities at the coexistence points, where they would jump from one branch to the other, as is the case when using the 3D Ising model \cite{Parotto:2018pwx,Karthein:2021nxe,Kahangirwe:2024cny}.
However, in our framework, we characterize the first order features beyond this jump and show the progression from the coexistence points to the spinodals via the super-heated hadron phase, or from the other side of the phase transition, to the super-cooled quark-gluon plasma phase. 
The typical features of the coexistence region can be seen for each of these quantities along the chosen trajectories.
In addition, we maintain the multi-valued nature of the thermodynamics in this region and can describe each of the three solutions here, two of which are complex solutions and become real for $\mu_B>\mu_{B,c}$.
Furthermore, the metastable branches that describe the extreme conditions of super-heating/cooling correspond to the same two solutions shown in blue and yellow.
In other words, the blue and yellow solutions describe the system up through metastability.
It is important to have access to these metastable states in the case of lower energy heavy-ion collisions, which could probe the existence of a first order phase transition.
On the other hand, between the spinodal points is the unstable region, where the system proceeds from one phase to the other most likely via spinodal decomposition \cite{Binder1987,Randrup:2009gp,Kapusta:2024nii}.
The features seen in the example isotherms of the pure Ising model in Fig. \ref{fig:nBIsingSymm} are also reflected here  in the full thermodynamic results.
In principle, one may also calculate further thermodynamic derivatives, corresponding to fluctuations.
Fluctuations in this mean field approach are expected to diverge at the critical point, albeit with mean field critical exponents.
As such, these fluctuations go beyond our purpose, and we leave this to a future work.

\begin{figure*}
    \centering 
    \begin{tabular}{c c}
 \includegraphics[width=0.47\textwidth]{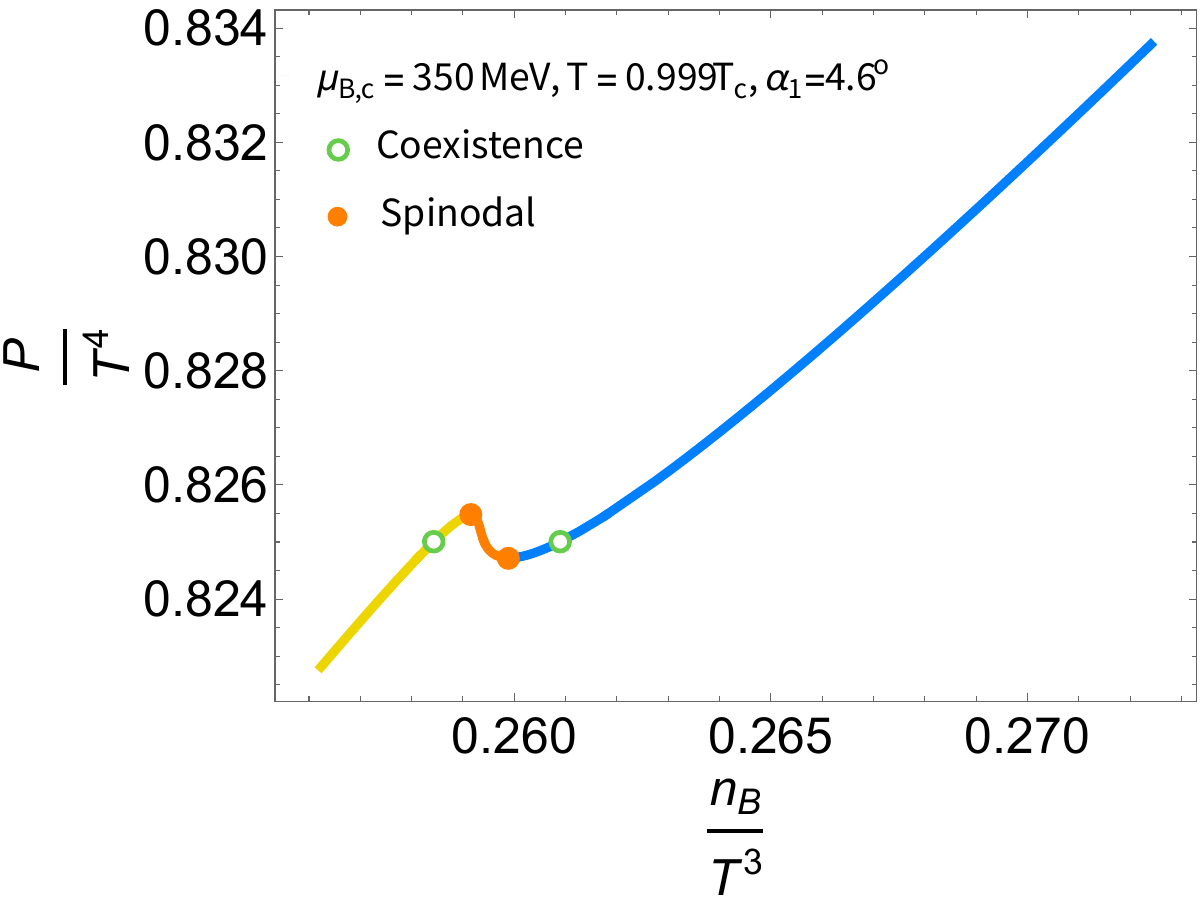}
 \includegraphics[width=0.49\textwidth]{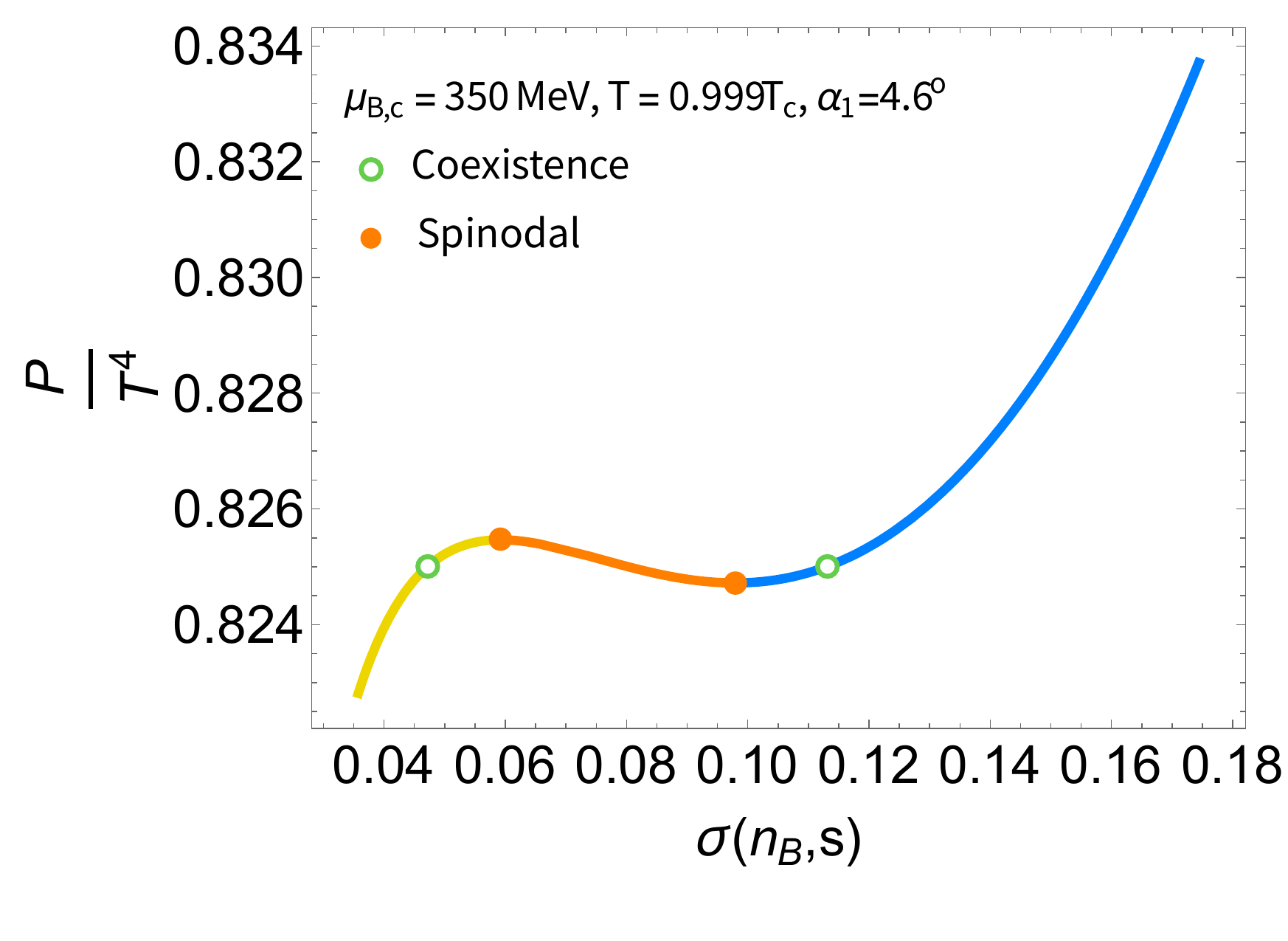} 
    \end{tabular}
    \caption{Full pressure after matching to lattice QCD at $\mu_B=0$ as a function of the baryon density, $n_B$, (left) and  a linear combination of entropy $s$ and baryon density $n_B$, $\sigma$, (right). 
    This mixed quantity becomes relevant due to the choice of angle $\alpha_1$, in which the Ising axes $r$ and $h$ are both contributing to the QCD variables $T$ and $\mu_B$. The coexistence points are indicated with the open green circles, while the spinodals are given by the filled orange circles.}
    \label{fig:pnBsigmaIsothermFullResult}
\end{figure*}

It is important to notice that the identification of the spinodal and coexistence points on these isotherms is possible only very close to the critical temperature, $T \lesssim T_c$, for this small angle parameter $\alpha_1$ where the size of the critical region is only beginning to grow.
\footnote{As discussed in Sec. \ref{sec:mapping}, small angle choices map $h \rightarrow T, ~ r \rightarrow \mu_B$ which leads to the spinodal features being mapped onto lines of fixed $\mu_B$ rather than isotherms.}
This is due to the small angle choice that corresponds to the relatively low value of $\mu_{B,c}=350$ MeV. 
As discussed in Sec. \ref{sec:spinodals_location}, for sufficiently small angles the isotherms may not cross both spinodal lines (see also Fig. \ref{fig:pFullThermo}).

In the current Section, for Figs. \ref{fig:pDerivIsothermsFullResult} and \ref{fig:pnBsigmaIsothermFullResult}, we consider an isothermal trajectory at temperature $T=145.9$, which is 99.9\% of the critical temperature.
For temperatures smaller than this, we do not obtain the typical features along an isotherm. Rather, the mapping has stretched the first order features such that the spinodal has moved to a different temperature, similarly to what is shown in Fig. \ref{fig:pFullThermo}.
These figures show that, by using the mean field Ising model, we gain access to the characterization of the first order phase transition region beyond that of the coexistence points.
In a future work, we forsee employing an equation of state for QCD that is valid at larger values of $\mu_B/T$, e.g. from Ref. \cite{Kahangirwe:2024cny}, opening up more of the phase diagram in order to place the critical point at larger $\mu_{B,c}$. 

We further study the behavior of the pressure along the first order phase transition line to highlight the features that can be accessed by utilizing the mean field Ising model.
For a magnetic system, such as the Ising model which we employ here, the order parameter is the magnetization $M$.
An analogous quantity in terms of QCD variables could be the baryon density $n_B$.
This would only be the case, if the coexistence line is parallel to the temperature axis, as in the Ising model.

We know from lattice QCD simulations, however, that this is not the case \cite{Bonati:2018nut,Bazavov:2018mes,Borsanyi:2020fev}, and the phase transition line in the $(T,\mu_B)$ plane is mostly a parabola with small corrections in the quartic term.
Thus, the order parameter is rather a linear combination of the baryon density along with other relevant quantities.
In this framework, utilizing the mean field Ising model as mapped to QCD, we have the Ising variables $r,~h$ mapped to $T,~\mu_B$ as shown in Eq. \eqref{eq:mapTmuB} and sketched in Fig. \ref{fig:mapTmuB}.
In order to take the analogy further, a flat phase transition line along a fixed value of $\mu_B$ would correspond to $\alpha_1=0$.
In this case the order parameter would be $n_B = (\partial P/\partial \mu_B)=(\partial h/\partial \mu_B)_r(\partial P/\partial h)$ due to the choice of mapping.
However, in general, additional terms arise when $\alpha_1 \neq 0$, as is the case when mapping along the chiral phase transition line as we demonstrate here.
As shown in Ref. \cite{Kapusta:2022pny}, one can write a more general linear combination of entropy and baryon density:
\begin{equation}
    \sigma(n_B,s) = \Big(\frac{\partial \mu_B}{\partial h}\Big)_r n_B +  \Big(\frac{\partial T}{\partial h}\Big)_r s,
\end{equation}
where $h(T,\mu_B)$ and $r(T,\mu_B)$ are defined in Eq. \eqref{eq:mapTmuB}.
Thus, upon applying the chain rule, we arrive at:
\begin{equation}
    \sigma(n_B,s) = w\, T_c (- \cos{\alpha_2} n_B +  \sin{\alpha_2} s),
\end{equation}
similar to what is found in Ref. \cite{Kapusta:2022pny} with the choice of mapping as in Eq. \eqref{eq:mapTmuB}.
Figure \ref{fig:pnBsigmaIsothermFullResult} shows the behavior of the pressure as a function of  the baryon density $n_B$ and also as a function of this general linear combination of $n_B$ and $s$, $\sigma$.
We see that, while both quantities $n_B$ and $\sigma$ exhibit the features of the first order phase transition, the linear combination of entropy $s$ and baryon density $n_B$ portrays the coexistence and spinodal points over a broader domain.
Thus, $\sigma$ displays the features which we have been interested in studying here, namely, the behavior along the metastable and unstable regions.

\section{Conclusions}\label{sec:concl}
In this study, we have shown that the mean field Ising model mapped to QCD allows access to the full features of the first order phase transition.
We obtained the pressure in the QCD phase diagram with three separate functional forms: one which is always real and a pair of complex conjugate solutions that become real when the phase transition becomes first order.
We characterized the multi-valued nature of each thermodynamical variable including the pressure, entropy density, baryon density, and energy density and gained full control over the functional forms of those three solutions.
In the case of all these quantities, we obtained full information beyond the discontinuity expected along the first order line for derivatives of the pressure.
Thus, we accessed the metastable phases of super-cooled quark gluon plasma and super-heated hadronic matter.
The limit of metastability given by the spinodal points was obtained and the spinodal lines were tracked in the QCD phase diagram.
Furthermore, we presented a discussion of the Ginzburg criterion, which determines the region of applicability of the mean field approach.
We showed the dependence of the location of the spinodal points on the non-universal mapping between the Ising model and QCD.
In the case of a small angle $\alpha_1$, the region covered by the spinodals is very wide-reaching in the phase diagram, and the spinodals open up to large values of chemical potentials.
This means that, once the system makes its way into the phase coexistence region, it may stay there for a longer time before transforming into the new phase.
However, as the angle increases, the spinodal curves turn down towards the $T=0$ axis, giving rise to a comparatively smaller coexistence region.
Furthermore, if the parameters of the mapping can be constrained by experimental efforts in the search for the QCD critical point, we will be able to determine the size of the spinodal region in the QCD phase diagram.

\section*{Acknowledgements}
The authors would like to thank Krishna Rajagopal, Mikhail Stephanov, and Volodymyr Vovchenko for fruitful discussions.
This  material is based upon work supported by the National Science Foundation under grants No. PHY-2208724, PHY-1654219 and PHY-2116686, and within the framework of the MUSES collaboration, under grant number No. OAC-2103680. This material is also based upon work supported by the U.S. Department of Energy, Office of Science, Office of Nuclear Physics, under Award Number
DE-SC0022023 and under contract number 
DE-AC02-05CH11231 as well as by the National Aeronautics and Space Agency (NASA) under Award Number 80NSSC24K0767.
J.M.K. is supported by an Ascending Postdoctoral Scholar Fellowship from the National Science Foundation under Award No. 2138063.

\appendix
\section{Additional Configurations of the Equation of State}
\label{sec:appendix}

The main goal of this paper is to highlight the features of the first order region that one can access by utilizing the mean field Ising model.
In particular, we aim to show that the choice of mean field is necessary in order to achieve a description of the spinodal lines as the universal choice of the 3D Ising model hides these features in the complex plane as Lee-Yang edge singularities \cite{An:2017brc}.
In order to do so, we have chosen the non-universal mapping parameters from Eq. \eqref{eq:mapTmuB} to be the same as the original publication from the BEST collaboration, which first established such a framework of mapping a critical point from the 3D Ising model onto the QCD phase diagram with consistency with the lattice results \cite{Parotto:2018pwx}.
In that study, the scaling parameters between Ising and QCD were taken to be $w=1$ and $\rho=2$.
In this appendix, we relax those constraints on those scaling parameters to further demonstrate the dependence on the mapping parameters and the flexibility of the framework to accommodate different sizes and shapes of the critical and first order regions.

\begin{figure}
    \centering
    \includegraphics[width=0.48\textwidth]{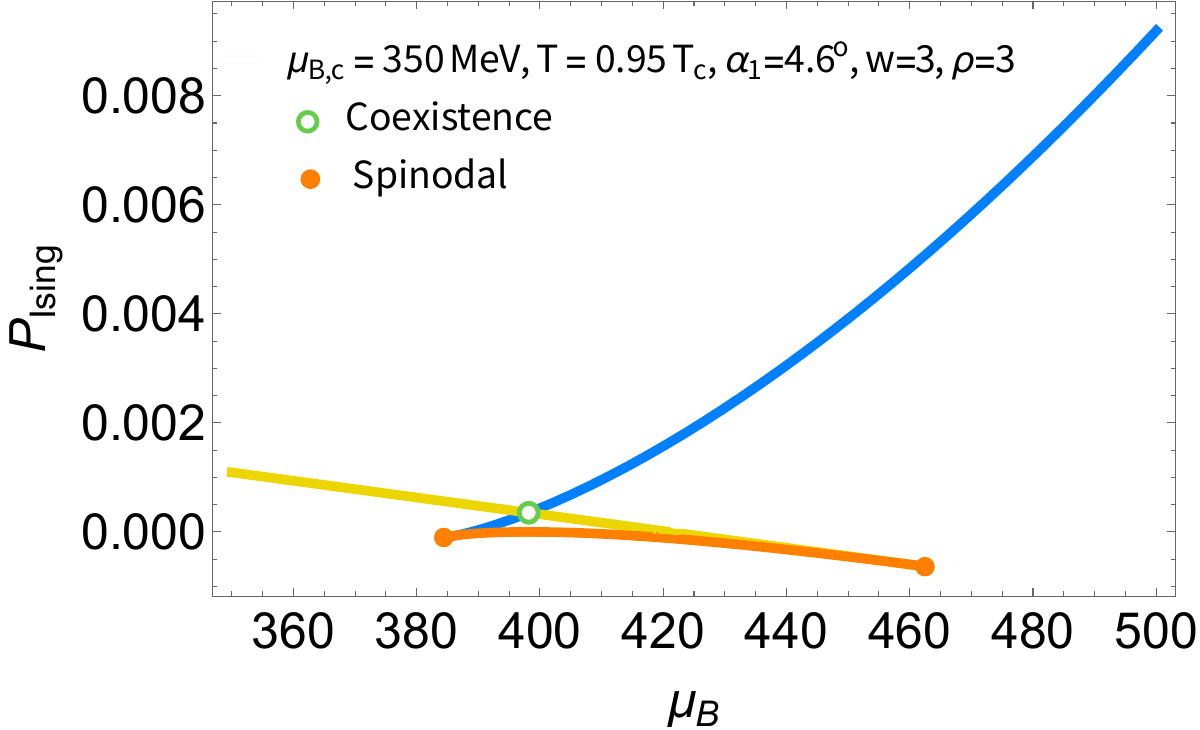}
    \caption{Ising pressure isotherm as mapped onto the QCD phase diagram for a critical point chosen to be the same as that of Refs. \cite{Parotto:2018pwx,Karthein:2021nxe} where $\mu_{B,c}=350$ MeV, but with increased scaling parameters $w=3, \rho=3$.}
    \label{fig:SpinodalIsotherms_OrigBEST_Incwrho}
\end{figure}

Fig. \ref{fig:SpinodalIsotherms_OrigBEST_Incwrho} shows that even for the shallow mapping at $\mu_{B,c}=350$ MeV, $\alpha_1=4.6^{\rm{o}}$ it is possible to achieve the expected behavior of the isotherms.
This is possible due to the choice of a different size and shape of the critical region given by modifying $w,~ \rho$.
For a larger choice of $w=3, \rho=3$, the size of the critical region is smaller, which allows for an isotherm to capture both spinodals even for such a small value of the angle $\alpha_1$.

\begin{figure*}
    \centering    \includegraphics[width=0.67\textwidth]{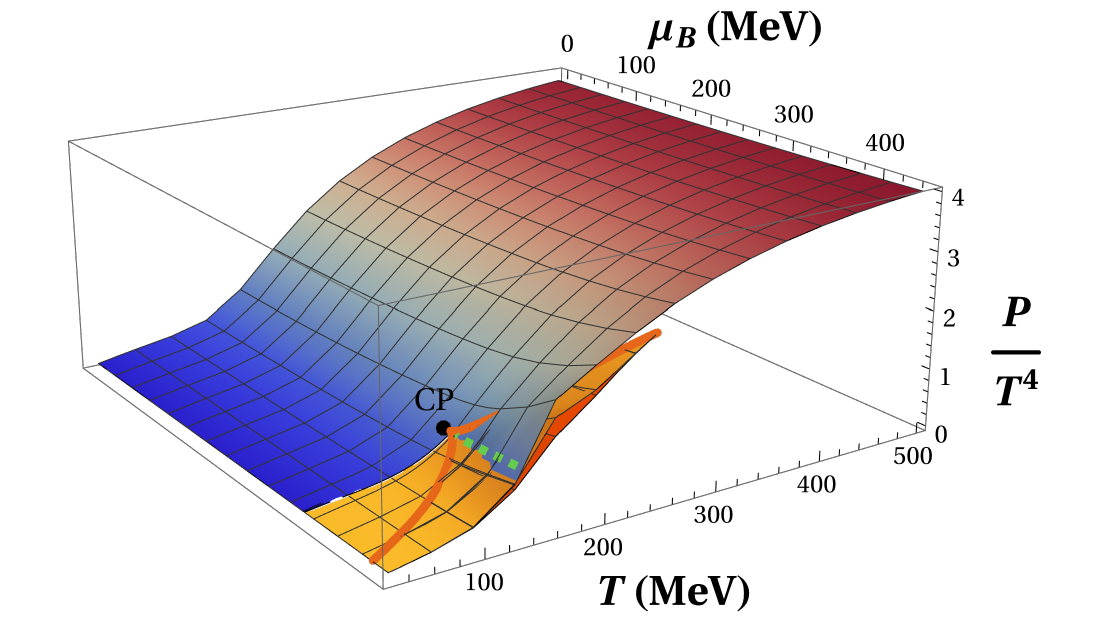} 
    \caption{Full pressure as a function of temperature and baryon chemical potential matching lattice QCD at $\mu_B=0$, with a critical point placed at $\mu_{B,c}=350$ MeV for comparison to Ref. \cite{Parotto:2018pwx}. The size of the critical region is increased by taking values of the scaling parameters $w=0.5, \rho=0.5$. Thus, the three solutions from the mean field Ising model are featured here prominently: one which is always real (red/blue gradient) and two complex conjugate solutions that are only real for $\mu_B>\mu_{B,c}$ (yellow and orange).}
    \label{fig:pFullThermoIncCritRegion}
\end{figure*}

On the other hand, we also study the result for the full pressure after the matching procedure as in Eq. \ref{eq:fullpress} for a larger critical region.
For the choice of mapping between Ising and QCD from the original formulation of the BEST EoS framework \cite{Parotto:2018pwx}, where the scaling parameters are $w=1, \rho=2$, the critical pressure is very small, as can be seen in Fig. \ref{fig:SpinodalIsotherms_OrigBEST}.
For this reason, although all three solutions appear, gradient red/blue, orange, and yellow, the features are quite subtle.
In order to enhance this effect, we choose to increase the size of the critical region with the choice of scaling parameters of $w=0.5, \rho=0.5$.
Figure \ref{fig:pFullThermoIncCritRegion} shows the full pressure in the phase diagram for this larger critical region.
In this case, we clearly see the three solutions present in the first order region by eye in this 3D plot.
However, we note that this increased critical region will lead to distorted isotherms, conversely to what is shown here in Appendix in Fig. \ref{fig:SpinodalIsotherms_OrigBEST_Incwrho} for a smaller critical region.

\bibliography{all}

\begin{thebibliography}{57}%
\makeatletter
\providecommand \@ifxundefined [1]{%
 \@ifx{#1\undefined}
}%
\providecommand \@ifnum [1]{%
 \ifnum #1\expandafter \@firstoftwo
 \else \expandafter \@secondoftwo
 \fi
}%
\providecommand \@ifx [1]{%
 \ifx #1\expandafter \@firstoftwo
 \else \expandafter \@secondoftwo
 \fi
}%
\providecommand \natexlab [1]{#1}%
\providecommand \enquote  [1]{``#1''}%
\providecommand \bibnamefont  [1]{#1}%
\providecommand \bibfnamefont [1]{#1}%
\providecommand \citenamefont [1]{#1}%
\providecommand \href@noop [0]{\@secondoftwo}%
\providecommand \href [0]{\begingroup \@sanitize@url \@href}%
\providecommand \@href[1]{\@@startlink{#1}\@@href}%
\providecommand \@@href[1]{\endgroup#1\@@endlink}%
\providecommand \@sanitize@url [0]{\catcode `\\12\catcode `\$12\catcode
  `\&12\catcode `\#12\catcode `\^12\catcode `\_12\catcode `\%12\relax}%
\providecommand \@@startlink[1]{}%
\providecommand \@@endlink[0]{}%
\providecommand \url  [0]{\begingroup\@sanitize@url \@url }%
\providecommand \@url [1]{\endgroup\@href {#1}{\urlprefix }}%
\providecommand \urlprefix  [0]{URL }%
\providecommand \Eprint [0]{\href }%
\providecommand \doibase [0]{http://dx.doi.org/}%
\providecommand \selectlanguage [0]{\@gobble}%
\providecommand \bibinfo  [0]{\@secondoftwo}%
\providecommand \bibfield  [0]{\@secondoftwo}%
\providecommand \translation [1]{[#1]}%
\providecommand \BibitemOpen [0]{}%
\providecommand \bibitemStop [0]{}%
\providecommand \bibitemNoStop [0]{.\EOS\space}%
\providecommand \EOS [0]{\spacefactor3000\relax}%
\providecommand \BibitemShut  [1]{\csname bibitem#1\endcsname}%
\let\auto@bib@innerbib\@empty
\bibitem [{\citenamefont {Aoki}\ \emph {et~al.}(2006)\citenamefont {Aoki},
  \citenamefont {Endrodi}, \citenamefont {Fodor}, \citenamefont {Katz},\ and\
  \citenamefont {Szabo}}]{Aoki:2006we}%
  \BibitemOpen
  \bibfield  {author} {\bibinfo {author} {\bibfnamefont {Y.}~\bibnamefont
  {Aoki}}, \bibinfo {author} {\bibfnamefont {G.}~\bibnamefont {Endrodi}},
  \bibinfo {author} {\bibfnamefont {Z.}~\bibnamefont {Fodor}}, \bibinfo
  {author} {\bibfnamefont {S.~D.}\ \bibnamefont {Katz}}, \ and\ \bibinfo
  {author} {\bibfnamefont {K.~K.}\ \bibnamefont {Szabo}},\ }\href {\doibase
  10.1038/nature05120} {\bibfield  {journal} {\bibinfo  {journal} {Nature}\
  }\textbf {\bibinfo {volume} {443}},\ \bibinfo {pages} {675} (\bibinfo {year}
  {2006})},\ \Eprint {http://arxiv.org/abs/hep-lat/0611014}
  {arXiv:hep-lat/0611014 [hep-lat]} \BibitemShut {NoStop}%
\bibitem [{\citenamefont {Bzdak}\ \emph {et~al.}(2020)\citenamefont {Bzdak},
  \citenamefont {Esumi}, \citenamefont {Koch}, \citenamefont {Liao},
  \citenamefont {Stephanov},\ and\ \citenamefont {Xu}}]{Bzdak:2019pkr}%
  \BibitemOpen
  \bibfield  {author} {\bibinfo {author} {\bibfnamefont {A.}~\bibnamefont
  {Bzdak}}, \bibinfo {author} {\bibfnamefont {S.}~\bibnamefont {Esumi}},
  \bibinfo {author} {\bibfnamefont {V.}~\bibnamefont {Koch}}, \bibinfo {author}
  {\bibfnamefont {J.}~\bibnamefont {Liao}}, \bibinfo {author} {\bibfnamefont
  {M.}~\bibnamefont {Stephanov}}, \ and\ \bibinfo {author} {\bibfnamefont
  {N.}~\bibnamefont {Xu}},\ }\href {\doibase 10.1016/j.physrep.2020.01.005}
  {\bibfield  {journal} {\bibinfo  {journal} {Phys. Rept.}\ }\textbf {\bibinfo
  {volume} {853}},\ \bibinfo {pages} {1} (\bibinfo {year} {2020})},\ \Eprint
  {http://arxiv.org/abs/1906.00936} {arXiv:1906.00936 [nucl-th]} \BibitemShut
  {NoStop}%
\bibitem [{\citenamefont {Du}\ \emph {et~al.}(2024)\citenamefont {Du},
  \citenamefont {Sorensen},\ and\ \citenamefont {Stephanov}}]{Du:2024wjm}%
  \BibitemOpen
  \bibfield  {author} {\bibinfo {author} {\bibfnamefont {L.}~\bibnamefont
  {Du}}, \bibinfo {author} {\bibfnamefont {A.}~\bibnamefont {Sorensen}}, \ and\
  \bibinfo {author} {\bibfnamefont {M.}~\bibnamefont {Stephanov}}\ }(\bibinfo
  {year} {2024})\ \Eprint {http://arxiv.org/abs/2402.10183} {arXiv:2402.10183
  [nucl-th]} \BibitemShut {NoStop}%
\bibitem [{\citenamefont {Parotto}\ \emph {et~al.}(2020)\citenamefont
  {Parotto}, \citenamefont {Bluhm}, \citenamefont {Mroczek}, \citenamefont
  {Nahrgang}, \citenamefont {Noronha-Hostler}, \citenamefont {Rajagopal},
  \citenamefont {Ratti}, \citenamefont {Sch\"afer},\ and\ \citenamefont
  {Stephanov}}]{Parotto:2018pwx}%
  \BibitemOpen
  \bibfield  {author} {\bibinfo {author} {\bibfnamefont {P.}~\bibnamefont
  {Parotto}}, \bibinfo {author} {\bibfnamefont {M.}~\bibnamefont {Bluhm}},
  \bibinfo {author} {\bibfnamefont {D.}~\bibnamefont {Mroczek}}, \bibinfo
  {author} {\bibfnamefont {M.}~\bibnamefont {Nahrgang}}, \bibinfo {author}
  {\bibfnamefont {J.}~\bibnamefont {Noronha-Hostler}}, \bibinfo {author}
  {\bibfnamefont {K.}~\bibnamefont {Rajagopal}}, \bibinfo {author}
  {\bibfnamefont {C.}~\bibnamefont {Ratti}}, \bibinfo {author} {\bibfnamefont
  {T.}~\bibnamefont {Sch\"afer}}, \ and\ \bibinfo {author} {\bibfnamefont
  {M.}~\bibnamefont {Stephanov}},\ }\href {\doibase
  10.1103/PhysRevC.101.034901} {\bibfield  {journal} {\bibinfo  {journal}
  {Phys. Rev. C}\ }\textbf {\bibinfo {volume} {101}},\ \bibinfo {pages}
  {034901} (\bibinfo {year} {2020})},\ \Eprint
  {http://arxiv.org/abs/1805.05249} {arXiv:1805.05249 [hep-ph]} \BibitemShut
  {NoStop}%
\bibitem [{\citenamefont {Karthein}\ \emph {et~al.}(2021)\citenamefont
  {Karthein}, \citenamefont {Mroczek}, \citenamefont {Nava~Acuna},
  \citenamefont {Noronha-Hostler}, \citenamefont {Parotto}, \citenamefont
  {Price},\ and\ \citenamefont {Ratti}}]{Karthein:2021nxe}%
  \BibitemOpen
  \bibfield  {author} {\bibinfo {author} {\bibfnamefont {J.~M.}\ \bibnamefont
  {Karthein}}, \bibinfo {author} {\bibfnamefont {D.}~\bibnamefont {Mroczek}},
  \bibinfo {author} {\bibfnamefont {A.~R.}\ \bibnamefont {Nava~Acuna}},
  \bibinfo {author} {\bibfnamefont {J.}~\bibnamefont {Noronha-Hostler}},
  \bibinfo {author} {\bibfnamefont {P.}~\bibnamefont {Parotto}}, \bibinfo
  {author} {\bibfnamefont {D.~R.~P.}\ \bibnamefont {Price}}, \ and\ \bibinfo
  {author} {\bibfnamefont {C.}~\bibnamefont {Ratti}},\ }\href {\doibase
  10.1140/epjp/s13360-021-01615-5} {\bibfield  {journal} {\bibinfo  {journal}
  {Eur. Phys. J. Plus}\ }\textbf {\bibinfo {volume} {136}},\ \bibinfo {pages}
  {621} (\bibinfo {year} {2021})},\ \Eprint {http://arxiv.org/abs/2103.08146}
  {arXiv:2103.08146 [hep-ph]} \BibitemShut {NoStop}%
\bibitem [{\citenamefont {An}\ \emph {et~al.}(2022)\citenamefont {An} \emph
  {et~al.}}]{An:2021wof}%
  \BibitemOpen
  \bibfield  {author} {\bibinfo {author} {\bibfnamefont {X.}~\bibnamefont {An}}
  \emph {et~al.},\ }\href {\doibase 10.1016/j.nuclphysa.2021.122343} {\bibfield
   {journal} {\bibinfo  {journal} {Nucl. Phys. A}\ }\textbf {\bibinfo {volume}
  {1017}},\ \bibinfo {pages} {122343} (\bibinfo {year} {2022})},\ \Eprint
  {http://arxiv.org/abs/2108.13867} {arXiv:2108.13867 [nucl-th]} \BibitemShut
  {NoStop}%
\bibitem [{\citenamefont {Kahangirwe}\ \emph {et~al.}(2024)\citenamefont
  {Kahangirwe}, \citenamefont {Bass}, \citenamefont {Bratkovskaya},
  \citenamefont {Jahan}, \citenamefont {Moreau}, \citenamefont {Parotto},
  \citenamefont {Price}, \citenamefont {Ratti}, \citenamefont {Soloveva},\ and\
  \citenamefont {Stephanov}}]{Kahangirwe:2024cny}%
  \BibitemOpen
  \bibfield  {author} {\bibinfo {author} {\bibfnamefont {M.}~\bibnamefont
  {Kahangirwe}}, \bibinfo {author} {\bibfnamefont {S.~A.}\ \bibnamefont
  {Bass}}, \bibinfo {author} {\bibfnamefont {E.}~\bibnamefont {Bratkovskaya}},
  \bibinfo {author} {\bibfnamefont {J.}~\bibnamefont {Jahan}}, \bibinfo
  {author} {\bibfnamefont {P.}~\bibnamefont {Moreau}}, \bibinfo {author}
  {\bibfnamefont {P.}~\bibnamefont {Parotto}}, \bibinfo {author} {\bibfnamefont
  {D.}~\bibnamefont {Price}}, \bibinfo {author} {\bibfnamefont
  {C.}~\bibnamefont {Ratti}}, \bibinfo {author} {\bibfnamefont
  {O.}~\bibnamefont {Soloveva}}, \ and\ \bibinfo {author} {\bibfnamefont
  {M.}~\bibnamefont {Stephanov}},\ }\href {\doibase
  10.1103/PhysRevD.109.094046} {\bibfield  {journal} {\bibinfo  {journal}
  {Phys. Rev. D}\ }\textbf {\bibinfo {volume} {109}},\ \bibinfo {pages}
  {094046} (\bibinfo {year} {2024})},\ \Eprint
  {http://arxiv.org/abs/2402.08636} {arXiv:2402.08636 [nucl-th]} \BibitemShut
  {NoStop}%
\bibitem [{\citenamefont {Nonaka}\ and\ \citenamefont
  {Asakawa}(2005)}]{Nonaka:2004pg}%
  \BibitemOpen
  \bibfield  {author} {\bibinfo {author} {\bibfnamefont {C.}~\bibnamefont
  {Nonaka}}\ and\ \bibinfo {author} {\bibfnamefont {M.}~\bibnamefont
  {Asakawa}},\ }\href {\doibase 10.1103/PhysRevC.71.044904} {\bibfield
  {journal} {\bibinfo  {journal} {Phys. Rev. C}\ }\textbf {\bibinfo {volume}
  {71}},\ \bibinfo {pages} {044904} (\bibinfo {year} {2005})},\ \Eprint
  {http://arxiv.org/abs/nucl-th/0410078} {arXiv:nucl-th/0410078} \BibitemShut
  {NoStop}%
\bibitem [{\citenamefont {Plumberg}\ \emph {et~al.}(2018)\citenamefont
  {Plumberg}, \citenamefont {Welle},\ and\ \citenamefont
  {Kapusta}}]{Plumberg:2018fxo}%
  \BibitemOpen
  \bibfield  {author} {\bibinfo {author} {\bibfnamefont {C.~J.}\ \bibnamefont
  {Plumberg}}, \bibinfo {author} {\bibfnamefont {T.}~\bibnamefont {Welle}}, \
  and\ \bibinfo {author} {\bibfnamefont {J.~I.}\ \bibnamefont {Kapusta}},\
  }\href {\doibase 10.22323/1.347.0157} {\bibfield  {journal} {\bibinfo
  {journal} {PoS}\ }\textbf {\bibinfo {volume} {CORFU2018}},\ \bibinfo {pages}
  {157} (\bibinfo {year} {2018})},\ \Eprint {http://arxiv.org/abs/1812.01684}
  {arXiv:1812.01684 [nucl-th]} \BibitemShut {NoStop}%
\bibitem [{\citenamefont {Kapusta}\ and\ \citenamefont
  {Welle}(2022)}]{Kapusta:2022pny}%
  \BibitemOpen
  \bibfield  {author} {\bibinfo {author} {\bibfnamefont {J.~I.}\ \bibnamefont
  {Kapusta}}\ and\ \bibinfo {author} {\bibfnamefont {T.}~\bibnamefont
  {Welle}},\ }\href {\doibase 10.1103/PhysRevC.106.044901} {\bibfield
  {journal} {\bibinfo  {journal} {Phys. Rev. C}\ }\textbf {\bibinfo {volume}
  {106}},\ \bibinfo {pages} {044901} (\bibinfo {year} {2022})},\ \Eprint
  {http://arxiv.org/abs/2205.12150} {arXiv:2205.12150 [nucl-th]} \BibitemShut
  {NoStop}%
\bibitem [{\citenamefont {Pisarski}\ and\ \citenamefont
  {Wilczek}(1984)}]{Pisarski:1983ms}%
  \BibitemOpen
  \bibfield  {author} {\bibinfo {author} {\bibfnamefont {R.~D.}\ \bibnamefont
  {Pisarski}}\ and\ \bibinfo {author} {\bibfnamefont {F.}~\bibnamefont
  {Wilczek}},\ }\href {\doibase 10.1103/PhysRevD.29.338} {\bibfield  {journal}
  {\bibinfo  {journal} {Phys. Rev. D}\ }\textbf {\bibinfo {volume} {29}},\
  \bibinfo {pages} {338} (\bibinfo {year} {1984})}\BibitemShut {NoStop}%
\bibitem [{\citenamefont {Rajagopal}\ and\ \citenamefont
  {Wilczek}(1993)}]{Rajagopal:1992qz}%
  \BibitemOpen
  \bibfield  {author} {\bibinfo {author} {\bibfnamefont {K.}~\bibnamefont
  {Rajagopal}}\ and\ \bibinfo {author} {\bibfnamefont {F.}~\bibnamefont
  {Wilczek}},\ }\href {\doibase 10.1016/0550-3213(93)90502-G} {\bibfield
  {journal} {\bibinfo  {journal} {Nucl. Phys. B}\ }\textbf {\bibinfo {volume}
  {399}},\ \bibinfo {pages} {395} (\bibinfo {year} {1993})},\ \Eprint
  {http://arxiv.org/abs/hep-ph/9210253} {arXiv:hep-ph/9210253} \BibitemShut
  {NoStop}%
\bibitem [{\citenamefont {Stephanov}\ and\ \citenamefont
  {Yin}(2018)}]{Stephanov:2017ghc}%
  \BibitemOpen
  \bibfield  {author} {\bibinfo {author} {\bibfnamefont {M.}~\bibnamefont
  {Stephanov}}\ and\ \bibinfo {author} {\bibfnamefont {Y.}~\bibnamefont
  {Yin}},\ }\href {\doibase 10.1103/PhysRevD.98.036006} {\bibfield  {journal}
  {\bibinfo  {journal} {Phys. Rev. D}\ }\textbf {\bibinfo {volume} {98}},\
  \bibinfo {pages} {036006} (\bibinfo {year} {2018})},\ \Eprint
  {http://arxiv.org/abs/1712.10305} {arXiv:1712.10305 [nucl-th]} \BibitemShut
  {NoStop}%
\bibitem [{\citenamefont {Herold}\ \emph {et~al.}(2022)\citenamefont {Herold},
  \citenamefont {Limphirat}, \citenamefont {Saikham}, \citenamefont {Nahrgang},
  \citenamefont {Reichert},\ and\ \citenamefont {Bleicher}}]{Herold:2022laa}%
  \BibitemOpen
  \bibfield  {author} {\bibinfo {author} {\bibfnamefont {C.}~\bibnamefont
  {Herold}}, \bibinfo {author} {\bibfnamefont {A.}~\bibnamefont {Limphirat}},
  \bibinfo {author} {\bibfnamefont {P.}~\bibnamefont {Saikham}}, \bibinfo
  {author} {\bibfnamefont {M.}~\bibnamefont {Nahrgang}}, \bibinfo {author}
  {\bibfnamefont {T.}~\bibnamefont {Reichert}}, \ and\ \bibinfo {author}
  {\bibfnamefont {M.}~\bibnamefont {Bleicher}},\ }\href {\doibase
  10.1103/PhysRevC.106.024901} {\bibfield  {journal} {\bibinfo  {journal}
  {Phys. Rev. C}\ }\textbf {\bibinfo {volume} {106}},\ \bibinfo {pages}
  {024901} (\bibinfo {year} {2022})},\ \Eprint
  {http://arxiv.org/abs/2204.00286} {arXiv:2204.00286 [hep-ph]} \BibitemShut
  {NoStop}%
\bibitem [{\citenamefont {Pradeep}\ \emph {et~al.}(2022)\citenamefont
  {Pradeep}, \citenamefont {Rajagopal}, \citenamefont {Stephanov},\ and\
  \citenamefont {Yin}}]{Pradeep:2022mkf}%
  \BibitemOpen
  \bibfield  {author} {\bibinfo {author} {\bibfnamefont {M.}~\bibnamefont
  {Pradeep}}, \bibinfo {author} {\bibfnamefont {K.}~\bibnamefont {Rajagopal}},
  \bibinfo {author} {\bibfnamefont {M.}~\bibnamefont {Stephanov}}, \ and\
  \bibinfo {author} {\bibfnamefont {Y.}~\bibnamefont {Yin}},\ }\href {\doibase
  10.1103/PhysRevD.106.036017} {\bibfield  {journal} {\bibinfo  {journal}
  {Phys. Rev. D}\ }\textbf {\bibinfo {volume} {106}},\ \bibinfo {pages}
  {036017} (\bibinfo {year} {2022})},\ \Eprint
  {http://arxiv.org/abs/2204.00639} {arXiv:2204.00639 [hep-ph]} \BibitemShut
  {NoStop}%
\bibitem [{\citenamefont {Pihan}\ \emph {et~al.}(2023)\citenamefont {Pihan},
  \citenamefont {Bluhm}, \citenamefont {Kitazawa}, \citenamefont {Sami},\ and\
  \citenamefont {Nahrgang}}]{Pihan:2022xcl}%
  \BibitemOpen
  \bibfield  {author} {\bibinfo {author} {\bibfnamefont {G.}~\bibnamefont
  {Pihan}}, \bibinfo {author} {\bibfnamefont {M.}~\bibnamefont {Bluhm}},
  \bibinfo {author} {\bibfnamefont {M.}~\bibnamefont {Kitazawa}}, \bibinfo
  {author} {\bibfnamefont {T.}~\bibnamefont {Sami}}, \ and\ \bibinfo {author}
  {\bibfnamefont {M.}~\bibnamefont {Nahrgang}},\ }\href {\doibase
  10.1103/PhysRevC.107.014908} {\bibfield  {journal} {\bibinfo  {journal}
  {Phys. Rev. C}\ }\textbf {\bibinfo {volume} {107}},\ \bibinfo {pages}
  {014908} (\bibinfo {year} {2023})},\ \Eprint
  {http://arxiv.org/abs/2205.12834} {arXiv:2205.12834 [nucl-th]} \BibitemShut
  {NoStop}%
\bibitem [{\citenamefont {Binder}(1984)}]{Binder1984}%
  \BibitemOpen
  \bibfield  {author} {\bibinfo {author} {\bibfnamefont {K.}~\bibnamefont
  {Binder}},\ }\href {\doibase 10.1103/PhysRevA.29.341} {\bibfield  {journal}
  {\bibinfo  {journal} {Phys. Rev. A}\ }\textbf {\bibinfo {volume} {29}},\
  \bibinfo {pages} {341} (\bibinfo {year} {1984})}\BibitemShut {NoStop}%
\bibitem [{\citenamefont {Chomaz}\ \emph {et~al.}(2004)\citenamefont {Chomaz},
  \citenamefont {Colonna},\ and\ \citenamefont {Randrup}}]{Chomaz:2003dz}%
  \BibitemOpen
  \bibfield  {author} {\bibinfo {author} {\bibfnamefont {P.}~\bibnamefont
  {Chomaz}}, \bibinfo {author} {\bibfnamefont {M.}~\bibnamefont {Colonna}}, \
  and\ \bibinfo {author} {\bibfnamefont {J.}~\bibnamefont {Randrup}},\ }\href
  {\doibase 10.1016/j.physrep.2003.09.006} {\bibfield  {journal} {\bibinfo
  {journal} {Phys. Rept.}\ }\textbf {\bibinfo {volume} {389}},\ \bibinfo
  {pages} {263} (\bibinfo {year} {2004})}\BibitemShut {NoStop}%
\bibitem [{\citenamefont {Sasaki}\ \emph {et~al.}(2007)\citenamefont {Sasaki},
  \citenamefont {Friman},\ and\ \citenamefont {Redlich}}]{Sasaki:2007db}%
  \BibitemOpen
  \bibfield  {author} {\bibinfo {author} {\bibfnamefont {C.}~\bibnamefont
  {Sasaki}}, \bibinfo {author} {\bibfnamefont {B.}~\bibnamefont {Friman}}, \
  and\ \bibinfo {author} {\bibfnamefont {K.}~\bibnamefont {Redlich}},\ }\href
  {\doibase 10.1103/PhysRevLett.99.232301} {\bibfield  {journal} {\bibinfo
  {journal} {Phys. Rev. Lett.}\ }\textbf {\bibinfo {volume} {99}},\ \bibinfo
  {pages} {232301} (\bibinfo {year} {2007})},\ \Eprint
  {http://arxiv.org/abs/hep-ph/0702254} {arXiv:hep-ph/0702254} \BibitemShut
  {NoStop}%
\bibitem [{\citenamefont {Skokov}\ and\ \citenamefont
  {Voskresensky}(2009)}]{Skokov:2009yu}%
  \BibitemOpen
  \bibfield  {author} {\bibinfo {author} {\bibfnamefont {V.~V.}\ \bibnamefont
  {Skokov}}\ and\ \bibinfo {author} {\bibfnamefont {D.~N.}\ \bibnamefont
  {Voskresensky}},\ }\href {\doibase 10.1016/j.nuclphysa.2009.07.012}
  {\bibfield  {journal} {\bibinfo  {journal} {Nucl. Phys. A}\ }\textbf
  {\bibinfo {volume} {828}},\ \bibinfo {pages} {401} (\bibinfo {year}
  {2009})},\ \Eprint {http://arxiv.org/abs/0903.4335} {arXiv:0903.4335
  [nucl-th]} \BibitemShut {NoStop}%
\bibitem [{\citenamefont {Randrup}(2009)}]{Randrup:2009gp}%
  \BibitemOpen
  \bibfield  {author} {\bibinfo {author} {\bibfnamefont {J.}~\bibnamefont
  {Randrup}},\ }\href {\doibase 10.1103/PhysRevC.79.054911} {\bibfield
  {journal} {\bibinfo  {journal} {Phys. Rev. C}\ }\textbf {\bibinfo {volume}
  {79}},\ \bibinfo {pages} {054911} (\bibinfo {year} {2009})},\ \Eprint
  {http://arxiv.org/abs/0903.4736} {arXiv:0903.4736 [nucl-th]} \BibitemShut
  {NoStop}%
\bibitem [{\citenamefont {Steinheimer}\ and\ \citenamefont
  {Randrup}(2012)}]{Steinheimer:2012gc}%
  \BibitemOpen
  \bibfield  {author} {\bibinfo {author} {\bibfnamefont {J.}~\bibnamefont
  {Steinheimer}}\ and\ \bibinfo {author} {\bibfnamefont {J.}~\bibnamefont
  {Randrup}},\ }\href {\doibase 10.1103/PhysRevLett.109.212301} {\bibfield
  {journal} {\bibinfo  {journal} {Phys. Rev. Lett.}\ }\textbf {\bibinfo
  {volume} {109}},\ \bibinfo {pages} {212301} (\bibinfo {year} {2012})},\
  \Eprint {http://arxiv.org/abs/1209.2462} {arXiv:1209.2462 [nucl-th]}
  \BibitemShut {NoStop}%
\bibitem [{\citenamefont {Steinheimer}\ \emph {et~al.}(2019)\citenamefont
  {Steinheimer}, \citenamefont {Pang}, \citenamefont {Zhou}, \citenamefont
  {Koch}, \citenamefont {Randrup},\ and\ \citenamefont
  {Stoecker}}]{Steinheimer:2019iso}%
  \BibitemOpen
  \bibfield  {author} {\bibinfo {author} {\bibfnamefont {J.}~\bibnamefont
  {Steinheimer}}, \bibinfo {author} {\bibfnamefont {L.}~\bibnamefont {Pang}},
  \bibinfo {author} {\bibfnamefont {K.}~\bibnamefont {Zhou}}, \bibinfo {author}
  {\bibfnamefont {V.}~\bibnamefont {Koch}}, \bibinfo {author} {\bibfnamefont
  {J.}~\bibnamefont {Randrup}}, \ and\ \bibinfo {author} {\bibfnamefont
  {H.}~\bibnamefont {Stoecker}},\ }\href {\doibase 10.1007/JHEP12(2019)122}
  {\bibfield  {journal} {\bibinfo  {journal} {JHEP}\ }\textbf {\bibinfo
  {volume} {12}},\ \bibinfo {pages} {122} (\bibinfo {year} {2019})},\ \Eprint
  {http://arxiv.org/abs/1906.06562} {arXiv:1906.06562 [nucl-th]} \BibitemShut
  {NoStop}%
\bibitem [{\citenamefont {An}\ \emph {et~al.}(2018)\citenamefont {An},
  \citenamefont {Mesterh\'azy},\ and\ \citenamefont {Stephanov}}]{An:2017brc}%
  \BibitemOpen
  \bibfield  {author} {\bibinfo {author} {\bibfnamefont {X.}~\bibnamefont
  {An}}, \bibinfo {author} {\bibfnamefont {D.}~\bibnamefont {Mesterh\'azy}}, \
  and\ \bibinfo {author} {\bibfnamefont {M.~A.}\ \bibnamefont {Stephanov}},\
  }\href {\doibase 10.1088/1742-5468/aaac4a} {\bibfield  {journal} {\bibinfo
  {journal} {J. Stat. Mech.}\ }\textbf {\bibinfo {volume} {1803}},\ \bibinfo
  {pages} {033207} (\bibinfo {year} {2018})},\ \Eprint
  {http://arxiv.org/abs/1707.06447} {arXiv:1707.06447 [hep-th]} \BibitemShut
  {NoStop}%
\bibitem [{\citenamefont {Wellenhofer}\ \emph {et~al.}(2015)\citenamefont
  {Wellenhofer}, \citenamefont {Holt},\ and\ \citenamefont
  {Kaiser}}]{Wellenhofer:2015qba}%
  \BibitemOpen
  \bibfield  {author} {\bibinfo {author} {\bibfnamefont {C.}~\bibnamefont
  {Wellenhofer}}, \bibinfo {author} {\bibfnamefont {J.~W.}\ \bibnamefont
  {Holt}}, \ and\ \bibinfo {author} {\bibfnamefont {N.}~\bibnamefont
  {Kaiser}},\ }\href {\doibase 10.1103/PhysRevC.92.015801} {\bibfield
  {journal} {\bibinfo  {journal} {Phys. Rev. C}\ }\textbf {\bibinfo {volume}
  {92}},\ \bibinfo {pages} {015801} (\bibinfo {year} {2015})},\ \Eprint
  {http://arxiv.org/abs/1504.00177} {arXiv:1504.00177 [nucl-th]} \BibitemShut
  {NoStop}%
\bibitem [{\citenamefont {Kapusta}\ \emph {et~al.}(2024)\citenamefont
  {Kapusta}, \citenamefont {Singh},\ and\ \citenamefont
  {Welle}}]{Kapusta:2024nii}%
  \BibitemOpen
  \bibfield  {author} {\bibinfo {author} {\bibfnamefont {J.~I.}\ \bibnamefont
  {Kapusta}}, \bibinfo {author} {\bibfnamefont {M.}~\bibnamefont {Singh}}, \
  and\ \bibinfo {author} {\bibfnamefont {T.}~\bibnamefont {Welle}},\
  }\href@noop {} {\  (\bibinfo {year} {2024})},\ \Eprint
  {http://arxiv.org/abs/2407.16963} {arXiv:2407.16963 [hep-ph]} \BibitemShut
  {NoStop}%
\bibitem [{\citenamefont {Fisher}(1967)}]{Fisher:1967}%
  \BibitemOpen
  \bibfield  {author} {\bibinfo {author} {\bibfnamefont {M.~E.}\ \bibnamefont
  {Fisher}},\ }\href {\doibase 10.1088/0034-4885/30/2/306} {\bibfield
  {journal} {\bibinfo  {journal} {Rept. Prog. Phys}\ }\textbf {\bibinfo
  {volume} {30}},\ \bibinfo {pages} {615} (\bibinfo {year} {1967})}\BibitemShut
  {NoStop}%
\bibitem [{\citenamefont {{Binder}}(1987)}]{Binder1987}%
  \BibitemOpen
  \bibfield  {author} {\bibinfo {author} {\bibfnamefont {K.}~\bibnamefont
  {{Binder}}},\ }\href {\doibase 10.1088/0034-4885/50/7/001} {\bibfield
  {journal} {\bibinfo  {journal} {Rept. Prog. Phys}\ }\textbf {\bibinfo
  {volume} {50}},\ \bibinfo {pages} {783} (\bibinfo {year} {1987})}\BibitemShut
  {NoStop}%
\bibitem [{\citenamefont {Schofield}(1969)}]{Schofield:1969zza}%
  \BibitemOpen
  \bibfield  {author} {\bibinfo {author} {\bibfnamefont {P.}~\bibnamefont
  {Schofield}},\ }\href {\doibase 10.1103/PhysRevLett.22.606} {\bibfield
  {journal} {\bibinfo  {journal} {Phys. Rev. Lett.}\ }\textbf {\bibinfo
  {volume} {22}},\ \bibinfo {pages} {606} (\bibinfo {year} {1969})}\BibitemShut
  {NoStop}%
\bibitem [{\citenamefont {Guida}\ and\ \citenamefont
  {Zinn-Justin}(1997)}]{Guida:1996ep}%
  \BibitemOpen
  \bibfield  {author} {\bibinfo {author} {\bibfnamefont {R.}~\bibnamefont
  {Guida}}\ and\ \bibinfo {author} {\bibfnamefont {J.}~\bibnamefont
  {Zinn-Justin}},\ }\href {\doibase 10.1016/S0550-3213(96)00704-3} {\bibfield
  {journal} {\bibinfo  {journal} {Nucl. Phys. B}\ }\textbf {\bibinfo {volume}
  {489}},\ \bibinfo {pages} {626} (\bibinfo {year} {1997})},\ \Eprint
  {http://arxiv.org/abs/hep-th/9610223} {arXiv:hep-th/9610223} \BibitemShut
  {NoStop}%
\bibitem [{\citenamefont {Landau}\ and\ \citenamefont
  {Lifshitz}(1980)}]{LandauLifshitz}%
  \BibitemOpen
  \bibfield  {author} {\bibinfo {author} {\bibfnamefont {L.}~\bibnamefont
  {Landau}}\ and\ \bibinfo {author} {\bibfnamefont {E.}~\bibnamefont
  {Lifshitz}},\ }in\ \href@noop {} {\emph {\bibinfo {booktitle} {{Statistical
  Physics, Part 1, Vol. 5}}}},\ Vol.~\bibinfo {volume} {5}\ (\bibinfo {year}
  {1980})\BibitemShut {NoStop}%
\bibitem [{\citenamefont {Bellwied}\ \emph {et~al.}(2015)\citenamefont
  {Bellwied}, \citenamefont {Borsanyi}, \citenamefont {Fodor}, \citenamefont
  {G\"unther}, \citenamefont {Katz}, \citenamefont {Ratti},\ and\ \citenamefont
  {Szabo}}]{Bellwied:2015rza}%
  \BibitemOpen
  \bibfield  {author} {\bibinfo {author} {\bibfnamefont {R.}~\bibnamefont
  {Bellwied}}, \bibinfo {author} {\bibfnamefont {S.}~\bibnamefont {Borsanyi}},
  \bibinfo {author} {\bibfnamefont {Z.}~\bibnamefont {Fodor}}, \bibinfo
  {author} {\bibfnamefont {J.}~\bibnamefont {G\"unther}}, \bibinfo {author}
  {\bibfnamefont {S.~D.}\ \bibnamefont {Katz}}, \bibinfo {author}
  {\bibfnamefont {C.}~\bibnamefont {Ratti}}, \ and\ \bibinfo {author}
  {\bibfnamefont {K.~K.}\ \bibnamefont {Szabo}},\ }\href {\doibase
  10.1016/j.physletb.2015.11.011} {\bibfield  {journal} {\bibinfo  {journal}
  {Phys. Lett. B}\ }\textbf {\bibinfo {volume} {751}},\ \bibinfo {pages} {559}
  (\bibinfo {year} {2015})},\ \Eprint {http://arxiv.org/abs/1507.07510}
  {arXiv:1507.07510 [hep-lat]} \BibitemShut {NoStop}%
\bibitem [{\citenamefont {Randrup}(2010)}]{Randrup:2010ax}%
  \BibitemOpen
  \bibfield  {author} {\bibinfo {author} {\bibfnamefont {J.}~\bibnamefont
  {Randrup}},\ }\href {\doibase 10.1103/PhysRevC.82.034902} {\bibfield
  {journal} {\bibinfo  {journal} {Phys. Rev. C}\ }\textbf {\bibinfo {volume}
  {82}},\ \bibinfo {pages} {034902} (\bibinfo {year} {2010})},\ \Eprint
  {http://arxiv.org/abs/1007.1448} {arXiv:1007.1448 [nucl-th]} \BibitemShut
  {NoStop}%
\bibitem [{\citenamefont {Vovchenko}\ \emph {et~al.}(2015)\citenamefont
  {Vovchenko}, \citenamefont {Anchishkin},\ and\ \citenamefont
  {Gorenstein}}]{Vovchenko:2015vxa}%
  \BibitemOpen
  \bibfield  {author} {\bibinfo {author} {\bibfnamefont {V.}~\bibnamefont
  {Vovchenko}}, \bibinfo {author} {\bibfnamefont {D.~V.}\ \bibnamefont
  {Anchishkin}}, \ and\ \bibinfo {author} {\bibfnamefont {M.~I.}\ \bibnamefont
  {Gorenstein}},\ }\href {\doibase 10.1103/PhysRevC.91.064314} {\bibfield
  {journal} {\bibinfo  {journal} {Phys. Rev. C}\ }\textbf {\bibinfo {volume}
  {91}},\ \bibinfo {pages} {064314} (\bibinfo {year} {2015})},\ \Eprint
  {http://arxiv.org/abs/1504.01363} {arXiv:1504.01363 [nucl-th]} \BibitemShut
  {NoStop}%
\bibitem [{\citenamefont {Cahn}\ and\ \citenamefont
  {Hilliard}(1959)}]{Cahn1959}%
  \BibitemOpen
  \bibfield  {author} {\bibinfo {author} {\bibfnamefont {J.~W.}\ \bibnamefont
  {Cahn}}\ and\ \bibinfo {author} {\bibfnamefont {J.~E.}\ \bibnamefont
  {Hilliard}},\ }\href@noop {} {\bibfield  {journal} {\bibinfo  {journal} {J.
  Chem. Phys.}\ }\textbf {\bibinfo {volume} {31}},\ \bibinfo {pages} {688}
  (\bibinfo {year} {1959})}\BibitemShut {NoStop}%
\bibitem [{\citenamefont {Abyzov}\ and\ \citenamefont
  {Schmelzer}(2007)}]{Abyzov2007}%
  \BibitemOpen
  \bibfield  {author} {\bibinfo {author} {\bibfnamefont {A.~S.}\ \bibnamefont
  {Abyzov}}\ and\ \bibinfo {author} {\bibfnamefont {J.~W.~P.}\ \bibnamefont
  {Schmelzer}},\ }\href {\doibase 10.1063/1.2774989} {\bibfield  {journal}
  {\bibinfo  {journal} {The Journal of Chemical Physics}\ }\textbf {\bibinfo
  {volume} {127}},\ \bibinfo {pages} {114504} (\bibinfo {year} {2007})},\
  \Eprint
  {http://arxiv.org/abs/https://pubs.aip.org/aip/jcp/article-pdf/doi/10.1063/1.2774989/14785303/114504\_1\_online.pdf}
  {https://pubs.aip.org/aip/jcp/article-pdf/doi/10.1063/1.2774989/14785303/114504\_1\_online.pdf}
  \BibitemShut {NoStop}%
\bibitem [{\citenamefont {Mishustin}(1999)}]{Mishustin:1998eq}%
  \BibitemOpen
  \bibfield  {author} {\bibinfo {author} {\bibfnamefont {I.~N.}\ \bibnamefont
  {Mishustin}},\ }\href {\doibase 10.1103/PhysRevLett.82.4779} {\bibfield
  {journal} {\bibinfo  {journal} {Phys. Rev. Lett.}\ }\textbf {\bibinfo
  {volume} {82}},\ \bibinfo {pages} {4779} (\bibinfo {year} {1999})},\ \Eprint
  {http://arxiv.org/abs/hep-ph/9811307} {arXiv:hep-ph/9811307} \BibitemShut
  {NoStop}%
\bibitem [{\citenamefont {Csernai}\ and\ \citenamefont
  {Kapusta}(1992)}]{Csernai:1992as}%
  \BibitemOpen
  \bibfield  {author} {\bibinfo {author} {\bibfnamefont {L.~P.}\ \bibnamefont
  {Csernai}}\ and\ \bibinfo {author} {\bibfnamefont {J.~I.}\ \bibnamefont
  {Kapusta}},\ }\href {\doibase 10.1103/PhysRevLett.69.737} {\bibfield
  {journal} {\bibinfo  {journal} {Phys. Rev. Lett.}\ }\textbf {\bibinfo
  {volume} {69}},\ \bibinfo {pages} {737} (\bibinfo {year} {1992})}\BibitemShut
  {NoStop}%
\bibitem [{\citenamefont {Langer}(2000)}]{Langer2000}%
  \BibitemOpen
  \bibfield  {author} {\bibinfo {author} {\bibfnamefont {J.~S.}\ \bibnamefont
  {Langer}},\ }\href@noop {} {\bibfield  {journal} {\bibinfo  {journal} {Annals
  of Physics}\ }\textbf {\bibinfo {volume} {281}},\ \bibinfo {pages} {941}
  (\bibinfo {year} {2000})}\BibitemShut {NoStop}%
\bibitem [{\citenamefont {Lee}\ and\ \citenamefont {Yang}(1952)}]{Lee1952}%
  \BibitemOpen
  \bibfield  {author} {\bibinfo {author} {\bibfnamefont {T.-D.}\ \bibnamefont
  {Lee}}\ and\ \bibinfo {author} {\bibfnamefont {C.-N.}\ \bibnamefont {Yang}},\
  }\href@noop {} {\bibfield  {journal} {\bibinfo  {journal} {Physical Review}\
  }\textbf {\bibinfo {volume} {87}},\ \bibinfo {pages} {410} (\bibinfo {year}
  {1952})}\BibitemShut {NoStop}%
\bibitem [{\citenamefont {Jones}(2002)}]{Jones2002}%
  \BibitemOpen
  \bibfield  {author} {\bibinfo {author} {\bibfnamefont {R.}~\bibnamefont
  {Jones}},\ }\href {https://books.google.com/books?id=Hl_HBPUvoNsC} {\emph
  {\bibinfo {title} {Soft Condensed Matter}}},\ Oxford Master Series in
  Physics\ (\bibinfo  {publisher} {OUP Oxford},\ \bibinfo {year}
  {2002})\BibitemShut {NoStop}%
\bibitem [{\citenamefont {Ginzburg}(1960)}]{Ginzburg1960}%
  \BibitemOpen
  \bibfield  {author} {\bibinfo {author} {\bibfnamefont {V.~L.}\ \bibnamefont
  {Ginzburg}},\ }\href@noop {} {\bibfield  {journal} {\bibinfo  {journal} {Sov.
  Phys. Solid State}\ ,\ \bibinfo {pages} {1824}} (\bibinfo {year}
  {1960})}\BibitemShut {NoStop}%
\bibitem [{\citenamefont {Als-Nielsen}\ and\ \citenamefont
  {Birgeneau}(1977)}]{Als-Nielsen1977}%
  \BibitemOpen
  \bibfield  {author} {\bibinfo {author} {\bibfnamefont {J.}~\bibnamefont
  {Als-Nielsen}}\ and\ \bibinfo {author} {\bibfnamefont {R.~J.}\ \bibnamefont
  {Birgeneau}},\ }\href@noop {} {\bibfield  {journal} {\bibinfo  {journal} {Am.
  J. Phys}\ }\textbf {\bibinfo {volume} {45}},\ \bibinfo {pages} {87} (\bibinfo
  {year} {1977})}\BibitemShut {NoStop}%
\bibitem [{\citenamefont {Zinn-Justin}(2002)}]{ZinnJustin}%
  \BibitemOpen
  \bibfield  {author} {\bibinfo {author} {\bibfnamefont {J.}~\bibnamefont
  {Zinn-Justin}},\ }\href {\doibase 10.1093/acprof:oso/9780198509233.001.0001}
  {\emph {\bibinfo {title} {{Quantum Field Theory and Critical Phenomena}}}}\
  (\bibinfo  {publisher} {Oxford University Press},\ \bibinfo {year}
  {2002})\BibitemShut {NoStop}%
\bibitem [{\citenamefont {Berdnikov}\ and\ \citenamefont
  {Rajagopal}(2000)}]{Berdnikov:1999ph}%
  \BibitemOpen
  \bibfield  {author} {\bibinfo {author} {\bibfnamefont {B.}~\bibnamefont
  {Berdnikov}}\ and\ \bibinfo {author} {\bibfnamefont {K.}~\bibnamefont
  {Rajagopal}},\ }\href {\doibase 10.1103/PhysRevD.61.105017} {\bibfield
  {journal} {\bibinfo  {journal} {Phys. Rev. D}\ }\textbf {\bibinfo {volume}
  {61}},\ \bibinfo {pages} {105017} (\bibinfo {year} {2000})},\ \Eprint
  {http://arxiv.org/abs/hep-ph/9912274} {arXiv:hep-ph/9912274} \BibitemShut
  {NoStop}%
\bibitem [{\citenamefont {Pradeep}\ and\ \citenamefont
  {Stephanov}(2019)}]{Pradeep:2019ccv}%
  \BibitemOpen
  \bibfield  {author} {\bibinfo {author} {\bibfnamefont {M.~S.}\ \bibnamefont
  {Pradeep}}\ and\ \bibinfo {author} {\bibfnamefont {M.}~\bibnamefont
  {Stephanov}},\ }\href {\doibase 10.1103/PhysRevD.100.056003} {\bibfield
  {journal} {\bibinfo  {journal} {Phys. Rev. D}\ }\textbf {\bibinfo {volume}
  {100}},\ \bibinfo {pages} {056003} (\bibinfo {year} {2019})},\ \Eprint
  {http://arxiv.org/abs/1905.13247} {arXiv:1905.13247 [hep-ph]} \BibitemShut
  {NoStop}%
\bibitem [{\citenamefont {Mroczek}\ \emph {et~al.}(2022)\citenamefont
  {Mroczek}, \citenamefont {Hjorth-Jensen}, \citenamefont {Noronha-Hostler},
  \citenamefont {Parotto}, \citenamefont {Ratti},\ and\ \citenamefont
  {Vilalta}}]{Mroczek:2022oga}%
  \BibitemOpen
  \bibfield  {author} {\bibinfo {author} {\bibfnamefont {D.}~\bibnamefont
  {Mroczek}}, \bibinfo {author} {\bibfnamefont {M.}~\bibnamefont
  {Hjorth-Jensen}}, \bibinfo {author} {\bibfnamefont {J.}~\bibnamefont
  {Noronha-Hostler}}, \bibinfo {author} {\bibfnamefont {P.}~\bibnamefont
  {Parotto}}, \bibinfo {author} {\bibfnamefont {C.}~\bibnamefont {Ratti}}, \
  and\ \bibinfo {author} {\bibfnamefont {R.}~\bibnamefont {Vilalta}},\
  }\href@noop {} {\  (\bibinfo {year} {2022})},\ \Eprint
  {http://arxiv.org/abs/2203.13876} {arXiv:2203.13876 [nucl-th]} \BibitemShut
  {NoStop}%
\bibitem [{\citenamefont {Pradeep}\ \emph {et~al.}(2024)\citenamefont
  {Pradeep}, \citenamefont {Sogabe}, \citenamefont {Stephanov},\ and\
  \citenamefont {Yee}}]{Pradeep:2024cca}%
  \BibitemOpen
  \bibfield  {author} {\bibinfo {author} {\bibfnamefont {M.~S.}\ \bibnamefont
  {Pradeep}}, \bibinfo {author} {\bibfnamefont {N.}~\bibnamefont {Sogabe}},
  \bibinfo {author} {\bibfnamefont {M.}~\bibnamefont {Stephanov}}, \ and\
  \bibinfo {author} {\bibfnamefont {H.-U.}\ \bibnamefont {Yee}},\ }\href@noop
  {} {\  (\bibinfo {year} {2024})},\ \Eprint {http://arxiv.org/abs/2402.09519}
  {arXiv:2402.09519 [nucl-th]} \BibitemShut {NoStop}%
\bibitem [{\citenamefont {Bonati}\ \emph {et~al.}(2015)\citenamefont {Bonati},
  \citenamefont {D'Elia}, \citenamefont {Mariti}, \citenamefont {Mesiti},
  \citenamefont {Negro},\ and\ \citenamefont {Sanfilippo}}]{Bonati:2015bha}%
  \BibitemOpen
  \bibfield  {author} {\bibinfo {author} {\bibfnamefont {C.}~\bibnamefont
  {Bonati}}, \bibinfo {author} {\bibfnamefont {M.}~\bibnamefont {D'Elia}},
  \bibinfo {author} {\bibfnamefont {M.}~\bibnamefont {Mariti}}, \bibinfo
  {author} {\bibfnamefont {M.}~\bibnamefont {Mesiti}}, \bibinfo {author}
  {\bibfnamefont {F.}~\bibnamefont {Negro}}, \ and\ \bibinfo {author}
  {\bibfnamefont {F.}~\bibnamefont {Sanfilippo}},\ }\href {\doibase
  10.1103/PhysRevD.92.054503} {\bibfield  {journal} {\bibinfo  {journal} {Phys.
  Rev. D}\ }\textbf {\bibinfo {volume} {92}},\ \bibinfo {pages} {054503}
  (\bibinfo {year} {2015})},\ \Eprint {http://arxiv.org/abs/1507.03571}
  {arXiv:1507.03571 [hep-lat]} \BibitemShut {NoStop}%
\bibitem [{\citenamefont {Bonati}\ \emph {et~al.}(2018)\citenamefont {Bonati},
  \citenamefont {D'Elia}, \citenamefont {Negro}, \citenamefont {Sanfilippo},\
  and\ \citenamefont {Zambello}}]{Bonati:2018nut}%
  \BibitemOpen
  \bibfield  {author} {\bibinfo {author} {\bibfnamefont {C.}~\bibnamefont
  {Bonati}}, \bibinfo {author} {\bibfnamefont {M.}~\bibnamefont {D'Elia}},
  \bibinfo {author} {\bibfnamefont {F.}~\bibnamefont {Negro}}, \bibinfo
  {author} {\bibfnamefont {F.}~\bibnamefont {Sanfilippo}}, \ and\ \bibinfo
  {author} {\bibfnamefont {K.}~\bibnamefont {Zambello}},\ }\href {\doibase
  10.1103/PhysRevD.98.054510} {\bibfield  {journal} {\bibinfo  {journal} {Phys.
  Rev. D}\ }\textbf {\bibinfo {volume} {98}},\ \bibinfo {pages} {054510}
  (\bibinfo {year} {2018})},\ \Eprint {http://arxiv.org/abs/1805.02960}
  {arXiv:1805.02960 [hep-lat]} \BibitemShut {NoStop}%
\bibitem [{\citenamefont {Bazavov}\ \emph {et~al.}(2019)\citenamefont {Bazavov}
  \emph {et~al.}}]{Bazavov:2018mes}%
  \BibitemOpen
  \bibfield  {author} {\bibinfo {author} {\bibfnamefont {A.}~\bibnamefont
  {Bazavov}} \emph {et~al.} (\bibinfo {collaboration} {HotQCD}),\ }\href
  {\doibase 10.1016/j.physletb.2019.05.013} {\bibfield  {journal} {\bibinfo
  {journal} {Phys. Lett. B}\ }\textbf {\bibinfo {volume} {795}},\ \bibinfo
  {pages} {15} (\bibinfo {year} {2019})},\ \Eprint
  {http://arxiv.org/abs/1812.08235} {arXiv:1812.08235 [hep-lat]} \BibitemShut
  {NoStop}%
\bibitem [{\citenamefont {Borsanyi}\ \emph {et~al.}(2020)\citenamefont
  {Borsanyi}, \citenamefont {Fodor}, \citenamefont {Guenther}, \citenamefont
  {Kara}, \citenamefont {Katz}, \citenamefont {Parotto}, \citenamefont
  {Pasztor}, \citenamefont {Ratti},\ and\ \citenamefont
  {Szabo}}]{Borsanyi:2020fev}%
  \BibitemOpen
  \bibfield  {author} {\bibinfo {author} {\bibfnamefont {S.}~\bibnamefont
  {Borsanyi}}, \bibinfo {author} {\bibfnamefont {Z.}~\bibnamefont {Fodor}},
  \bibinfo {author} {\bibfnamefont {J.~N.}\ \bibnamefont {Guenther}}, \bibinfo
  {author} {\bibfnamefont {R.}~\bibnamefont {Kara}}, \bibinfo {author}
  {\bibfnamefont {S.~D.}\ \bibnamefont {Katz}}, \bibinfo {author}
  {\bibfnamefont {P.}~\bibnamefont {Parotto}}, \bibinfo {author} {\bibfnamefont
  {A.}~\bibnamefont {Pasztor}}, \bibinfo {author} {\bibfnamefont
  {C.}~\bibnamefont {Ratti}}, \ and\ \bibinfo {author} {\bibfnamefont {K.~K.}\
  \bibnamefont {Szabo}},\ }\href {\doibase 10.1103/PhysRevLett.125.052001}
  {\bibfield  {journal} {\bibinfo  {journal} {Phys. Rev. Lett.}\ }\textbf
  {\bibinfo {volume} {125}},\ \bibinfo {pages} {052001} (\bibinfo {year}
  {2020})},\ \Eprint {http://arxiv.org/abs/2002.02821} {arXiv:2002.02821
  [hep-lat]} \BibitemShut {NoStop}%
\bibitem [{\citenamefont {Halasz}\ \emph {et~al.}(1998)\citenamefont {Halasz},
  \citenamefont {Jackson}, \citenamefont {Shrock}, \citenamefont {Stephanov},\
  and\ \citenamefont {Verbaarschot}}]{Halasz:1998qr}%
  \BibitemOpen
  \bibfield  {author} {\bibinfo {author} {\bibfnamefont {A.~M.}\ \bibnamefont
  {Halasz}}, \bibinfo {author} {\bibfnamefont {A.~D.}\ \bibnamefont {Jackson}},
  \bibinfo {author} {\bibfnamefont {R.~E.}\ \bibnamefont {Shrock}}, \bibinfo
  {author} {\bibfnamefont {M.~A.}\ \bibnamefont {Stephanov}}, \ and\ \bibinfo
  {author} {\bibfnamefont {J.~J.~M.}\ \bibnamefont {Verbaarschot}},\ }\href
  {\doibase 10.1103/PhysRevD.58.096007} {\bibfield  {journal} {\bibinfo
  {journal} {Phys. Rev. D}\ }\textbf {\bibinfo {volume} {58}},\ \bibinfo
  {pages} {096007} (\bibinfo {year} {1998})},\ \Eprint
  {http://arxiv.org/abs/hep-ph/9804290} {arXiv:hep-ph/9804290} \BibitemShut
  {NoStop}%
\bibitem [{\citenamefont {Grefa}\ \emph {et~al.}(2021)\citenamefont {Grefa},
  \citenamefont {Noronha}, \citenamefont {Noronha-Hostler}, \citenamefont
  {Portillo}, \citenamefont {Ratti},\ and\ \citenamefont
  {Rougemont}}]{Grefa:2021qvt}%
  \BibitemOpen
  \bibfield  {author} {\bibinfo {author} {\bibfnamefont {J.}~\bibnamefont
  {Grefa}}, \bibinfo {author} {\bibfnamefont {J.}~\bibnamefont {Noronha}},
  \bibinfo {author} {\bibfnamefont {J.}~\bibnamefont {Noronha-Hostler}},
  \bibinfo {author} {\bibfnamefont {I.}~\bibnamefont {Portillo}}, \bibinfo
  {author} {\bibfnamefont {C.}~\bibnamefont {Ratti}}, \ and\ \bibinfo {author}
  {\bibfnamefont {R.}~\bibnamefont {Rougemont}},\ }\href@noop {} {\  (\bibinfo
  {year} {2021})},\ \Eprint {http://arxiv.org/abs/2102.12042} {arXiv:2102.12042
  [nucl-th]} \BibitemShut {NoStop}%
\bibitem [{\citenamefont {Hippert}\ \emph {et~al.}(2023)\citenamefont
  {Hippert}, \citenamefont {Grefa}, \citenamefont {Manning}, \citenamefont
  {Noronha}, \citenamefont {Noronha-Hostler}, \citenamefont {Portillo~Vazquez},
  \citenamefont {Ratti}, \citenamefont {Rougemont},\ and\ \citenamefont
  {Trujillo}}]{Hippert:2023bel}%
  \BibitemOpen
  \bibfield  {author} {\bibinfo {author} {\bibfnamefont {M.}~\bibnamefont
  {Hippert}}, \bibinfo {author} {\bibfnamefont {J.}~\bibnamefont {Grefa}},
  \bibinfo {author} {\bibfnamefont {T.~A.}\ \bibnamefont {Manning}}, \bibinfo
  {author} {\bibfnamefont {J.}~\bibnamefont {Noronha}}, \bibinfo {author}
  {\bibfnamefont {J.}~\bibnamefont {Noronha-Hostler}}, \bibinfo {author}
  {\bibfnamefont {I.}~\bibnamefont {Portillo~Vazquez}}, \bibinfo {author}
  {\bibfnamefont {C.}~\bibnamefont {Ratti}}, \bibinfo {author} {\bibfnamefont
  {R.}~\bibnamefont {Rougemont}}, \ and\ \bibinfo {author} {\bibfnamefont
  {M.}~\bibnamefont {Trujillo}},\ }\href@noop {} {\  (\bibinfo {year}
  {2023})},\ \Eprint {http://arxiv.org/abs/2309.00579} {arXiv:2309.00579
  [nucl-th]} \BibitemShut {NoStop}%
\bibitem [{\citenamefont {DeWolfe}\ \emph {et~al.}(2011)\citenamefont
  {DeWolfe}, \citenamefont {Gubser},\ and\ \citenamefont
  {Rosen}}]{DeWolfe:2010he}%
  \BibitemOpen
  \bibfield  {author} {\bibinfo {author} {\bibfnamefont {O.}~\bibnamefont
  {DeWolfe}}, \bibinfo {author} {\bibfnamefont {S.~S.}\ \bibnamefont {Gubser}},
  \ and\ \bibinfo {author} {\bibfnamefont {C.}~\bibnamefont {Rosen}},\ }\href
  {\doibase 10.1103/PhysRevD.83.086005} {\bibfield  {journal} {\bibinfo
  {journal} {Phys. Rev. D}\ }\textbf {\bibinfo {volume} {83}},\ \bibinfo
  {pages} {086005} (\bibinfo {year} {2011})},\ \Eprint
  {http://arxiv.org/abs/1012.1864} {arXiv:1012.1864 [hep-th]} \BibitemShut
  {NoStop}%
\bibitem [{\citenamefont {Guenther}\ \emph {et~al.}(2017)\citenamefont
  {Guenther}, \citenamefont {Bellwied}, \citenamefont {Borsanyi}, \citenamefont
  {Fodor}, \citenamefont {Katz}, \citenamefont {Pasztor}, \citenamefont
  {Ratti},\ and\ \citenamefont {Szab\'o}}]{Guenther:2017hnx}%
  \BibitemOpen
  \bibfield  {author} {\bibinfo {author} {\bibfnamefont {J.~N.}\ \bibnamefont
  {Guenther}}, \bibinfo {author} {\bibfnamefont {R.}~\bibnamefont {Bellwied}},
  \bibinfo {author} {\bibfnamefont {S.}~\bibnamefont {Borsanyi}}, \bibinfo
  {author} {\bibfnamefont {Z.}~\bibnamefont {Fodor}}, \bibinfo {author}
  {\bibfnamefont {S.~D.}\ \bibnamefont {Katz}}, \bibinfo {author}
  {\bibfnamefont {A.}~\bibnamefont {Pasztor}}, \bibinfo {author} {\bibfnamefont
  {C.}~\bibnamefont {Ratti}}, \ and\ \bibinfo {author} {\bibfnamefont {K.~K.}\
  \bibnamefont {Szab\'o}},\ }\href {\doibase 10.1016/j.nuclphysa.2017.05.044}
  {\bibfield  {journal} {\bibinfo  {journal} {Nucl. Phys. A}\ }\textbf
  {\bibinfo {volume} {967}},\ \bibinfo {pages} {720} (\bibinfo {year}
  {2017})},\ \Eprint {http://arxiv.org/abs/1607.02493} {arXiv:1607.02493
  [hep-lat]} \BibitemShut {NoStop}%
\end{thebibliography}%

\end{document}